\documentclass[12pt]{article}
\pdfoutput=1

\usepackage{color}
\usepackage{amssymb,amsmath,bm}
\usepackage{epsf}
\usepackage{epsfig}
\usepackage{afterpage}
\usepackage{longtable}
\usepackage[dvipsnames]{xcolor}
\usepackage[linktoc=page,bookmarks=false,colorlinks=false,linkbordercolor=RoyalBlue,citebordercolor=ForestGreen,urlbordercolor=CornflowerBlue]{hyperref}
\usepackage{latexsym,mathrsfs}
\usepackage[normalem]{ulem} % for strikeout \sout
\usepackage[compress]{cite}
\usepackage{graphicx}
\usepackage{url}
\usepackage{paralist}
\usepackage{slashed}
\usepackage{multirow}

\usepackage[hypcap]{caption, subcaption}

\usepackage{booktabs}
\usepackage{listings}
\usepackage{textcomp}

\usepackage[titles]{tocloft}
\renewcommand{\arraystretch}{1.25}

\allowdisplaybreaks[1]

\numberwithin{equation}{section}
\def \refeq#1{(\ref{#1})}
\def \refsec#1{Section~\ref{#1}}
\def \refSec#1{Section~\ref{#1}}
\def \refapp#1{Appendix~\ref{#1}}
\def \reffig#1{Figure~\ref{#1}}
\def \reftab#1{Table~\ref{#1}}

\def\Slash#1{#1 \hskip-0.59em /}
\def\slash2#1{#1 \hskip-0.45em /}

\DeclareMathOperator{\re}{Re}
\DeclareMathOperator{\im}{Im}

\definecolor{darkgreen}{rgb}{0.0,0.6,0.0}

\newcounter{MBQ}

\newcounter{RSQ}

\newcommand{\calA}{\mathcal{A}}
\newcommand{\calN}{\mathcal{N}}
\newcommand{\calO}{\mathcal{O}}

\newcommand{\Op}{{Q}}
\newcommand{\OpI}{\mathcal{O}}
\newcommand{\OpII}{\mathcal{J}}

\newcommand{\EWnorm}{\mathcal{N}_{\Delta B = 1}}

\newcommand{\muEW}{\mu_W}
\newcommand{\mub}{\mu_b}

\newcommand{\alS}{\alpha_s}
\newcommand{\alE}{\alpha_\text{em}}
\newcommand{\LambQCD}{\Lambda_\text{QCD}}

\newcommand \oL[1]{{\overline{#1}}}

\newcommand{\eps}{\eps}
\newcommand \lamB{\lambda}

\newcommand{\SCETI}{SCET$_\text{I}$}
\newcommand{\SCETII}{SCET$_\text{II}$}

\def \np{{n_+}}
\def \nm{{n_-}}
\def \vb{{v}}
\newcommand{\nms}{\slashed{n}_-}
\newcommand{\nps}{\slashed{n}_+}

\begin{document}

\allowdisplaybreaks

\begin{titlepage}

\begin{flushright}
{\small
TUM-HEP-1212/19 \\
arXiv:1907.07011\\
November 20, 2022
}
\end{flushright}

\vspace{1cm}
\begin{center}
{\Large\bf\boldmath
  Power-enhanced leading-logarithmic QED \\[0.2cm] 
  corrections to $B_q \to \mu^+\mu^-$
}
\\[8mm]
{\sc 
  Martin~Beneke$^{a}$, 
  Christoph~Bobeth$^{a}$ and 
  Robert~Szafron$^{a}$
}
\\[0.6cm]

$^{a}${\it Physik Department T31,\\
James-Franck-Stra\ss e~1, 
Technische Universit\"at M\"unchen,\\
D--85748 Garching, Germany\\[0.2cm]}
\end{center}

\vspace{0.6cm}
\begin{abstract}
\vskip0.2cm\noindent
We provide a systematic treatment of the previously discovered power-enhanced QED
corrections to the leptonic decay $B_q\to \mu^+\mu^-$ ($q = d, s$) in the framework
of soft-collinear effective theory (SCET).  Employing two-step matching on \SCETI{}
and \SCETII{}, and the respective renormalization group equations, we sum the
leading-logarithmic QED corrections and the mixed QED--QCD corrections 
to all orders in the couplings for the matrix element of the semileptonic weak
effective operator~$\Op_9$. We propose a treatment of the $B$-meson decay constant
and light-cone distribution amplitude in the presence of process-specific QED 
corrections. Finally we include ultrasoft photon radiation and provide updated
values of the non-radiative and radiative branching fractions of $B_q \to \mu^+\mu^-$
decay that include the double-logarithmic QED and QCD corrections.
\end{abstract}
\end{titlepage}

\setcounter{page}{0}
\thispagestyle{empty}
\newpage

{%\small

% suppress subsections in table of contents
\setcounter{tocdepth}{2}

\tableofcontents

}

%\newpage

%--------+---------+---------+---------+---------+---------+---------+---------+
%
%
%
%--------+---------+---------+---------+---------+---------+---------+---------+

\section{Introduction}

The purely leptonic $B$-meson decays $B_u \to \ell\bar\nu_\ell$ and $B_{d,s} \to
\ell^+\ell^-$ ($\ell = e, \mu, \tau$) are among the most valuable probes of the
quark-mixing parameters in the Standard Model (SM), namely the parameters of the
Cabibbo-Kobayashi-Maskawa (CKM) matrix. The charged-current mediated tree-level
decays $B_u \to \ell\bar\nu_\ell$ give direct access to the CKM element $|V_{ub}|$,
whereas the flavour-changing neutral-current-mediated $B_{d,s} \to \ell^+\ell^-$
decays allow to determine the combination $|V_{tb}^{} V_{td,ts}^*|$ up to a 
perturbatively calculable short-distance factor that depends on the top-quark
mass \cite{Inami:1980fz}. Moreover the helicity suppression of the decay rate 
leads to a high sensitivity to scalar- and pseudo-scalar interactions beyond
the SM.

Their importance derives from the fact that the nonperturbative hadronic bound-state
effects of the $B_q$ mesons due to the strong interaction~(QCD) appear in theoretical
predictions at leading order (LO) in electromagnetic (QED) interactions only in 
the form of the $B$-meson decay constant $f_{B_q}$. The most recent lattice-QCD 
values of $f_{B_{u,d}}$ and $f_{B_s}$ of the FNAL/MILC Collaboration 
\cite{Bazavov:2017lyh} have now reached the relative precision of about $0.7\,\%$
and $0.5\,\%$, respectively. 
It is expected that this precision will be confirmed by other lattice groups 
and reduced even further in the future thus paving the way to very precise
determinations of CKM parameters in the SM. Such a degree of theoretical control
on the QCD hadronic uncertainties in FCNC flavour physics is currently only
available for $K\to \pi \nu\bar\nu$ decays \cite{Buras:2006gb} and will be 
for the mass differences $\Delta M_q$ in neutral $B$-meson mixing once lattice
calculations achieve the required precision.

Given the small uncertainties due to $f_{B_q}$, it is mandatory to control
all other corrections, which arise from several energy scales spanned by the SM, 
at the percent level. Such control is already achieved for
perturbatively calculable higher-order QCD and electroweak (EW) corrections
in the framework of the effective theory (EFT) of electroweak interactions of
the SM for $\Delta B = 1$ decays \cite{Bobeth:2013uxa}. This comprises $i)$~the 
decoupling of the heavy $W$ and $Z$ bosons and the top quark at the electroweak scale 
$\mu_W \sim m_W$ for $b\to u \ell \bar\nu_\ell$~\cite{Sirlin:1977sv, Sirlin:1981ie,
Marciano:1993sh} and $b\to q \ell^+\ell^-$~\cite{Bobeth:2013tba, Hermann:2013kca}
and $ii)$~the resummation of large logarithms under evolution of QCD and QED down
to the scale $\mu_b \sim m_b$ of the order of the bottom-quark mass using
renormalization-group (RG) improved perturbation theory \cite{Bobeth:2003at,
Huber:2005ig}. 

%%%%%%%%%%%%%%%%%%%%%%%%%%%%%%%%%%%%%%%%%%%%%%%%%%%%%%%%%%%%%%%%%%%%%%%%%%%%%%%%
\begin{figure}
  \centering
  \includegraphics[width=0.95\textwidth]{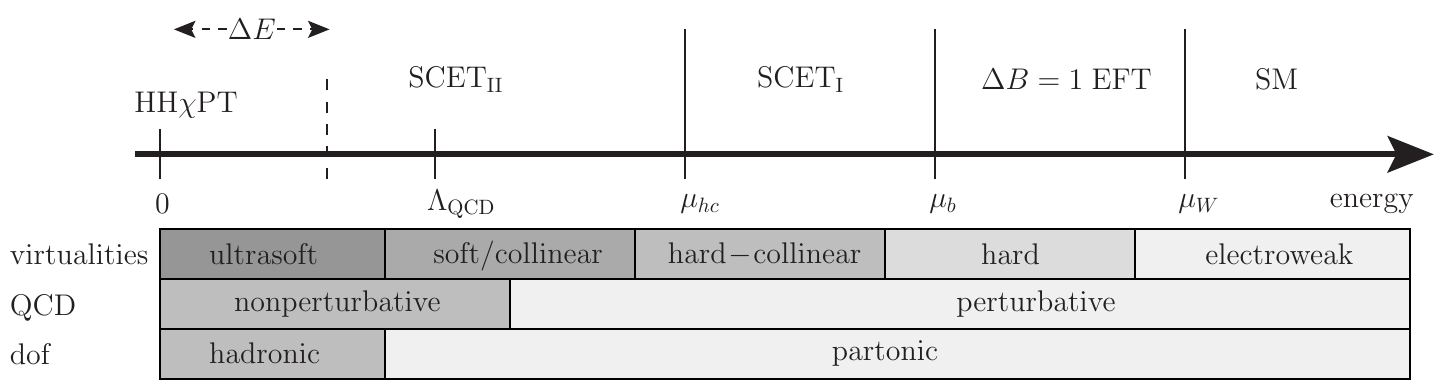}
\caption{\small Scheme of the multiple scales and the respective tower of effective
  theories applicable to $B_q \to \ell^+\ell^-$ transitions and more generally also
  other $b$-hadron decays. See text for more explanations. The range of $\Delta E$
  is indicated for the case $\Delta E \ll \LambQCD$ that we consider here. 
  The degrees of freedom (dof) are hadronic at low energies in HH$\chi$PT. 
}
\label{fig:scale-scheme}
\end{figure}
%%%%%%%%%%%%%%%%%%%%%%%%%%%%%%%%%%%%%%%%%%%%%%%%%%%%%%%%%%%%%%%%%%%%%%%%%%%%%%%%

On the other hand, a consistent simultaneous treatment of QCD and QED corrections is
lacking for scales below $\mu_b$. On general grounds, it is well understood
that only a suitably defined decay rate $\Gamma[B_q \to \ell^+\ell^-] + 
\Gamma[B_q \to \ell^+\ell^- + n\,\gamma(E_\gamma < \Delta E)]$ that
includes real and virtual photon radiation is infrared-finite and well-defined.
It is subject to the experimental setup in the form of a photon-energy
cutoff $\Delta E$ that requires to include in theoretical predictions 
an arbitrary number of additional undetected real photons with energy
$E_\gamma < \Delta E$. The soft-photon emission from the final-state leptons is
currently simulated in experimental analyses \cite{Aaij:2013aka, Chatrchyan:2013bka,
CMS:2014xfa, Aaboud:2016ire, Aaij:2017vad, Aaboud:2018mst} with tools like 
{PHOTOS} \cite{Golonka:2005pn}, such that the measured branching fraction is
interpreted as the non-radiative one \cite{Buras:2012ru}. Further, the
soft initial-state radiation has been estimated to be very small based on 
heavy-hadron chiral perturbation theory (HH$\chi$PT) \cite{Aditya:2012im} provided
$\Delta E \lesssim 60$~MeV. Thus the present knowledge
of QED corrections below the scale $\mu_b$ is restricted to very low
(ultrasoft) scales $\mu_{us} \ll \LambQCD$ below the 
QCD confinement scale, where virtual photons cannot resolve partons in the 
$B_q$ meson. Moreover, it relies entirely on a description in terms of hadronic 
degrees of freedom (i.e. mesons), which, although it permits a perturbative
treatment of QED effects, requires in principle the knowledge of low-energy
constants (LEC). The LECs include the impact of the dynamics above the ultrasoft 
scales, but conceptually little is known of the consistent theoretical treatment 
of the scales up to $\mu_b$ to reliably control the theoretical uncertainties
to the percent level. Although one might work perturbatively in a partonic
picture even below scales $\mu_b$, at least at the (hard-collinear) scale 
$\mu_{hc} \sim 1$~GeV, a nonperturbative regime sets in below $\mu_{hc}$ that 
still requires to use the partonic picture because photons continue to resolve the 
constituents of the hadrons. In the nonperturbative regime, QED corrections need the
evaluation of non-local time-ordered products of the electromagnetic quark
currents. This spoils naive factorization of the QED and QCD effects based
on the soft-photon approximation. A more elaborate treatment based on
effective field theory (EFT) approach is necessary to perform the systematic
expansion of the higher-order QED matrix elements in powers of $\LambQCD/m_b$.   
The theoretical treatment will also depend on the actual magnitude of
$\Delta E$ and its place within the hierarchy of the above scales.
The above discussion is summarized schematically in \reffig{fig:scale-scheme}.
The nonperturbative matching to HH$\chi$PT and hence the hadronic picture at 
very low virtualities is optional if one parameterizes the low-energy 
physics in terms of matrix elements of the previous EFT, \SCETII{}. However, 
in this case the point-like coupling of ultrasoft photons 
to mesons is not manifest.

The first step towards a systematic treatment of QED effects below the scale 
$\mu_b$ has been taken in \cite{Beneke:2017vpq} exploiting the special kinematic 
situation of the $B_q \to \mu^+\mu^-$ decays. The final-state muons are energetic, 
low mass (``collinear'') modes. Their dynamics at scales below $\mu_b$ is 
described by soft-collinear effective theory (SCET). In a two-step decoupling, 
similar to the treatment of QCD effects in heavy-to-light form factors 
and hadronic decays (see, for instance, the review~\cite{Beneke:2015wfa}), first 
hard virtualities $\mathcal{O}(m_b^2)$ and subsequently hard-collinear
virtualities $\mathcal{O}(m_b \Lambda)$ are removed perturbatively to 
arrive at SCET$_{\rm II}$ that describes muons with (collinear or soft) virtualities
of at most $\mathcal{O}(\Lambda^2)$. The scale 
$\Lambda \sim \mathcal{O}(100\,\text{MeV})$ represents 
a typical scale for the muon mass, the spectator quark mass and at the same time
hadronic bound-state effects $\LambQCD$. 

The one-loop calculation of electromagnetic corrections below the scale 
$\mu_b$ performed in \cite{Beneke:2017vpq} resulted in the expression (notation 
explained there)
\begin{equation}
  \label{eq:mainresult}
\begin{aligned}
  i \mathcal{A} & = m_\ell f_{B_q} \calN \, C_{10} \,[\bar{\ell} \gamma_5 \ell]
  + \frac{\alE}{4\pi} Q_\ell Q_q\, 
    m_\ell\, m_{B_q} f_{B_q} \calN \,[\bar{\ell} (1+\gamma_5) \ell]
\\ & \times\,\Bigg\{
  \int_0^1 du \,(1-u)\,C_9^{\rm eff}(u m_b^2)\,
  \int_0^\infty\frac{d\omega}{\omega}\, \phi_{+}(\omega) 
  \left[\ln\frac{m_b\omega}{m_\ell^2}+\ln\frac{u}{1-u}\right]
\\ & \hspace*{0.8cm}
  - \,Q_\ell  C_7^{\rm eff} 
  \int_0^\infty\frac{d\omega}{\omega}\, \phi_{+}(\omega)
  \left[
     \ln^2\frac{m_b\omega}{m_\ell^2}
   - 2\ln\frac{m_b\omega}{m_\ell^2}+\frac{2\pi^2}{3} \right]
  \Bigg\} + \ldots 
\end{aligned}
\end{equation}
for the $B_s \to \mu^+\mu^-$ decay amplitude. A surprising feature of 
the electromagnetic correction in this expression is that in the expansion 
$\Lambda/m_b$ it is {\em power-enhanced} by a factor of $m_b/\Lambda$
relative to the well-known amplitude in the absence of QED effects, thereby 
partially compensating the suppression with the electromagnetic coupling $\alE$. 
The virtual photon exchange between the final-state leptons and the spectator 
quark in the $B_q$ meson leads to a non-local annihilation over distances 
$(m_b \Lambda)^{-1/2}$ inside the $B_q$ meson, different from the local 
annihilation through weak currents. Whereas the latter is described by $f_{B_q}$, 
the former involves the $B$-meson light-cone distribution amplitude (LCDA) 
$\phi_{+}(\omega)$, showing that strong interaction effects cannot be 
solely described in terms of $f_{B_q}$ once QED effects below the scale $\mu_b$ 
are included. The power-enhanced QED contribution involves two competing terms 
in the curly brackets, one from the semileptonic operator $\Op_9$ and one 
from the dipole operator $\Op_7$. Both terms are further enhanced by large 
logarithms $\ln(m_b\omega/m_\ell^2) \sim \ln (m_b \Lambda/m_\ell^2)$, and 
interfere destructively, which reduces the size of the power enhancement. 
It was also found that in $b\to u\ell\bar\nu_\ell$ the structure of the
semi-leptonic weak currents does not give rise to such a power enhancement in 
$B_u \to \mu\bar\nu_\mu$.

In the present work, the SCET interpretation underlying the above result, 
which was only briefly mentioned in \cite{Beneke:2017vpq}, is provided
in detail, together with the EFT treatment of QED and the summation of 
logarithms. The SCET approach to QED differs from standard QCD applications 
in several details and factorization theorems for QED effects are not well 
established, unlike the case for the pure QCD corrections. Two crucial 
differences are the presence of masses for leptons that regularize the 
collinear divergences in QED, and the presence of electromagnetically charged 
external states. Additionally, the soft-photon cutoff is typically below the 
scale of lepton masses, and thus real collinear photon radiation may be excluded,
while virtual collinear corrections can be still present. An additional challenge
is related to the proper treatment of QED radiation from light quarks, where
nonperturbative QCD has to be consistently treated. 
Here we use SCET to resum the leading logarithms for the 
power-enhanced contribution, which arises entirely from virtual 
effects between the hard and soft/collinear scales. We focus on the 
contribution of the semileptonic operator $\Op_9$, since one of the 
two logarithms enhancing the  dipole operator $\Op_7$ term is not a 
standard RG logarithm, in which case the summation with SCET methods 
is presently not understood. However, from the numerical point of view, 
our main finding is that higher-order QED logarithms appear to be 
negligibly small. The principal effect of resummation arises from QCD 
evolution on top of the one-loop QED effect shown above. This observation 
will allow us to also estimate the effect of resummation on the contribution
of the dipole operator.

As a by-product of this investigation, we find that hadronic matrix 
elements in the presence of QED are less universal than is usually assumed. 
For example, the nonperturbative matrix elements defining ``the'' 
$B$-meson decay constant and the LCDA depend on the charges and directions 
of the outgoing energetic particles through light-like electromagnetic 
Wilson lines.

The outline of the paper is as follows. After a short introduction to the 
conventions for the $\Delta B = 1$ EFT of $b\to q \ell^+\ell^-$ decays in 
\refsec{sec:DeltaB1-EFT}, we introduce the power counting set by the external 
kinematics of $B_q \to \ell^+\ell^-$ decays in \refsec{sec:kinematics} and
provide the power-counting of the SCET fields in \refsec{sec:SCET-defs}. 
\refSec{sec:heuristic} briefly recapitulates and interprets the findings of the
fixed-order calculation \cite{Beneke:2017vpq} relevant to the SCET approach and
provides a short outlook on the various contributions in SCET, discussed in the
main part later. We proceed with the decoupling of hard virtualities and the
RG evolution in \SCETI{} in \refsec{sec:SCET-1} and further the decoupling of
hard-collinear virtualities and the RG evolution in \SCETII{} in \refsec{sec:SCET-2}. 
The definition of the $B$-meson decay constant and LCDA in the presence of 
QED corrections is discussed in \refsec{sec:fB-LCDA-defs}. The factorization of the
power-enhanced amplitude is presented in \refsec{sec:factorization} and the
combination with the leading amplitude together with the ultrasoft parts
given in \refsec{sec:decay-width}. Eventually we present the numerical
impact of QED corrections and updated calculations of the non-radiative and radiative
branching fractions in \refsec{sec:numeric}.
Technical details on SCET conventions and definitions as well as the construction
of SCET operators have been relegated to appendices.

%--------+---------+---------+---------+---------+---------+---------+---------+
%
%
%
%--------+---------+---------+---------+---------+---------+---------+---------+

\section{Preliminaries}

%
%
%--------+---------+---------+---------+---------+---------+---------+---------+
\subsection[$\Delta B = 1$ effective theory for $b\to q\ell^+\ell^-$]
{\boldmath  $\Delta B = 1$ effective theory for $b\to q\ell^+\ell^-$}
\label{sec:DeltaB1-EFT}

The effective theory for $|\Delta B| = 1$ decays $b\to q\ell^+\ell^-$ with
$q = d, s$ in the framework of the SM,
\begin{align}
  \label{eq:DF1-EFT-Lagr}
  {\cal L}_{\Delta B = 1} &
  = \EWnorm \left[ 
    \sum_{i=1}^{10} C_i(\mub)\, \Op_i 
    + \frac{V_{ub}^{} V_{uq}^*}{V_{tb}^{} V_{tq}^*} 
      \sum_{i=1}^2 C_i(\mu_b) \big( \Op_i^u - \Op_i^c \big)
    \right] + \text{h.c.}\,, 
\end{align}
includes operators $\Op_i$, which are charged-current ($i=1,2$), QCD-penguin
operators $(i = 3,\ldots,6)$, dipole operators $(i=7,8$) and semileptonic
operators $(i = 9, 10)$. These operators are sufficient for the treatment of
the QED effects in $B_q\to \ell^+\ell^-$ discussed in this paper. We follow the 
operator definitions of \cite{Chetyrkin:1996vx}  and give only those of the three most
relevant operators for our purposes
\begin{align}
  \label{eq:def-Op-7}
  \Op_7 & 
  = \frac{e}{(4\pi)^2} \oL{m}_b \,
    \big[\bar{q} \sigma^{\mu\nu} P_R b\big] F_{\mu\nu},
\\
  \label{eq:def-Op-9}
  \Op_9 & 
  = \frac{\alE}{4\pi} 
    \big(\bar{q} \gamma^\mu P_L b\big) 
    \sum_\ell \bar{\ell} \gamma_\mu \ell ,
\\
  \label{eq:def-Op-10}
  \Op_{10} & 
  = \frac{\alE}{4\pi}
    \big(\bar{q} \gamma^\mu P_L b \big)
    \sum_\ell \bar{\ell} \gamma_\mu \gamma_5 \ell ,
\end{align}
where $\alE \equiv e^2/(4\pi)$ and $\oL{m}_b$ denotes the running $\oL{\text{MS}}$
$b$-quark mass. The overall normalization factor is $\EWnorm \equiv 2 \sqrt{2} G_F
V_{tb}^{} V_{ts}^*$. The term proportional to $V_{ub}^{} V_{uq}^*$ enters $B_q \to
\ell^+\ell^-$ only through the QED correction. The Wilson coefficients $C_i(\mub)$
and running quark masses need to be evaluated at the renormalization scale
$\mub \sim m_b$ of the order of the $b$-quark mass. In the SM they include
NNLO QCD matching corrections \cite{Bobeth:1999mk, Hermann:2013kca} at the 
electroweak scale $\muEW \sim m_W$ of the order of the $W$-boson mass, and 
$C_{10}$ further includes the NLO EW matching corrections \cite{Bobeth:2013tba}. 
The resummation of large logarithms between the scales $\mu_W$ and $\mu_b$ has
been taken into account to the corresponding
order following \cite{Bobeth:2003at, Huber:2005ig}, 
see also \cite{Bobeth:2013tba} for further details. Especially the inclusion
of NLO EW corrections \cite{Bobeth:2013tba} to $C_{10}$ requires 
care in the choice of the numerical input of the electroweak parameters. It
must respect the adopted renormalization scheme as for example $m_W$ is not
an independent parameter any more.
 
%
%
%--------+---------+---------+---------+---------+---------+---------+---------+
\subsection[Kinematics of $B_q\to\ell^+\ell^-$ and power counting]
{\boldmath  Kinematics of $B_q\to\ell^+\ell^-$ and power counting}
\label{sec:kinematics}

The two-body decay $B_q(p_B) \to \ell^+(p_\oL{\ell}) \ell^-(p_\ell)$ implies lepton
energies $E_\ell^{} = E_\oL{\ell} = m_{B_q}/2$, such that for light leptons $\ell = e,
\mu$ the hierarchy $m_\ell \ll E_\ell$ implies that the leptons are actually 
``collinear'' particles. At the partonic level,
\begin{align}
  b(p_b) + q(l_q) \;\; & \to \;\; \ell^+(p_\oL{\ell}) + \ell^-(p_\ell) ,
\end{align}
the mesonic bound state restricts the initial-state quarks to be soft. 
Writing $p_b = m_b v+l_b$, both quarks move inside the $B_q$ meson with  
soft residual momenta $l_b, \,l_q \sim \LambQCD$
of the order of the strong binding energy $\LambQCD$. In the decomposition of
$p_b$, $\vb$ is a normalized time-like vector, $\vb^2 = 1$, which can be
interpreted as the four-velocity of the $B_q$~meson. The soft scaling of the 
residual $b$- and light-quark momenta can be expressed as
\begin{align}
  l_b, \,l_q  \sim m_b\, \lamB_s^2  
\end{align}
in terms of the small dimensionless quantity
\begin{align}
  \lamB_s &
  = \sqrt{\frac{\LambQCD}{m_b}} \ll 1  &
  & \mbox{for} & \LambQCD \approx (0.2 - 0.4)~\mbox{GeV}.
\end{align}
In this picture both quarks are bound 
in the $B_q$ and annihilate via the $\Delta B = 1$ operators \refeq{eq:DF1-EFT-Lagr}.
The energy stored in the $b$-quark mass is released in the form of the energetic
lepton pair, which is emitted back-to-back in the $B_q$ rest frame thereby singling
out a particular direction. This direction can be described by a pair of light-like
vectors $n_+^2 = n_-^2 = 0$ and $\np \cdot \nm = 2$ 
and any four-vector can be decomposed as
\begin{align}
  p^\mu &
  = (\np p)\frac{n_-^\mu}{2} + p^\mu_\perp + (\nm p)\frac{n_+^\mu}{2} .
\end{align}
The components $p \sim (\np p,\, p^\mu_\perp,\, \nm p)$ of the lepton momenta
then exhibit the scaling 
\begin{align}
  p_\ell      & \sim m_b (1,\; \lamB_\ell^2,\; \lamB_\ell^4) , &
  p_\oL{\ell} & \sim m_b (\lamB_\ell^4,\; \lamB_\ell^2,\; 1) ,  
\end{align}
referred to as collinear and anti-collinear, respectively.
Here we introduced the small dimensionless quantity
\begin{align}
  \lamB_\ell &
  = \sqrt{\frac{m_\ell}{m_b}} \ll 1  &
  & \mbox{for} & \ell & = e, \mu.
\end{align}
The two cases of $\ell = e$ and $\ell = \mu$ are quite different, given that 
\begin{align}
  \lamB_e & \ll \lamB_s , &
  \lamB_\mu & \approx \lamB_s \equiv \lamB \,.
\end{align}
Subsequently we focus on $\ell = \mu$. We note that experimental
prospects are best for the decays $B_q \to \mu^+\mu^-$, in particular for the
CKM-enhanced mode $q = s$. The following different virtualities are set
by the kinematic invariants
\begin{align}
  p_b^2 \; \sim \; p_\ell \cdot p_{\bar\ell} \; \sim \; p_b \cdot p_{\ell,\bar\ell} 
  & \;\; \sim \;\;  m_b^2 ,
\\
  p_b \cdot l_q \; \sim \; l_q \cdot p_{\ell, \bar\ell} 
  & \;\; \sim \;\; m_b \Lambda ,
\\
  l_q^2 \; \sim \; p_\ell^2 \; \sim \; p_{\bar\ell}^2 
  & \;\; \sim \;\; \Lambda^2 ,
\end{align}
where $\Lambda = (m_\mu,\LambQCD)$ stands for either of the two small scales, 
the muon mass $m_\mu$ or $\LambQCD$, which we assume to be parametrically of same
size. Besides the hard virtuality $m_b^2$ and the
soft and collinear virtuality $\Lambda^2$ there is also the 
hard-collinear virtuality $m_b \Lambda$. In consequence we will go through
a two-step matching of EFTs,
\begin{align*}
  \hskip0.5cm \text{full QED}\hskip0.7cm & & & 
  \to & \text{\SCETI{}} & & & 
  \to & \text{\SCETII{}}
\\
  \hskip-0.5cm \text{hard:} \; \mu_b^2 \sim m_b^2 & & & & 
  \text{hard-collinear:}  & \; \mu_{hc}^2 \sim m_b \Lambda & & &
  \text{soft/collinear:}  & \; \mu_s^2 \sim \mu_c^2 \sim \Lambda^2 
\end{align*}
involving two versions of SCET. We note that given the symmetry of the 
final state under an exchange of $n_+$ and $n_-$, whenever a (hard-) collinear 
contribution exists the corresponding (hard-) anti-collinear contribution 
from the configuration with lepton and anti-lepton interchanged is implied.

The decay rate into the exclusive final state $\ell^+ \ell^-$ discussed  
up to now is not infrared (IR) safe in the presence of QED. 
The IR-safe definition includes the emission of
real photons with energies below a certain value $\Delta E$.
Throughout we will restrict the discussion to the case of $\Delta E 
\ll \Lambda$ that is we assume $\Delta E$ to be below the soft 
and collinear scale of \SCETII{}. Therefore only virtual corrections need to 
be considered above and at the scale $\Lambda$, 
and for the most part of the paper we therefore focus on the non-radiative 
amplitude. Ultra-soft photons, i.e. photons with virtuality much smaller 
than $\mu_{s,c}^2\sim \Lambda^2$, will be taken into account at the very end 
when we put together the final expression for the QED-corrected decay 
width. 

%
%
%--------+---------+---------+---------+---------+---------+---------+---------+
\subsection{\boldmath SCET: Definitions and conventions}
\label{sec:SCET-defs}

A systematic approach to the construction of \SCETI{} operator bases was 
discussed in \cite{Beneke:2003pa}. In this paper, we apply the same method, and 
we follow the same conventions and those of \cite{Beneke:2017ztn} 
when possible. Capital
letters $C$ ($\oL{C}$) refer to hard-collinear (anti-hard-collinear) \SCETI{}
fields, respectively, which we assume to contain both, the hard-collinear \SCETI{} 
modes and the collinear \SCETII{} modes. Collinear (anti-collinear) fields in 
\SCETII{} are denoted by the index $c$ ($\oL{c}$); these fields contain only collinear
modes and thus the power-counting of the \SCETII{} fields is homogeneous. The
index $s$ denotes
the soft fields. The $\lamB$ scaling of the heavy $b$-quark, light
spectator quark as well as the lepton fields in \SCETI{} and \SCETII{} is summarized
in \reftab{tab:fields-scaling}.

The masses of leptons and light quarks scale like $\lamB^2$. Accordingly, in
\SCETI{} collinear mass terms are part of the power-suppressed collinear Lagrangian,
while in \SCETII{} they are included in the leading-power collinear Lagrangian.
Mass factors may also appear explicitly in the operators. More details on the
relevant parts of the SCET Lagrangian are given in \refapp{app:SCET-Lag}. For 
definitions of renormalization constants we refer to \refapp{app:SCET-renorm}.

%%%%%%%%%%%%%%%%%%%%%%%%%%%%%%%%%%%%%%%%%%%%%%%%%%%%%%%%%%%%%%%%%%%%%%%%%%%%%%%%
\begin{table}
\centering
\renewcommand{\arraystretch}{1.3}
\begin{tabular}{|c|c|ccc|cc|ccc|}
\hline
  Field  
& heavy quark
& \multicolumn{3}{c|}{light quark}
& \multicolumn{2}{c|}{leptons}
& \multicolumn{3}{c|}{photon (gluon)}
\\
& $h_\vb$
& $\chi_{C}$    & $\chi_c$      & $q_s$
& $\ell_{C}$    & $\ell_c$
& $A_C\, (G_C)$ & $A_c\, (G_c)$ & $A_s\, (G_s)$

\\
\hline
  Scaling
& $\lamB^3$
& $\lamB$ & $\lamB^2$ & $\lamB^3$
& $\lamB$ & $\lamB^2$
& $(1, \lamB, \lamB^2)$ & $(1, \lamB^2, \lamB^4)$ & $ \lamB^2 (1, 1, 1)$
\\
\hline
\end{tabular}
\renewcommand{\arraystretch}{1.0}
\caption{\label{tab:fields-scaling}
  \small
  Fields and their power counting in SCET. In addition there are anti-hard-collinear 
  ($\chi_\oL{C}$, $\ell_\oL{C}$) and anti-collinear 
  ($\chi_\oL{c}$, $\ell_\oL{c}$) quark and lepton fields with the same scaling
  as their (hard)-collinear counterparts. The components for the photon field are
  $(\np A,\,  A_\perp^\mu,\, \nm A)$ and the gauge-invariant building blocks
  $\calA_{C \perp}^\mu$ and $\calA_{c \perp}^\mu$ scale as the 
  $\perp$-components of the fields $A_{C\perp} \sim \lamB$ and $A_{c\perp} \sim \lamB^2$,
  respectively. The light quark and lepton masses scale as $m_{q,\ell} \sim \lamB^2$.
}
\end{table}
%%%%%%%%%%%%%%%%%%%%%%%%%%%%%%%%%%%%%%%%%%%%%%%%%%%%%%%%%%%%%%%%%%%%%%%%%%%%%%%%

%--------+---------+---------+---------+---------+---------+---------+---------+
\subsection{\boldmath Heuristic discussion}
\label{sec:heuristic}

Before we begin the detailed formal discussion of resummation and factorization
in SCET, we recapitulate and interpret the main finding \refeq{eq:mainresult}
of the one-loop calculation \cite{Beneke:2017vpq} in the framework of SCET. 

\begin{figure}
\begin{center}
\vspace{0.3cm}\hskip0cm
\includegraphics[width=0.95\textwidth]{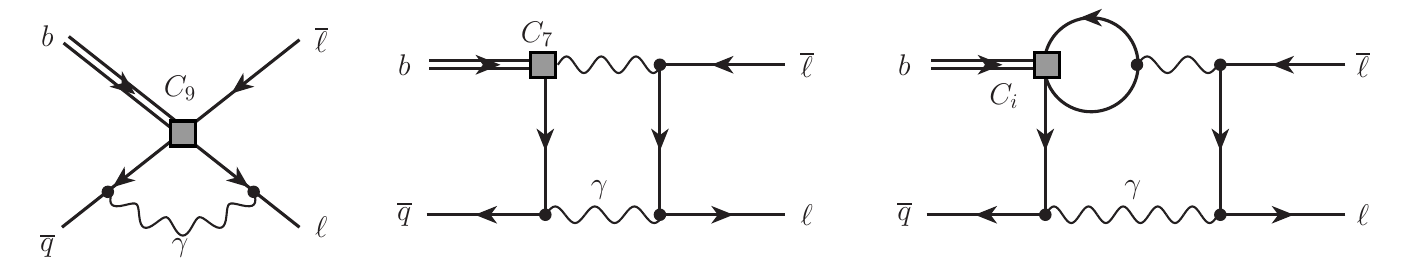}
\end{center}
\caption{\small Feynman diagrams that contain the power-enhanced electromagnetic
  correction. Symmetric diagrams with order of vertices on the leptonic line
  interchanged are not displayed.}
  \label{fig:oneloopdiagrams}
\end{figure}

The starting point is the one-loop virtual photon correction to the matrix 
elements of the operators $\Op_{7,9,10}$ at the scale $\mu_b$. The analysis based
on the method of expansion by regions \cite{Beneke:1997zp, Jantzen:2011nz} shows
that only the diagrams where the photon is exchanged between the soft spectator
quark and either of the final-state leptons can be power-enhanced, and that 
the power-enhancement cannot originate from the hard loop-momentum region. 
Examples are shown by the first two diagrams in~\reffig{fig:oneloopdiagrams}. 
The calculation of these diagrams in full QED, solving first the integrals analytically
in full generality\footnote{The analytic solutions of the one-loop integrals
were also obtained with ``Package X'' \cite{Patel:2015tea,Patel:2016fam}.} and
performing the expansion in $\lamB$ only afterwards confirms this result. The
one-loop expression contains logarithms of the ratio of hard-collinear over collinear
virtualities, $\ln(\mu_{hc}/\mu_c)$, for insertions of $\Op_9$ and even
double-logarithms $\ln^2(\mu_{hc}/\mu_{c,s})$ for~$\Op_7$. Note that the virtual
corrections do not lift the helicity suppression of the leptonic 
$B_q\to \ell^+\ell^-$ decays.

%%%%%%%%%%%%%%%%%%%%%%%%%%%%%%%%%%%%%%%%%%%%%%%%%%%%%%%%%%%%%%%%%%%%%%%%%%%%%%%%
%%%%%%%%%%%%%%%%%%%%%%%%%%%%%%%%%%%%%%%%%%%%%%%%%%%%%%%%%%%%%%%%%%%%%%%%%%%%%%%%

The SCET approach is used here to factorize the short-distance contributions
perturbatively to the leading non-vanishing order in the expansion in $\lamB$
and to resum the arising logarithms. Since the one-loop power-enhanced terms
do not arise from the hard region, the matching from full QED to \SCETI{} 
operators relevant to these terms proceeds at tree-level. Thereby the field
content of the semileptonic and dipole operators changes
\begin{align}
  \Op_{9,10} & \sim [\oL{q} \ldots b] [\oL{\ell} \ldots \ell] &
  \to &&
  \OpI_i & \sim [\oL{\chi}_{C,\,\oL{C}} \ldots h_\vb] [\oL{\ell}_C \ldots \ell_\oL{C}] 
  \; \sim \;\lamB^6 ,
  \label{eq:SCETIO9}
\\
  \Op_7  & \sim [\oL{q} \ldots b] F^{\mu\nu} &
  \to &&
  \OpI_i & \sim [\oL{\chi}_{C,\,\oL{C}} \ldots h_\vb] \calA_{\oL{C}\perp,\, C\perp}^\mu 
  \;\;\; \sim \; \lamB^5 ,
  \label{eq:SCETIO7}
\end{align}
where the $b$-quark is represented by a heavy-quark $h_\vb$ in HQET and the 
spectator quark is (anti-) hard-collinear $\chi_{C,\oL{C}}$, whereas the
lepton $\oL{\ell}_C$ is hard-collinear and the anti-lepton $\ell_\oL{C}$ is
anti-hard-collinear. In the case of $\Op_7$ the photon $\calA_{\oL{C}\perp}$
in \refeq{eq:SCETIO7} is anti-hard-collinear for hard-collinear $\chi_C$ and 
vice versa. $\OpI_i$ from \refeq{eq:SCETIO9} also appears in the matching 
of $\Op_7$. The scaling
of these operators in $\lambda$ follows from the scaling of the fields as
summarized in \reftab{tab:fields-scaling}. The large logarithms between the
hard and hard-collinear scales are then resummed with the aid of RG equations 
(RGEs) in \SCETI{}, as will be shown below. These logarithms appear only in
higher orders, i.e. they dress the diagrams shown~\reffig{fig:oneloopdiagrams}.

Let us briefly remark on the two-loop diagram in~\reffig{fig:oneloopdiagrams}, which
is generated by the four-quark operators $\Op_{1-6}$ in the effective Lagrangian 
\refeq{eq:DF1-EFT-Lagr}. It is well-known from $B\to X_s \ell^+\ell^-$ decays that
the quark loop can be fully absorbed into effective Wilson coefficients 
$C_9^\text{eff}(q^2)$ and $C_7^\text{eff}$, so that these diagrams should be
considered as one-loop QED corrections, as has been done in \refeq{eq:mainresult}.
This is implicitly understood when we refer to tree-level matching of $\Op_{7,9,10}$.

The second matching step from \SCETI{} to \SCETII{} produces the one-loop
logarithms. In the case of $Q_{9,10}$ (first diagram in~\reffig{fig:oneloopdiagrams}) 
there is a hard-collinear and a collinear momentum region. The first belongs to a
one-loop matching coefficient, while the second must be reproduced by the matrix
element of a \SCETII{} operator. The \SCETI{} operator $\OpI_i$ from~\refeq{eq:SCETIO9}
contains a $C$-antiquark, which is converted into the external soft spectator
anti-quark through the subleading-power \SCETI{} interaction
$\mathcal{L}_{\xi q}^{(1)}$ \cite{Beneke:2002ph}, see \refeq{eq:SCET-Lag-xiq1}, 
by emission of a transverse hard-collinear or collinear photon~$\calA_{C\perp}$.
The relevant \SCETII{} operators are
\begin{equation}
  \label{eq:heuristic-op-1}
  [\oL{\chi}_C \ldots h_\vb] [\oL{\ell}_C \ldots \ell_\oL{C}] 
  \quad \to \quad 
  \OpII_{\calA\chi}^{B1},\; \OpII_{m\chi}^{A1}\,,
\end{equation} 
where 
\begin{align}
  \OpII_{\calA\chi}^{B1} &
  \sim [\oL{q}_s (i \nm \overleftarrow \partial)^{-1}\ldots h_\vb] 
       [\oL{\ell}_c  \calA_{c\perp}\ldots \ell_\oL{c}] 
  \;\; \sim \; \lamB^{10} ,
  \label{eq:SCETIIOAchiB1}
\\[0.2cm]
  \OpII_{m\chi}^{A1} &
  \sim m_\ell\, [\oL{q}_s (i \nm \overleftarrow \partial)^{-1} \nms P_L h_\vb]
                [\oL{\ell}_c P_R \ell_\oL{c}]
  \; \sim \; \lamB^{10} .
  \label{eq:SCETIIOmchiA1}
\end{align}
To reproduce the $C_9^\text{eff}$ term in \refeq{eq:mainresult}, the matching to
the first of these operators is needed at tree-level. Its one-loop \SCETII{} matrix
element accounts for the collinear region. The matching coefficient of the second
operator is needed at the one-loop level to reproduce the hard-collinear region.
This leads to two important observations. First, the power-enhanced contribution
to $B_q\to \ell^+\ell^-$ decays requires a power-suppressed interaction in SCET, 
because the usual, non-enhanced $B_q\to \ell^+\ell^-$ amplitude involving 
$\Op_{10}$ (first term on the right-hand side of \refeq{eq:mainresult})
is in fact doubly suppressed due to helicity conservation {\em and} the 
point-like annihilation of the heavy quark with a soft anti-quark. Second, even in the collinear 
loop, the anti-quark propagator has hard-collinear virtuality -- only the lepton
and photon propagators have collinear virtuality. This enables the perturbative
calculation of the collinear contribution including the non-logarithmic terms.

%%%%%%%%%%%%%%%%%%%%%%%%%%%%%%%%%%%%%%%%%%%%%%%%%%%%%%%%%%%%%%%%%%%%%%%%%%%%%%%%
\begin{figure}
  \centering
  \includegraphics[width=0.98\textwidth]{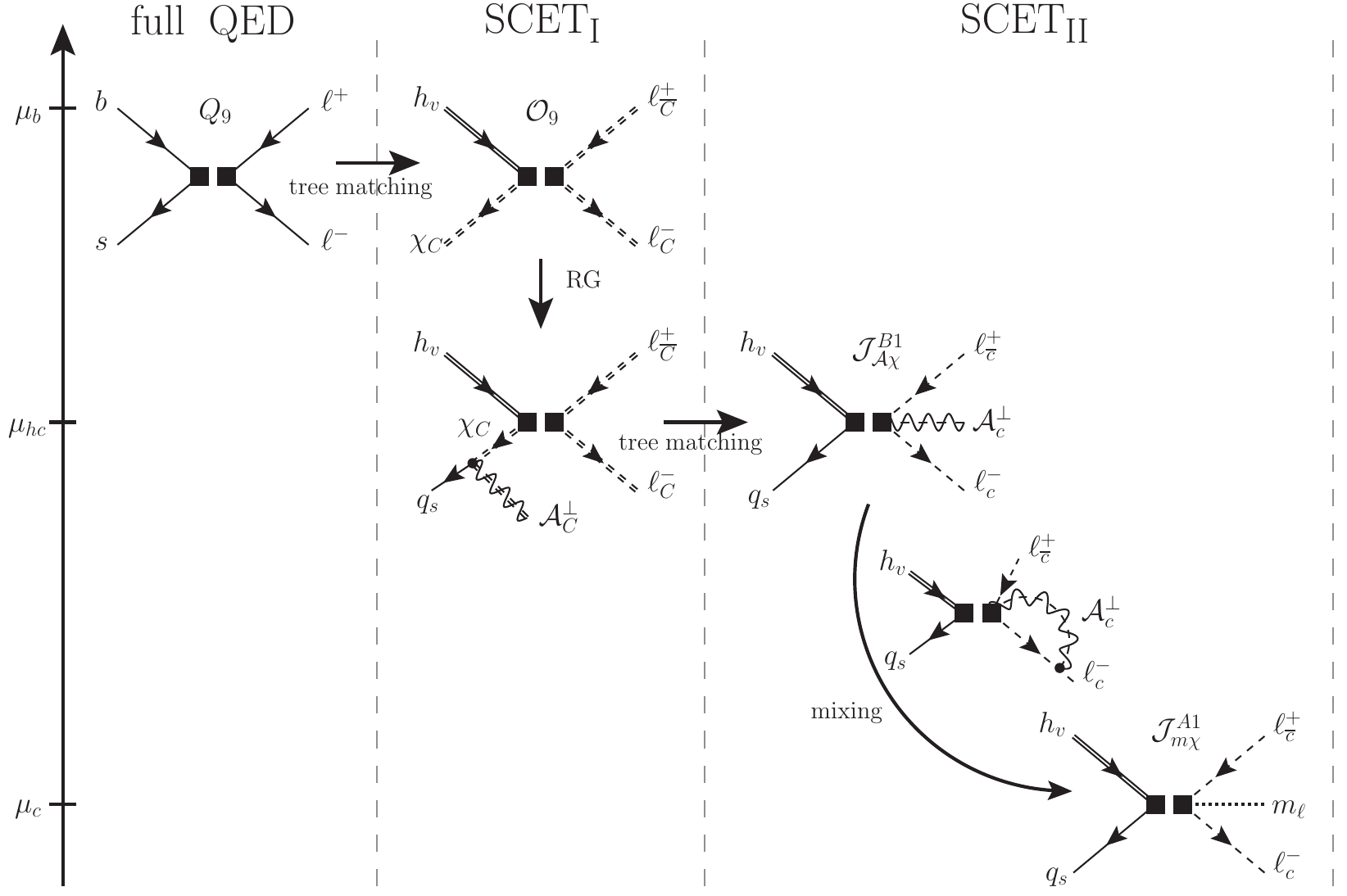}
\vskip0.2cm
\caption{\small The scheme shows the tree-level matching steps from
  full QED $\to$ \SCETI{} $\to$ \SCETII{} at the two scales
  $\mu_b$ and $\mu_{hc}$ horizontally from left to right. Vertically
  the RG evolution in \SCETI{} involves only self-mixing, whereas
  in \SCETII{} a mixing takes place of the operators $\OpII_{\calA\chi}^{B1}$
  into $\OpII_{m\chi}^{A1}$. The notation and scaling of the
  SCET fields are given in \reftab{tab:fields-scaling}. The propagators of
  the fields are chosen as double-solid for the heavy quark $h_\vb$,
  double-dashed for hard-collinear fermions in \SCETI{}, whereas 
  single-dashed for collinear fermions in \SCETII{}. The double- and
  single-dashed lines accompanied by a wavy line depict the hard-collinear
  and collinear photon fields in \SCETI{} and \SCETII{}, respectively.
  The single-solid line depicts the soft spectator quark~$q_s$. 
  The dotted line in $\OpII_{m\chi}^{A1}$ indicates that this operator
  contains a factor of the lepton mass $m_\ell$.
}
\label{fig:heuristic-scheme}
\end{figure}
%%%%%%%%%%%%%%%%%%%%%%%%%%%%%%%%%%%%%%%%%%%%%%%%%%%%%%%%%%%%%%%%%%%%%%%%%%%%%%%%

Note that in \SCETI{} $\to$ \SCETII{} matching, $C$-fields in the \SCETI{} operator
change to fields with collinear virtualities (denoted by $c$) in \SCETII{},
thereby increasing the power of the \SCETII{} operators in $\lamB$. In the above
two \SCETII{} operators we included the inverse soft derivative in their
definition to explicitly indicate the correct scaling of the operator.\footnote{Later
on, we will move this enhancement factor to the coefficient function 
\cite{Hill:2002vw}, which is more convenient for calculations. This 
factor is responsible for the appearance of the $1/\omega$ moment 
of the $B$-meson LCDA in~\refeq{eq:mainresult}.} One might have 
expected that an operator $[\oL{q}_s \gamma_\mu^\perp P_L h_\vb][\oL{\ell}_c 
\gamma^\mu_\perp \gamma_5 \ell_\oL{c}] \sim \lamB^{10}$ is generated by 
tree-level leading-power matching, but this operator has no 
overlap with the pseudo-scalar $B$-meson in the process 
$B_q \to \ell^+\ell^-$. An additional helicity suppression 
of $m_\ell \sim \lamB^2$ is required. In fact, also the operator 
\refeq{eq:SCETIIOAchiB1} has no overlap with the external states of 
$B_q \to \ell^+\ell^-$ because of the additional 
photon field $\calA_{c\perp}$. However, this operator mixes
under QED renormalization with $\OpII_{m\chi}^{A1}$, which 
has the correct chiral properties, but nevertheless scales as $\lamB^{10}$ 
due the compensating  $\lamB^{-2}$ power from the inverse soft derivative. 
It is precisely the anomalous dimension of this operator-mixing that 
reproduces the logarithms $\ln(\mu_{hc}/\mu_c)$ in the one-loop QED 
correction \refeq{eq:mainresult}. Below we will employ the RGEs
of \SCETII{} to resum these logarithms. 

The two-step matching of the operators $\Op_{9,10}$ and the RG evolution 
are schematically summarized in \reffig{fig:heuristic-scheme}. In the 
remainder of this work, we will derive the resummed result in detail for 
these operators.

Before proceeding, we comment on why we do not discuss the summation of logarithms
for the electromagnetic dipole operator $\Op_7$. The relevant diagram is now the
second one in~\reffig{fig:oneloopdiagrams}. While at first sight the hard-collinear
and collinear regions appear similar to the case discussed above, one finds that
the additional photon propagator attached to the dipole operator vertex causes an
endpoint-singularity as $u$, that is, the virtuality $u m_b^2$ of the photon,
goes to zero in the hard-collinear and collinear convolution integrals
for the box diagram. In this limit, the hard photon from the electromagnetic 
dipole operator becomes hard-collinear. The singularity is cancelled by a soft
contribution (virtuality $\Lambda^2\sim m_\ell^2$), where the leptons in the final
state interact with each other through the exchange of a soft lepton~\cite{Beneke:2017vpq}.
The relevance of soft-fermion exchange is interesting by itself since it is beyond
the standard analysis of logarithmically enhanced terms in QED. Moreover, the
endpoint or rapidity divergence encountered here is of a form that defies known
methods to sum such logarithms, since the breakdown of soft-collinear factorization
arises from a singular matching coefficient, rather than the soft or collinear
propagators themselves. A very similar phenomenon has subsequently been encountered
in \cite{Alte:2018nbn,Beneke:2019kgv}. The double logarithm in the $C_7^\text{eff}$ term in
\refeq{eq:mainresult} arises from this additional endpoint divergence. At the one-loop
order, the endpoint singularity can be regularized by a non-dimensional regulator
\cite{Beneke:2003pa},
which renders all integrals well-defined, with the result given in \refeq{eq:mainresult}.
We also verified this logarithm from the expansion of the full one-loop amplitude,
without using the split-up into regions, as mentioned above. However, it is currently
not known how to write down RGEs for suitably defined renormalized objects
for this situation, and hence resummation cannot be performed. 

%--------+---------+---------+---------+---------+---------+---------+---------+
%
%
%
%--------+---------+---------+---------+---------+---------+---------+---------+

\section{\boldmath \SCETI{}}
\label{sec:SCET-1}

%
%
%--------+---------+---------+---------+---------+---------+---------+---------+
\subsection{Operators}
\label{sec:SCET-1-ops}

The first decoupling step involves integrating out the hard modes of 
the light quark and lepton fields, as well as all other fields, in the 
matching on the \SCETI{} operators. For processes described by \SCETII{} 
a complication in the construction of the relevant operators in the
intermediate \SCETI{} appears, namely operators of different $\lambda$
scaling may contribute to the same order in $\lambda$ after matching to 
\SCETII{}~\cite{Beneke:2003pa}. The power-enhanced contribution requires 
only a single type of \SCETI{} four-fermion operator where the 
light quark is either hard-collinear or anti-hard-collinear. In position
space, denoted by a tilde, they read for a hard-collinear light quark
\begin{align}
  \widetilde{\OpI}_9 (s, t) &
  = \, g^\perp_{\mu\nu} \,
    \big[\oL{\chi}_C (s\np)\, \gamma_\perp^\mu P_L \, h_\vb(0)   \big]
    \big[\oL{\ell}_C (t\np)\, \gamma_\perp^\nu \, \ell_\oL{C}(0) \big] ,
\\
  \widetilde{\OpI}_{10} (s, t) &
  = i \varepsilon^\perp_{\mu \nu} 
    \big[\oL{\chi}_C (s\np)\, \gamma_\perp^\mu P_L \, h_\vb(0)   \big]
    \big[\oL{\ell}_C (t\np)\, \gamma_\perp^\nu \, \ell_\oL{C}(0) \big];
\intertext{for a anti-hard-collinear quark}
  \widetilde{\OpI}_\oL{9} (s, t) &
  = \, g^\perp_{\mu\nu} \,
    \big[\oL{\chi}_\oL{C} (s\nm)\, \gamma_\perp^\mu P_L \, h_\vb(0)\big]
    \big[\oL{\ell}_C (0)\, \gamma_\perp^\nu \, \ell_\oL{C}(t\nm)   \big] ,
\\
  \widetilde{\OpI}_\oL{10} (s, t) &
  = i \varepsilon^\perp_{\mu \nu} 
    \big[\oL{\chi}_\oL{C} (s\nm)\, \gamma_\perp^\mu P_L \, h_\vb(0)\big]
    \big[\oL{\ell}_C (0)\, \gamma_\perp^\nu \, \ell_\oL{C}(t\nm)   \big] .
\end{align} % CB: checked
The definitions of $g^\perp_{\mu\nu}$ and $\varepsilon^\perp_{\mu \nu}$ are given in 
\refapp{app:SCET}. In the classification scheme of \cite{Beneke:2002ph,
Beneke:2017ztn} these are operators of the B1-type with two hard-collinear 
(or anti-hard-collinear) fields in one of the directions, and of the 
A0-type in the opposite direction.  
The operators $i = \oL{9}, \oL{10}$ contain an anti-hard-collinear light
quark field $\chi_\oL{C}$ instead of a $\chi_C$ in operators
$i = 9, 10$. 

The Fourier-transformed \SCETI{} operators are defined as
\begin{align}
  \OpI_{i}(u) &
  = \np p_C \int \frac{dr}{2\pi} \, e^{-iu\, r (\np p_C)} \,
    %\exp\left[-iu\, r (\np p_C) \right]
    \widetilde{\OpI}_{i}(0,r) \,.
\end{align} % CB: checked
Hard-collinear momentum conservation has been used to drop the dependence
on the total hard-collinear momentum $\np p_C = \np (p_\chi + p_\ell)$ 
on the left-hand side and the first argument of $\widetilde{\OpI}_{i}$ 
is set to zero. The variable $u$ should be interpreted as the fraction
$\np p_\ell/\np p_C$ of $\np p_C$ carried by the lepton field, while the 
hard-collinear light anti-quark has momentum fraction $\oL{u} \equiv 
(1 - u) = \np p_\chi/\np p_C$.  For the operators 
$\widetilde{\OpI}_\oL{i}$ similar definitions apply after 
replacing $n_+$ by $n_-$.

The \SCETI{} Wilson coefficients of these operators, the so-called 
``hard functions'', are introduced in momentum space as
\begin{align}
  \mathcal{L}^\text{I}_{\Delta B = 1} &
  = \sum_i \int du\, H_i(u, \mu)\, \OpI_i(u) .
\end{align} % CB: checked
They are found by matching full QED+QCD $\to$ \SCETI{} at the hard scale
$\mu =\mu_b \sim \mathcal{O}(m_b)$ as described in \refsec{sec:SCET1-matching} 
below.

A complete basis of four-fermion operators when naive dimensional regularization
with anti-commuting $\gamma_5$ is employed would include in addition also 
operators with Dirac matrices vanishing in four dimensions, the so-called 
evanescent operators. However, the logarithms that we aim to sum in this paper
in \SCETI{} are derived from one-loop anomalous dimensions, which are given
by the pole parts in $1/\epsilon$, where $\epsilon = (4-D)/2$ in terms of
the number of space-time dimension $D$,
of the one-loop diagrams that are independent of the definition of evanescent
operators. 

%
%
%--------+---------+---------+---------+---------+---------+---------+---------+
\subsection{Matching}
\label{sec:SCET1-matching}

For the leading logarithmic accuracy it is sufficient to perform only the
tree-level matching of $\Op_9$ and $\Op_{10}$ operators on the \SCETI{} 
operators $\OpI_i$. One-loop matching is needed for the four-quark 
operators $\Op_i$ ($i = 1, \ldots, 6$), which can be included as is
commonly done by the replacement $C_9 \to C_9^\text{eff}$~\cite{Buras:1994dj}
as mentioned above. The hard matching condition at the scale $\mu_b$ is given by 
\begin{align}
  \EWnorm \sum_k C_k (\mu_{b})\, \Op_k & 
  = \sum_i \int du\, H_i(u, \mu_b)\, \OpI_i(u) .
\label{eq:SCETImatchrel}
\end{align}
Evaluating this equation in the appropriate matrix element with 
a hard-collinear (anti-hard-collinear) light quark state, we 
find, at tree-level
\begin{equation}
  \label{eq:SCETI-hard-matching}
\begin{aligned}
  H_9(u, \mu_b) & 
  = \calN \, C_9^\text{eff} (u, \mu_b) ,
  \qquad &  
  H_\oL{9} & = H_9 , 
\\
  H_{10}(u, \mu_b) & 
  = \calN \, C_{10} (\mu_b) , 
  \qquad &  
  H_\oL{10} & = H_{10} \,.
\end{aligned}
\end{equation}
Here 
\begin{equation}
   \calN \equiv \EWnorm \frac{\alE(\mu_b)}{4\pi}\,,
\end{equation}
and 
\begin{equation}
  \label{eq:C_9^eff}
\begin{aligned}
  C_9^\text{eff} (u, \mu_b) & 
  = C_9(\mu_b) + Y(u s_{\ell\oL{\ell}},\, \mu_b) 
\\[0.2cm] &
  - \frac{V_{ub}^{} V_{uq}^*}{V_{tb}^{} V_{tq}^*} \left(\frac{4}{3} C_1 + C_2 \right)
    \big[h(0, u s_{\ell\oL{\ell}}) - h(m_c, u s_{\ell\oL{\ell}})\big],
\end{aligned}
\end{equation} % CB: checked
with the dilepton invariant mass $s_{\ell\oL{\ell}} \equiv (\np p_\ell)\, 
(\nm p_\oL{\ell})$. We use the definition of the function $Y(u s_{\ell\oL{\ell}})$ 
from~\cite{Beneke:2001at}. The function $h(m_q, q^2)$~\cite{Bobeth:1999mk} 
depends on the light quark masses $m_{u,d}$ that are set to zero, or the
charm-quark mass $m_c$.

%
%
%--------+---------+---------+---------+---------+---------+---------+---------+
\subsection{RG evolution}
\label{sec:SCET1-rge}

The RGE in \SCETI{} governs the evolution of the matching coefficient
$H_i(u, \mu)$ from the hard scale $\mu_b$ to the hard-collinear scale $\mu_{hc}$.
The renormalization constants and the anomalous dimensions of the operators 
$\OpI_i$ can be computed similarly to the ones for $N$-jet operators 
\cite{Beneke:2017ztn, Beneke:2018rbh}, with the addition of a soft heavy-quark
field. Our conventions follow \cite{Beneke:2017ztn} and are summarized in 
\refapp{app:SCET-renorm}. We take into account both QCD and QED effects. The
evolution of the hard function is determined by 
\begin{align}
  \label{eq:SCET1-rge}
  \frac{d H_i (u,\mu)}{d \ln \mu} &
  = \Gamma_\text{cusp}^\text{I} 
      \left(\ln \frac{m_{B_q}}{\mu} - \frac{i\pi}{2} \right) H_i (u,\mu)
    + \int \! du' \, \Gamma_i(u',u) H_i (u',\mu) \,.
\end{align}
The $B$-meson mass in the cusp logarithm arises from the kinematic constraint
$s_{\ell\oL{\ell}} = m_{B_q}^2$. 
The imaginary parts arise from $\ln[ -(s_{\ell\oL{\ell}} + i 0^+)/\mu^2] = 
\ln( m^2_{B_q}/\mu^2) - i \pi$, and will be neglected throughout, as they do not
contribute at the leading logarithmic accuracy. For the summation of the leading
logarithms (LL) we require the one-loop cusp anomalous dimension
\begin{align}
  \label{eq:SCET1-ADM-cusp}
  \Gamma_\text{cusp}^\text{I} (\alS, \alE) &
  = \Gamma_c(\alE) + \Gamma_s(\alS, \alE) ,
\end{align} % CB: checked
that has been split for later convenience into a part $\Gamma_c \propto Q_\ell^2$ and
the remainder $\Gamma_s$ that includes also the QCD cusp anomalous dimension, 
\begin{align}
  \label{eq:cusp-ADMs}
  \Gamma_c & = \frac{\alE}{\pi} 2 Q_\ell^2 , &
  \Gamma_s & = \frac{\alS}{\pi} C_F + \frac{\alE}{\pi} Q_q (2 Q_\ell + Q_q) ,
\end{align} % CB: checked
expressed in terms of the electric quark and lepton charges, $Q_q$ and $Q_\ell$,
respectively, and the QCD Casimir $C_F=4/3$. At the next-to-leading logarithmic (NLL) accuracy one would also include
\begin{align}
  \label{eq:SCET1-ADM-remd}
  \Gamma_i(x, y) & 
  = \frac{\alS C_F}{4\pi} \big[4 \ln (1-x) -5\big] \delta(x-y)
    + \frac{\alE}{4\pi} \,\gamma_i(x,y) , &
  (i & = 9, 10) ,
\end{align} % CB: checked
and the two-loop cusp part. The function $\gamma_i(x, y)$ is provided for 
completeness in \refeq{eq:SCET1-ADM-remainder}. The general solution of 
the RGE \refeq{eq:SCET1-rge} when only the cusp anomalous dimension 
is kept (and the imaginary part neglected) reads
\begin{align}
  \label{eq:RG-SCET1-H}
  H_i(u, \mu) & 
  = \exp\left[ \int_{\mu_b}^\mu \frac{d\mu'}{\mu'}\, 
      \Gamma_\text{cusp}^\text{I}(\mu') \ln \frac{m_{B_q}}{\mu'} 
    \right] H_i(u, \mu_b),
\end{align}
and amounts to a global, momentum-fraction independent rescaling of 
the hard functions $H_i(u, \mu)$ by a Sudakov factor. This property  
is particular to the LL approximation. From NLL accuracy, when 
the non-cusp anomalous dimension $\Gamma_i(u',u)$ is included, 
the QCD logarithms lead to a momentum-fraction dependent rescaling 
from the $\ln (1-x)$ term in~\refeq{eq:SCET1-ADM-remd}, while the 
QED corrections governed by $\gamma_i$ reshuffle the momentum 
fractions carried by the spectator quark and lepton.

The integral in \refeq{eq:RG-SCET1-H} can in general be evaluated only numerically.
When the running of the strong coupling $\alS$ is included, but the one 
of $\alE$ as well as the influence of $\alE$ on the QCD running is neglected, 
we obtain the solution given in~\refeq{eq:general-RG-sol-SCETI}. However, 
our aim is to sum leading logarithms in QED to all orders. When the solution 
is written in the form of \refeq{eq:RG-SCET1-H} ``LL accuracy'' is defined 
by including all terms of the form $\log\times (\alpha\log)^n$ for any $n$ 
{\em in the exponent}, where $\alpha$ can be $\alS$ or $\alE$. The 
``double logarithmic'' approximation corresponds to keeping only the first 
term $n=1$ in the LL series. 

In the LL approximation the one-loop cusp anomalous dimension is the 
sum of a QCD and a QED term (not to be confused with the split into 
$\Gamma_c$ and $\Gamma_s$ above, which will be useful later). The exponential 
factorizes into a QCD and a QED contribution. Even in this approximation 
it is convenient to perform the integrals numerically, when the coupling 
runs through flavour thresholds. We shall use such numerical solutions 
in the final numerical results in \refsec{sec:numeric}. 
For the purpose of discussion, we present the 
analytic solution, when flavour thresholds in the interval $[\mu,\mu_b]$ 
are neglected,
\begin{equation}
  \label{eq:SCET1-RG-ana-sol}
  \frac{H_i(u,\mu)}{H_i(u, \mu_b)} 
  = \exp\!\bigg[
      \frac{4\pi\,}{\alS(\mu_b)} \frac{C_F}{\beta_0^2} \,g_0(\eta_s) \bigg]\,
    \exp\!\bigg[
      \frac{4\pi}{\alE(\mu_b)} 
      \frac{\big[2 Q_\ell^2 + Q_q(2 Q_\ell + Q_q)\big]}{\beta_{0,\rm em}^2} \,
      g_0(\eta_{\rm em}) \bigg],
\end{equation}
where $g_0(x) = 1 - x + \ln x$ and  $\eta_i$ stands for  
$\eta_i(\mu_b,\mu) \equiv \alpha_i(\mu_b)/\alpha_i(\mu)$ 
with $i = s,\, \text{em}$. 
To obtain this expression from \refeq{eq:RG-SCET1-H} we replace $m_{B_q}$ in 
the cusp logarithm by $\mu_b$ and neglect the non-enhanced logarithm 
$\ln \,(m_{B_q}/\mu_b)$. The ambiguity in choosing the precise value of 
the hard-matching scale $\mu_b\sim m_b$ is resolved only beyond the LL 
approximation.

Neglecting the running of the QED coupling in \refeq{eq:SCET1-RG-ana-sol}
amounts to the QED double-logarithmic (DL) approximation and the approximation 
$g_0(x) = -(1-x)^2/2 +\mathcal{O}((1-x)^3)$. In the above and similar 
expressions below we can always switch between the LL (left) and DL (right) 
QED resummation by the replacement 
\begin{equation}
  \label{eq:LLvsDL}
  \exp\!\bigg[\frac{4\pi}{\alE(\mu_1)} \frac{\mathcal{Q}}
  {\beta_{0,\rm em}^2} \, g_0(\eta_{\rm em}) \bigg]
  \quad \longleftrightarrow \quad
  \exp\bigg[-\frac{\alE}{2\pi} \mathcal{Q}\,
               \ln^2\frac{\mu_1}{\mu_2} \bigg] ,
\end{equation} % CB: checked
where now $\eta_{\rm em}$ denotes $\eta_{\rm em}(\mu_1,\mu_2) = 
\alE(\mu_1)/\alE(\mu_2)$, and $\mathcal{Q}$ stands 
for the appropriate charge factor.

For later purposes it is convenient to pull out the part of
the exponent with $\Gamma_c \propto \alE Q_\ell^2$ as follows
\begin{equation}
  \label{eq:sol-RGE-Hi}
  H_i(u,\mu) 
  = \exp\big[ S_{\ell}(\mu_b,\, \mu)\, 
            + \, S_{q}(\mu_b,\, \mu) \big]\, H_i(u, \mu_b) ,
\end{equation}
thereby introducing the Sudakov exponents
\begin{align}
  \label{eq:def-S_ell}
  S_{\ell}(\mu_b,\, \mu) & = 
    \frac{4\pi}{\alE(\mu_b)} \frac{2 Q_\ell^2}{\beta_{0,\rm em}^2} g_0(\eta_{\rm em})
    \; \stackrel{\rm DL}{\longrightarrow} \; - \frac{\Gamma_c}{2} \ln^2 \frac{\mu_b}{\mu} ,
\\
  \nonumber
  S_{q}(\mu_b,\, \mu) & = 
    \frac{4\pi\,}{\alS(\mu_b)} \frac{C_F}{\beta_0^2} \,g_0(\eta_s) + 
      \frac{4\pi}{\alE(\mu_b)} \frac{\big[Q_q(2 Q_\ell + Q_q)\big]}{\beta_{0,\rm em}^2} \,
      g_0(\eta_{\rm em}) 
\\
  \label{eq:def-S_qI} 
  & \stackrel{\rm DL}{\longrightarrow} 
    \frac{4\pi C_F}{\alS(\mu_b)\beta_0^2} \left(1 - \eta_s + \ln\eta_s \right)
  - \frac{\alE}{2\pi} \big[Q_q(2 Q_\ell + Q_q)\big] \ln^2 \frac{\mu_b}{\mu}\,.
\end{align} % CB: checked

%--------+---------+---------+---------+---------+---------+---------+---------+
%
%
%
%--------+---------+---------+---------+---------+---------+---------+---------+

\section{\SCETII{}}
\label{sec:SCET-2}

The above equations are used to evolve the \SCETI{} operators 
$\widetilde{\OpI}_{9,10}$, $\widetilde{\OpI}_{\oL{9},\oL{10}}$ to the 
hard-collinear scale $\mu_{hc}$, at which the hard-collinear modes 
with virtuality $\mathcal{O}(m_b\Lambda)$ are removed and the \SCETI{} 
operators are matched to \SCETII{}. 

An important distinction between \SCETI{} and \SCETII{} for the problem at 
hand is the treatment of the lepton mass. Parametrically the muon mass is of 
the same order as the soft/collinear scale $m_\mu \sim \LambQCD$. Thus the 
lepton mass terms are part of the leading-power collinear Lagrangian in 
\SCETII{}, see \refeq{eq:SCET-Lag-xi0}. 
In consequence the muon mass is retained in the denominators of the collinear
lepton propagators and serves as a regulator of the collinear divergences. 

To develop an idea of operator matching to \SCETII{}, we recall that the 
\SCETI{} operators $\OpI_i$ contain hard-collinear light quark fields,
while the $B$-meson contains only soft fields.  
The hard-collinear field in the \SCETI{} operators must be converted
into a soft quark field through emission of a (hard-) collinear photon by
the power-suppressed \SCETI{} Lagrangian $\mathcal{L}_{\xi q}^{(1)}$ 
(definition in \refeq{eq:SCET-Lag-xiq1}) to obtain a non-vanishing
overlap with the $B$-meson state.
Therefore,
we match the time-ordered product of the \SCETI{} operators $\OpI_i$ 
with $\mathcal{L}_{\xi q}^{(1)}$ to \SCETII{} operators. 
The tree-level matching relation is depicted in the second line 
labelled ``tree matching'' in \reffig{fig:heuristic-scheme}.
Starting from the one-loop order, pure four-fermion operators without 
collinear photons also appear (not shown). 
The systematic construction of the \SCETII{} operator basis is 
substantially more complicated than for \SCETI{}, since one must control 
the degree of non-locality of soft fields~\cite{Beneke:2003pa}. 
In the following we discuss the operators, their renormalization and 
matching coefficients relevant to LL resummation. Some further details 
are provided in Appendices~\ref{app:SCET} and~\ref{sec:operator}.

%
%
%--------+---------+---------+---------+---------+---------+---------+---------+
\subsection{Operators}

We note that, quite generally, in \SCETII{} operators also the soft fields are 
delocalized along the direction 
of the light-cone. The small component $\nm p$ of the hard-collinear mode, which is 
integrated out, is of the same order as the soft momentum, hence the soft 
field can be at any position in the $\nm$ direction. The roles of $\nm$ and $\np$ 
are reversed when the anti-hard-collinear mode is integrated out.

A power-counting analysis similar to the one performed in~\cite{Beneke:2003pa}
for heavy-to-light meson form factors shows that only two different \SCETII{} 
operators for each collinear direction are relevant to the power-enhanced 
correction from the $Q_9$ operator. The two \SCETII{} operators mix under 
renormalization. The technical arguments can be found in 
\refapp{sec:operator}. In position space, the two operators are defined 
as 
\begin{align}
  \label{eq:def-SCET-II-ops-1}
  \widetilde{\OpII}_{m\chi}^{A1} (v) &
  = \oL{q}_s (v \nm) Y(v \nm, 0) \frac{\nms}{2} P_L h_\vb (0) 
    \big[Y_+^\dagger\, Y_- \big](0) 
    \,\big[\oL{\ell}_c(0) (4 m_\ell P_R) \, \ell_\oL{c}(0)\big],
\\ \nonumber
  \widetilde{\OpII}_{\calA\chi}^{B1} (v, t) &
  = \oL{q}_s (v \nm)Y(v \nm, 0) \frac{\nms}{2} P_L h_\vb (0) 
    \big[Y_+^\dagger\, Y_- \big](0) 
    \,\big[\oL{\ell}_c (0) (g_{\mu\nu}^\perp + i\varepsilon_{\mu \nu}^\perp) 
         \calA_{c\perp}^\mu (t\np) \gamma_\perp^\nu \ell_\oL{c}(0) \big] 
\\ & 
  \label{eq:def-SCET-II-ops-2}  
  = \oL{q}_s (v \nm)Y(v \nm, 0) \frac{\nms}{2} P_L h_\vb (0) 
    \big[Y_+^\dagger\, Y_- \big](0) 
    \,\big[\oL{\ell}_c (0) (2 \slashed\calA_{c\perp} (t\np) P_R) \ell_\oL{c}(0) \big] .
\end{align} % CB: checked
For $\widetilde{\OpII}_{\calA\chi}^{B1}$ the second line provides 
an equivalent representation which makes the chirality of the leptons explicit.
The analogous operators generated from the matching of 
$\OpI_{\oL{9}, \oL{10}}$ are defined as
\begin{align}
  \widetilde{\OpII}_{m\oL{\chi}}^{A1} (v) &
  = \oL{q}_s (v \np)Y(v \np, 0) \frac{\nps}{2} P_L h_\vb (0) 
    \big[Y_+^\dagger\, Y_- \big](0)
    \,\big[\oL{\ell}_c (0) (4 m_\ell P_R) \, \ell_\oL{c} (0) \big] ,
\\ \nonumber
  \widetilde{\OpII}_{\calA\oL{\chi}}^{B1} (v, t) &
  = \oL{q}_s (v \np) Y(v \np, 0) \frac{\nps}{2} P_L h_\vb (0) 
    \big[Y_+^\dagger\, Y_- \big](0)
    \,\big[\oL{\ell}_c (0) (g_{\mu\nu}^\perp - i\varepsilon_{\mu \nu}^\perp) 
         \calA_{\oL{c}\perp}^\mu (t\nm) \gamma_\perp^\nu \ell_\oL{c}(0) \big] 
\\ &   \label{eq:def-SCET-II-ops-4}
  = \oL{q}_s (v \np)Y(v \np, 0) \frac{\nps}{2} P_L h_\vb (0)  
    \big[Y_+^\dagger\, Y_- \big](0) 
    \,\big[\oL{\ell}_c (0) (2 P_R \slashed\calA_{\oL{c}\perp} (t\nm)) 
\ell_\oL{c}(0) \big] .
\end{align} % CB: checked
The A1-type operators are constructed from leading-power building blocks
and multiplied by a factor of the lepton mass where the factor of $4$ is 
introduced for convenience. The B1-type operators contain the (anti-) 
collinear photon field $\calA_{c\perp\, (\oL{c}\perp)}^\mu$. Both operators 
have the same $\lamB$ scaling. The product of Wilson lines 
$[Y_+^\dagger Y_-] (0) \equiv Y_+^\dagger(0) Y_-(0)$ appears after 
decoupling of soft photons from the collinear
and anti-collinear leptons in \SCETI{}, see also \refeq{eq:decoupling-trafo}. 
These electromagnetic Wilson lines are defined as
\begin{align}
  \label{eq:SCETII-def-soft-Y}
  Y_\pm (x) & 
  = \exp\left[- i e \,Q_\ell \int_0^{\infty} \!\! ds \, n_\mp A_s (x+ sn_\mp)
  \right] .
\end{align} % CB: checked
For the quark current the usual finite-distance Wilson line
\begin{align}
  \label{eq:SCETII-def-soft-finite-Y}
  Y (x,y) & 
  = \exp\left[ i e \, Q_q \int_y^x \!\! dz_\mu \,  A^\mu_s (z) \right] \;
    \mathcal{P} \exp\left[  i g_s \int_y^x \!\! dz_\mu \, G^\mu_s (z) \right] 
\end{align} % CB: checked
appears, which is necessary to maintain the QCD and QED gauge invariance
of non-local operators. Here $\mathcal{P}$ is the path-ordering operator and 
$G^\mu_s = G_s^{\mu A} T^A$ is the soft gluon field. The integral is evaluated 
along the straight line connecting the points $x$ and $y$. 
We define the Fourier transforms
\begin{align}
  \OpII_i^{A1}\left(\omega\right)&
  = \int\! \frac{dv}{2\pi} 
    \, e^{i\, \omega \, v} \, \widetilde{\OpII}_i^{A1} (v) \,,
\label{eq:JA1FT}\\
\OpII_i^{B1}\left(\omega,w\right)	&
 = (n\cdot p) \int\! \frac{dv}{2\pi} \, e^{i\, \omega\, v} 
 \int\! \frac{dt}{2\pi}
     \, e^{- i \oL{w} \,t (n\cdot p)} \, \widetilde{\OpII}_i^{B1} (v, t) 
\label{eq:JB1FT}
\end{align} % CB: checked
of the operators, where $w$ corresponds to the collinear momentum fraction 
carried by the lepton,
and $\omega$ may be interpreted as the soft momentum of the light quark along
the $\np$ or $\nm$ direction, depending on the operator. Further $(n\cdot p) = 
\np p_c = \np (p_\ell + p_{\calA_{c\perp}})$ for $i = \calA\chi$ and 
$(n\cdot p) = \nm p_\oL{c} = \nm (p_\oL{\ell} + p_{\calA_{\oL{c}\perp}})$ for
$i = \calA\oL{\chi}$, respectively. In this way, after taking the matrix element, 
$w = \np p_\ell/\np p_c$ denotes the momentum fraction of $\np p_c$ carried 
by the collinear lepton and analogously for the anti-collinear case with
appropriate replacements $\np \to \nm$ and $c \to \oL{c}$.  
We further defined $\oL{w}\equiv 1-w$.

%
%
%--------+---------+---------+---------+---------+---------+---------+---------+
\subsection{Renormalization}
\label{sec:SCET2-renormalization}

The \SCETII{} operators \refeq{eq:def-SCET-II-ops-1}--\refeq{eq:def-SCET-II-ops-4}
are composed of soft, collinear and anti-collinear field products
\begin{align}
  \label{eq:SCET2-op-fact}
  \widetilde{\OpII}_i &
  = \widehat{\OpII}_{i,s} \otimes \widehat{\OpII}_{i,c} \otimes \widehat{\OpII}_{i,\oL{c}}\,,
\end{align}
where the $\otimes$ symbol indicates potential summation/contractions over spinorial
and/or Lorentz indices. In \SCETII{}, the soft, collinear and anti-collinear fields 
do not interact with one another, which implies that the matrix elements of the 
\SCETII{} operators factorize accordingly into matrix elements of the separate 
factors on the right-hand side of \refeq{eq:SCET2-op-fact} in the respective 
soft, collinear and anti-collinear Hilbert space.  The UV counterterms can also be 
defined separately for each sector. However, a rearrangement is necessary due to the
factorization anomaly as discussed below.  The renormalization of \SCETII{} 
operators then proceeds similarly to the \SCETI{} case, 
see~\cite{Beneke:2017ztn, Beneke:2018rbh, Beneke:2019kgv}.
We next discuss the 
renormalization of each sector separately and then present the combined result
for the  \SCETII{} operators.
%
%--------+---------+---------+---------+---------+---------+---------+---------+
\subsubsection{Soft sector}

\begin{figure}
\centering
  \begin{subfigure}[t]{0.42\textwidth}
    \centering
    \includegraphics[width=1.0\textwidth]{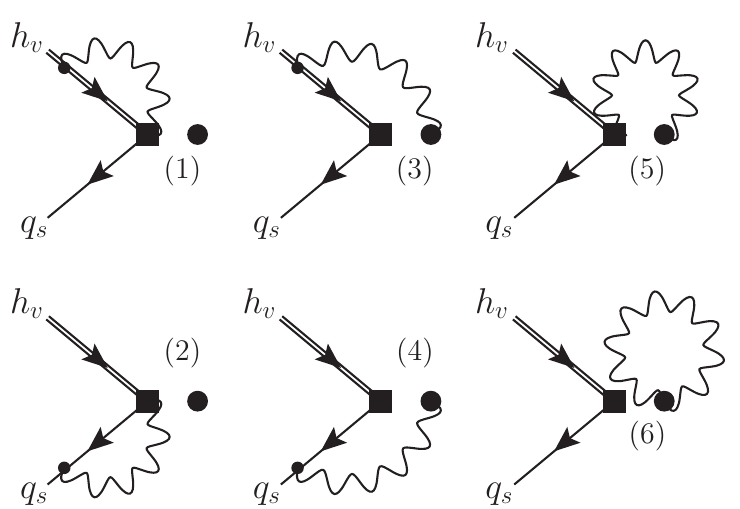} 
    \caption{}
    \label{fig:SCET2-adm-soft}
  \end{subfigure}
  \begin{subfigure}[t]{0.18\textwidth}
    \centering
    \includegraphics[width=0.99\textwidth]{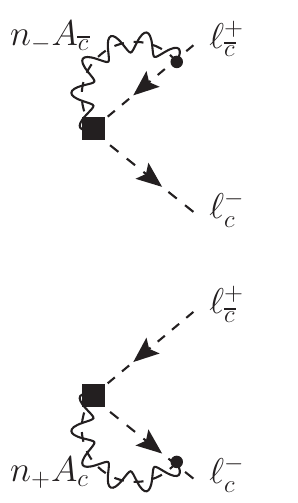}
    \caption{}
    \label{fig:SCET2-adm-coll}
  \end{subfigure}
  \begin{subfigure}[t]{0.38\textwidth}
    \centering
    \includegraphics[width=0.48\textwidth]{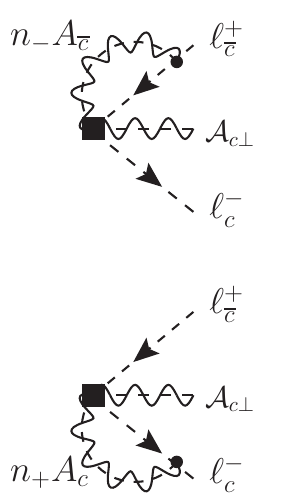}
    \hskip -0.04\textwidth
    \includegraphics[width=0.48\textwidth]{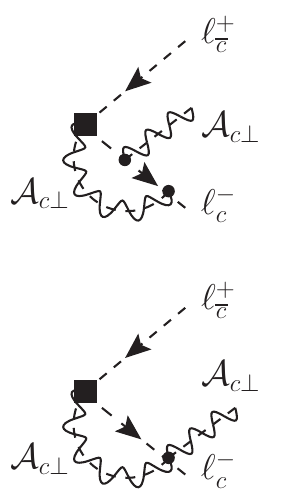}
    \caption{}
    \label{fig:SCET2-adm-coll-aChi}
  \end{subfigure}
\caption{\label{fig:SCET2-adm}
  \small
  The diagrams in \reffig{fig:SCET2-adm-soft} show the parts of 
  the \SCETII{} operators $\OpII_i$ $(i = m\chi, \calA\chi)$ with soft photon exchange
  (wavy lines) from $i)$ the Wilson lines in the soft fields $h_\vb$ (double line) 
  and $q_s$ (single line) depicted by the square and $ii)$ the product of soft
  Wilson lines $Y_+^\dagger(0) Y_-(0)$ depicted by the solid blob. They contribute
  to the $Z_s^\text{QED}$ (diagrams $1-5$), and $Z_{\overline{c}}$, $Z^c_{m\chi}$ and $Z^c_{\calA\chi}$
   (diagram 6 $\propto Q_\ell^2$). 
  \reffig{fig:SCET2-adm-coll} shows diagrams relevant for $Z_{\overline{c}}$, $Z^c_{m\chi}$ 
and   \reffig{fig:SCET2-adm-coll-aChi} diagrams relevant for 
$Z^c_{\calA\chi}$.
  The dashed lines depict the lepton and anti-lepton and the wavy-dashed lines the
  (anti-) collinear photons from the corresponding Wilson lines.
}
\end{figure}

The soft part of the operators  $\OpII_{m\chi}^{A1}$ and $\OpII_{\calA\chi}^{B1}$, 
\begin{align}
  \widehat{\OpII}_s(v) &
  = \oL{q}_s (v \nm) Y(v \nm, 0) \frac{\nms}{2} P_L h_\vb (0) 
         \big[Y_+^\dagger\, Y_- \big](0) ,
\end{align}
is common to both. We thus omit the subindex $i$, and write $\widehat{\OpII}_{i,s} = 
\widehat{\OpII}_s$. The discussion for $\OpII_{m\oL{\chi}}^{A1}$ and 
$\OpII_{\calA\oL{\chi}}^{B1}$ proceeds analogously after exchanging  
$n_- \leftrightarrow \np$ in the $\oL{q}_s [\ldots ]h_v$ part of the 
operator. The QED one-loop diagrams due to soft photons from the 
soft Wilson lines contributing to the renormalization of  $\widehat{\OpII}_s(v)$ 
are shown in \reffig{fig:SCET2-adm-soft}. Not shown is the vertex diagram 
from photon exchange between the heavy and light quark, and the field 
renormalization contribution. The QCD one-loop diagrams are the same as those 
that appear in the calculation of the renormalization of the leading-twist 
$B$-meson light-cone distribution amplitude \cite{Lange:2003ff}. 

To find the UV poles in dimensional regularization, we evaluate the operator 
between a heavy-light quark state and the vacuum, and regulate the infrared (IR) 
divergences by taking the external lines slightly off-shell. 
See \refapp{app:SCET-2-adm} for more details. Calculating in Feynman gauge, 
the dependence on the off-shell IR regulators cancels except for the tadpole-type 
diagram~(6) of~\reffig{fig:SCET2-adm-soft}. The remaining IR-regulator dependence 
is cancelled by the diagrams in \reffig{fig:SCET2-adm-coll} and 
\reffig{fig:SCET2-adm-coll-aChi} with collinear and anti-collinear photons 
$\np A_c$ and $\nm A_\oL{c}$, respectively. While at first sight, this 
appears to be in conflict with the factorization of the soft and (anti-) collinear 
sectors, we can subtract the overlap between soft and collinear and
anti-collinear regions by defining and renormalizing the soft operator
\begin{align}
  \label{eq:def-soft-op}
  \widetilde{\OpII}_s (v) & 
  \equiv \frac{\widehat{\OpII}_s(v)}
         {\big< 0\big|\big[Y_+^\dagger\, Y_- \big](0)\big| 0 \big>}\,.
\end{align}
For the operators $i = m\oL{\chi},\, \calA \oL{\chi}$ we proceed 
in complete analogy using the respective soft field product. 
The operator \refeq{eq:def-soft-op} is divided by 
the vacuum expectation value of the gauge-invariant product of Wilson lines
\begin{align}
  \big< 0\big|\big[Y_+^\dagger\, Y_- \big](0)\big| 0 \big> & 
  \equiv R_+ R_- .
\end{align}
At the one-loop order, this subtraction simply removes the tadpole diagram~(6) 
in~\reffig{fig:SCET2-adm-soft} from the soft operator. Beyond one-loop it 
ensures that the UV counterterm for the soft operator is independent of the 
IR regulator as is required for consistent operator renormalization. Further 
it ensures that the renormalization of the soft sector does not depend on the 
structure of the (anti-) collinear parts of the \SCETII{} operators, but only 
on the total charge of the final state associated to the (anti-) collinear 
direction. 

%%%%%%%%%%%%%%%%%%%%%%%%%%%%%%%%%%%%%%%%%%%%%%%%%%%%%%%%%%%%%%%%%%%%%%%%%%%%%%%%
%%%%%%%%%%%%%%%%%%%%%%%%%%%%%%%%%%%%%%%%%%%%%%%%%%%%%%%%%%%%%%%%%%%%%%%%%%%%%%%%

Using separate IR regulators for collinear and anti-collinear fields, we 
further factorized the vacuum expectation value of the Wilson lines into 
factors $R_+$ and
$R_-$, which depend only on the collinear and anti-collinear IR regulators,
respectively. This split can always be performed. At the one-loop order 
one obtains the sum of two terms, each of which depends only on one of the 
regulators; beyond, the one-loop IR divergence exponentiates. 
There is a freedom in the choice of splitting the product $R_+ R_-$ into 
the separate factors $R_+$ and $R_-$, which affects the definition of the 
collinear and anti-collinear renormalization constants discussed 
below.\footnote{We note the similarity to the factorization of the soft function 
in the definition of the transverse momentum-dependent parton distribution 
functions, see, for example, \cite{Echevarria:2012js}.} We adopt the symmetric 
convention, such that $R_+$ equals $R_-$ upon exchanging $\np\leftrightarrow \nm$.
This corresponds to rearranging the \SCETII{} operator \refeq{eq:SCET2-op-fact} as
\begin{align}
  \widetilde{\OpII}_i &
  \; = \; \frac{\widehat{\OpII}_{s}}{R_+ R_-} 
  \otimes R_+ \widehat{\OpII}_{i,c} 
  \otimes R_- \widehat{\OpII}_{i,\oL{c}}
  \; \equiv \; \widetilde{\OpII}_{s} 
  \otimes \widetilde{\OpII}_{i,c} 
  \otimes \widetilde{\OpII}_{i,\oL{c}} \,,
\end{align}
where now in the soft, collinear and anti-collinear factors 
$\widetilde{\OpII}$ can be renormalized consistently in contrast 
to the original $\widehat{\OpII}$.

We denote by $Z_s$ the UV renormalization factor of the Fourier
transform $\OpII_s (\omega) = \int \frac{dv}{2\pi} e^{i \omega v}
\widetilde{\OpII}_s (v)$ of the soft operator. At the one-loop order, 
$Z_s^{(1)}$ is the sum 
\begin{align}
  \label{eq:SCET-2-soft-RC}
  Z_s^{(1)} & 
  = \frac{\alE}{4\pi} Z^\text{QED}_s
  + \frac{\alS}{4\pi} Z^\text{QCD}_s 
\end{align} % CB: checked
of the QED and QCD contribution. The expressions for $Z^\text{QED}_s$ 
and  $Z^\text{QCD}_s$  are given in \refeq{eq:Z_s^QED} and 
\refeq{eq:Z_s^QCD}, respectively. As explained above, the tadpole 
diagram 6 of~\reffig{fig:SCET2-adm-soft} cancels with the corresponding diagram in
the denominator of \refeq{eq:def-soft-op}. With the help of \refeq{eq:gamma},
the corresponding anomalous dimension reads\footnote{See erratum at the end of the paper.}
\begin{eqnarray}
  \label{eq:Gamma_s-LCDA}
  \Gamma^s(\omega, \omega') 
  & = & \Big[ - \Gamma_s \ln \frac{\omega}{\mu}  
     - 5 \left(\frac{\alS}{4\pi} C_F + \frac{\alE}{4\pi} Q_q^2 \right)
  \Big] \,\delta(\omega - \omega')
\nonumber \\ 
  && - \,4 \left[\frac{\alS}{4 \pi} C_F + \frac{\alE}{4 \pi} 
  Q_q(Q_q + Q_\ell) 
   \right] F(\omega, \omega') \,,
\end{eqnarray} % CB: checked
where $F(\omega, \omega')$ is given in \refeq{eq:F(om,om')}. The anomalous 
dimension contains the cusp part $\Gamma_s$, which appeared already in the 
anomalous dimension \refeq{eq:cusp-ADMs} of the \SCETI{} operators. However, 
here it enters with the opposite 
sign and is multiplied by $\ln (\omega/\mu)$. Note that the QED part of
the anomalous dimension is proportional to the light-quark charge $Q_q$.

The soft operator fulfils the RGE\footnote{The minus sign on 
the right-hand side appears in accordance with the 
general convention \refeq{eq:def-RG-op_and_C} for the RGEs of 
operators and coefficient functions.}
\begin{align}
\label{eq:softRGE}
  \frac{d}{d \ln \mu} \OpII_s (\omega; \mu) &
  = - \int\! d \omega' \, \Gamma^s(\omega, \omega')\, \OpII_s (\omega'; \mu) \,,
\end{align}
which at the LL accuracy, i.e.~keeping only the cusp part of the
anomalous dimension, admits a solution of the form  
\begin{align}
\label{eq:softRGEU}
  \OpII_s (\omega; \mu) & 
  = U_s(\mu,\mu_s; \omega)\, \OpII_s (\omega; \mu_s) \,.
\end{align} % CB: checked
The LL soft RG evolution factor $U_s$ from an initial soft scale 
$\mu_s\sim\omega$ to $\mu$ is given by
\begin{align}
  \nonumber
  U_{s}(\mu,\mu_s;\omega) & =
  \exp\!\bigg[
 \frac{4\pi}{\alS(\mu_s)} \frac{C_F}{\beta_0^2}
 \Big(g_0(\eta_s) 
      + \frac{\alS(\mu_s)}{2\pi}  \beta_0 \ln\eta_s \,\ln\frac{\omega}{\mu_s}
 \Big) \bigg]
\\
  \label{eq:softRGEUsolution}
  \times \exp\! & \bigg[
 \frac{4\pi}{\alE(\mu_s)} \frac{Q_q\left(2 Q_\ell+Q_q\right)}{\beta_{0,\rm em}^2}
    \Big( g_0(\eta_{\rm em}) 
        + \frac{\alE(\mu_s)}{2\pi}  \beta_{0, \rm em} \ln\eta_{\rm em} \, \ln\frac{\omega}{\mu_s}
    \Big) \bigg]\qquad
\\ 
  & \xrightarrow{\text{DL}} 
  \exp\left[\frac{\Gamma_s}{2} \left(\ln^2\frac{\omega}{\mu_s}-\ln^2\frac{\omega}{\mu} \right) \right],
\end{align}
where for soft evolution $\eta_i$ stands for  
$\eta_i(\mu_s,\mu) \equiv \alpha_i(\mu_s)/\alpha_i(\mu)$ 
with $i = s,\, \text{em}$. In the last line we have taken the 
double-logarithmic approximation of the QCD and QED factor, but we will 
not make use of this approximation later on. 

The above result can be simplified by noting that 
$\omega, \mu_s \sim \Lambda$, hence $\ln\,(\omega/\mu_s)$ is never a large 
logarithm. Similar to the solution of the \SCETI{} evolution equation, we 
may drop such $\mathcal{O}(1)$ logarithms. In the present case, this renders 
the evolution factor independent 
of the momentum variable $\omega$, resulting in 
\begin{align}
  U_s(\mu,\mu_s) & =
  \exp\!\bigg[
      \frac{4\pi\,}{\alS(\mu_s)} \frac{C_F}{\beta_0^2} \,g_0(\eta_s) \bigg]\,
    \exp\!\bigg[
      \frac{4\pi}{\alE(\mu_s)} \frac{Q_q(2 Q_\ell + Q_q)}
      {\beta_{0,\rm em}^2} \, g_0(\eta_{\rm em}) \bigg]
  \nonumber
\\
  \label{eq:SCET2-softRG-ana-sol}
  & \xrightarrow{\text{DL}} 
  \exp\!\bigg[
      \frac{4\pi\,}{\alS(\mu_s)} \frac{C_F}{\beta_0^2} \,g_0(\eta_s) \bigg]\,
  \exp\!\bigg[-\frac{\alE}{2\pi} \big[Q_q(2 Q_\ell + Q_q)\big] \ln^2 \frac{\mu_s}{\mu}
      \bigg]\,.\quad
\end{align}
We note the same form as \refeq{eq:SCET1-RG-ana-sol} for the hard-collinear 
evolution, except now the evolution starts at $\mu_s$, and there is no 
$Q_\ell^2$ term in the anomalous dimension.

%
%--------+---------+---------+---------+---------+---------+---------+---------+
\subsubsection{Anti-collinear sector}

The anti-collinear sector is the same for both operators and given by the
anti-collinear lepton field $\widehat{\OpII}_\oL{c} = 
\ell_\oL{c}(0)$.\footnote{We note that 
this refers to the field including the anti-collinear Wilson line, 
which is invariant under anti-collinear gauge transformations, 
see \refeq{eq:def-ell_C}.} We define the anti-collinear operator
\begin{align}
\label{eq:Jcbar}
  \OpII_{\oL{c}} &
  \equiv \OpII_{i,\oL{c}} 
  = R_- \ell_\oL{c}(0) \,,
\end{align}
including the $R_-$ factor from the soft subtraction. The operator has a 
single open spinor index which is omitted for simplicity, as
the anomalous dimension is diagonal.

The one-loop diagrams needed to compute the anomalous dimension of the above
operator correspond to the anti-collinear part of the 
two diagrams in \reffig{fig:SCET2-adm-coll} and 
\reffig{fig:SCET2-adm-coll-aChi} involving $\nm A_\oL{c}$. The 
factor $R_-$, which originates from the soft tadpole diagram (6), ensures the 
cancellation of the off-shell IR regulator in the UV divergent part. 
We introduce the renormalization
constant $Z_\oL{c}$ associated with the UV counterterm, for which the 
one-loop result is given in \refeq{eq:Z^1_ac}.

The anti-collinear part obeys the RG equation  
\begin{align}
  \frac{d}{d\ln\mu}\OpII_{\oL{c}}(\mu) &
  = - \Gamma^\oL{c} \, \OpII_{\oL{c}}(\mu) \,,
\end{align}
with the one-loop anomalous dimension 
\begin{align}
  \label{eq:SCET2-ADM_ac}
  \Gamma^\oL{c} & 
  = \frac{\Gamma_c}{2} \left(\ln \frac{m_{B_q}}{\mu} - \frac{i \pi}{2} \right) 
  - \frac{\alE}{4\pi} 3 Q_\ell^2 \,, 
\end{align} % CB: checked
and the cusp anomalous dimension $\Gamma_c$ previously defined in 
\refeq{eq:cusp-ADMs}. The solution to LL accuracy is 
\begin{align}
  \label{eq:SCET2-U-ac}
  \OpII_{\oL{c}}(\mu) & 
  = U_{\oL{c}}(\mu,\mu_c)\, \OpII_{\oL{c}}(\mu_c) \,,
\end{align} % CB: checked
with 
\begin{align}
  \label{eq:SCET2_Uc}
  U_{\oL{c}}(\mu,\mu_c) & =
  \exp\!\bigg[
    -\frac{4\pi}{\alE(\mu_c)} \frac{Q_\ell^2}{\beta_{0,\rm em}^2}
    \Big( g_0(\eta_{\rm em}) 
        + \frac{\alE(\mu_c)}{2\pi}  \beta_{0,\rm em} \ln\eta_{\rm em} \, 
    \ln\frac{m_{B_q}}{\mu_c}
    \Big) \bigg]
\\ 
   & \xrightarrow{\text{DL}}
   \exp\left[-\frac{\Gamma_c}{4} \left(
      \ln^2\frac{m_{B_q}}{\mu_c} - \ln^2\frac{m_{B_q}}{\mu} \right) \right].
\end{align} % CB: checked
Here $\eta_{\rm em}$ is  $\eta_{\rm em}(\mu_c,\mu) = 
\alE(\mu_c)/\alE(\mu)$ for (anti-) collinear evolution. 
Note that we cannot neglect the 
first term in the exponent in this case, since $\ln\,(m_{B_q}/\mu_c)$ is a 
large logarithm. 

%
%--------+---------+---------+---------+---------+---------+---------+---------+
\subsubsection{Collinear sector}

The collinear part of operators \refeq{eq:def-SCET-II-ops-1} and 
\refeq{eq:def-SCET-II-ops-2} consists of the two operators
\begin{align}
  \label{eq:opcoll1}
  \OpII^{A1}_{c} &
  \equiv \OpII^{A1}_{m\chi,c} 
  = R_+\, \oL{\ell}_c(0)\, 4 m_\ell P_R \,,
\\
  \label{eq:opcoll2}
  \OpII^{B1}_{c}(w) &
  \equiv \OpII^{B1}_{\calA\chi,c} 
  = R_+\, (\np p) \int \frac{dt}{2\pi} \, e^{-i \oL{w}t \,n_+ p}\,
   \oL{\ell}_c (0)\, 2 \slashed \calA_{c\perp}(t \np ) P_R \,,
\end{align} % CB checked
with $\oL{w} \equiv 1-w$, 
which mix under renormalization. Similar to the anti-collinear part, the 
factor $R_+$ must be included to cancel the
IR regulator dependence in the anomalous dimension.  The $2 \times 2$ 
renormalization matrix has the structure
\begin{align}
  \begin{pmatrix}
    \left[ \OpII^{A1}_{c} \right]_\text{ren} \\[0.2cm]
    \left[ \OpII^{B1}_{c} \right]_\text{ren}
  \end{pmatrix} &
  = \begin{pmatrix}
    Z^c_{m\chi}\,    &  0 \\[0.2cm]
    Z^c_{\calA\chi,\, m\chi} & Z^c_{\calA\chi}
  \end{pmatrix} \otimes_{w'} \begin{pmatrix}
   \left[ \OpII^{A1}_{c} \right]_\text{bare} \\[0.2cm]
    \left[ \OpII^{B1}_{c} \right]_\text{bare}
  \end{pmatrix} ,
\end{align} % CB checked
where $\otimes_{w'}$ indicates the presence of the convolution with respect to
the collinear momentum fraction. Both operators have in common the collinear
lepton field $\oL{\ell}_c(0)$, for which the associated one-loop diagrams due
to the collinear Wilson lines are the diagrams in 
\reffig{fig:SCET2-adm-coll} and \reffig{fig:SCET2-adm-coll-aChi}  
which involve $\np A_c$. Further,
the operator $\OpII_{c}^{A1}$ contains an explicit factor of $m_\ell$ in the
$\oL{\text{MS}}$ scheme, which we assign to the collinear sector as can be
motivated by the one-loop matching calculation for this operator in
\refsec{sec:SCET-2-matching}. The operator $\OpII^{B1}_{c} $
contains the additional collinear photon field $\calA_{c\perp}^\mu (t\np)$,
which gives two more one-loop diagrams shown 
in \reffig{fig:SCET2-adm-coll-aChi}. The one-loop result of the 
diagonal elements $Z_{m\chi}^{c,(1)}$ and
$Z^{c,(1)}_{\calA\chi}$ are given in \refeq{eq:Z^1_chi} and 
\refeq{eq:Z^1_Achi}, respectively. 

For massless fermions in \SCETI{}, the mixing of B1-type operators into 
A1-type operators is absent. In \SCETII{} with non-zero fermion mass, 
we find the non-vanishing one-loop off-shell collinear matrix element of 
B1-type operator shown as the middle diagram  in the column labelled 
``\SCETII{}'' in \reffig{fig:heuristic-scheme}. Its divergent part is 
proportional to the tree-level matrix element of the  
mass-suppressed A1-type operator. Explicitly, the matrix element is given by 
\begin{align}
  \np p \int\! \frac{dt}{2\pi} & 
  \, e^{-i\oL{w} t\, (\np p)} \,
  \big\langle \ell(p) \big|\oL{\ell}_c (0) \, \calA_{c \perp}^\mu (t\np)
  \big| 0 \big\rangle 
\nonumber \\
  & = - \frac{\alE}{4\pi} Q_\ell\,
  \oL{w} \left[ \frac{1}{\epsilon} + \ln\frac{\mu^2}{\oL{w}\left(m_\ell^2 - p^2 w \right)}
  \right] m_{\ell}\, \oL{u}_{c}(p) \gamma_\perp^\mu \,,
\label{eq:collME}
\end{align}  % CB: checked
yielding the mixing counterterm 
\begin{align}
  \label{eq:ZAmp}
  Z_{\calA^\mu\chi_\alpha,\, m\chi_\beta}^{c(1)} (w) &
  = \frac{\alE}{4\pi} \frac{Q_\ell}{\epsilon} \oL{w} 
    \left[\gamma_\perp^\mu \right]_{\alpha\beta} 
\end{align}  % CB checked
for the general case with open Dirac and Lorentz indices. Contracting them 
with $2 \gamma^{\perp \mu} P_R$, the \SCETII{} mixing counterterm pertinent 
to the operators \refeq{eq:opcoll1} and \refeq{eq:opcoll2} is 
\begin{align}
  \label{eq:ZAm}
  Z_{\calA\chi,\, m\chi}^{c(1)} ( w) &
  = \frac{\alE}{4\pi} \frac{Q_\ell}{\epsilon} \oL{w}  \,.
\end{align}  % CB checked

The renormalization of the collinear fields leads to the coupled system of RGEs,
\begin{align}
  \label{eq:SCET2-rge-coll}
  \frac{d}{d\ln\mu} \begin{pmatrix}
    \OpII^{A1}_{c}(\mu)  \\[0.2cm]
    \OpII^{B1}_{c}(w; \mu) 
  \end{pmatrix} &
  = - \begin{pmatrix}
    \Gamma^c_{m\chi}\,    &  0 \\[0.2cm]
    \Gamma^c_{\calA\chi,\, m\chi} & \Gamma^c_{\calA\chi}
  \end{pmatrix} \otimes_{w'} \begin{pmatrix}
    \OpII^{A1}_{c}(\mu) \\[0.2cm]
     \OpII^{B1}_{c} (w'; \mu)
  \end{pmatrix} ,
\end{align} % CB: checked
with the one-loop collinear anomalous dimensions given 
by\footnote{A very similar operator mixing calculation 
appears in the SCET analysis of power-suppressed two-jet operators 
sourced by a new heavy particle \cite{Alte:2018nbn}. In this 
application, the insertion of an external Higgs field operator 
corresponds to the lepton mass factor. The off-diagonal anomalous 
dimension in  \cite{Alte:2018nbn} misses the factor $\oL{w}$, 
because the one-particle reducible diagram with the Higgs 
insertion on the external leg was not included.} 
\begin{eqnarray}
  \label{eq:SCET2-ADM-mChi}
  \Gamma_{m\chi}^c & 
  =& \frac{\Gamma_c}{2} \left(\ln \frac{m_{B_q}}{\mu} - \frac{i \pi}{2} \right)
  + \frac{\alE}{4\pi} 3 Q_\ell^2 \,,
\\
  \label{eq:SCET2-ADM-amChi}
  \Gamma_{A\chi,\, m\chi}^c (w) & 
  =& \frac{\alE}{4\pi} \, 2 Q_\ell \, \oL{w} \,,
\\ \nonumber
  \Gamma_{A\chi}^c (w, w') & 
  =& \delta(w - w') \left[
     \frac{ \Gamma_c}{2} \left(\ln \frac{m_{B_q}}{\mu} - \frac{i \pi}{2} \right)
          + \frac{\alE}{4\pi} \, Q_\ell^2 (4 \ln w - 6)  
    \right]
\\ &&  \label{eq:SCET2-ADM-aChi}
   - \,\frac{\alE}{4\pi} \, 2 Q_\ell^2 \,
     \gamma_{\calA\chi,\, \calA\chi} (w, w') \,.
\end{eqnarray}  % CB checked
The non-cusp anomalous dimension $\gamma_{\calA\chi,\, \calA\chi} (w, w')$ 
is provided for completeness in \refeq{eq:gam-calAchi}. The 
opposite sign of the non-cusp term in \refeq{eq:SCET2-ADM-mChi}
compared to \refeq{eq:SCET2-ADM_ac} arises from the anomalous dimension 
of the $\oL{\text{MS}}$ lepton mass in the definition of 
$\OpII^{A1}_{c}$.

At LL accuracy, keeping only the cusp anomalous dimension terms, 
the system of RGEs~\refeq{eq:SCET2-rge-coll}
is easily solved first for $\OpII^{A1}_{c}(\mu)$, and subsequently for 
$\OpII^{B1}_{c}(w,\mu)$, yielding 
\begin{align}
\OpII^{A1}_{c}(\mu) & = U_{c}(\mu,\mu_c)\, \OpII^{A1}_{c}(\mu_c) \,, 
\label{eq:SCET2-RG-sol-A1}\\
\label{eq:SCET2-RG-sol-B1}
\OpII^{B1}_{c}(w; \mu) &
  = U_{c}(\mu,\mu_c)\,\bigg[\OpII^{B1}_{c}(w; \mu_c) 
     -  \,   \frac{Q_\ell \, \oL{w}}{\beta_{0,\rm em}} \ln\eta_{\rm em}\,
     \OpII^{A1}_{c}(\mu_c) \bigg] \,.
\end{align} 
Here  $\eta_{\rm em}$ equals  $\eta_{\rm em}(\mu_c,\mu) = 
\alE(\mu_c)/\alE(\mu)$ and $U_{c}(\mu,\mu_c)= U_\oL{c}(\mu,\mu_c)$ 
defined in \refeq{eq:SCET2-U-ac} with LL accuracy, because the cusp  
part of the anomalous dimensions $\Gamma^c_{m\chi}$, $\Gamma^c_{\calA\chi}$ 
is the same as of $\Gamma^\oL{c}$ in \refeq{eq:SCET2-ADM_ac}. 

Naively, the second term in the bracket in
\refeq{eq:SCET2-RG-sol-B1} appears suppressed as it contains $\alE$ times 
a single logarithm. However, the tree-level matrix element of the operator 
$\OpII^{B1}_{c}(\mu_c)$ vanishes for $B_q\to \mu^+\mu^-$. Hence, 
the second term is actually the leading term, as outlined in 
\reffig{fig:heuristic-scheme}. The matrix element of the B1-operator 
contributes at the one-loop order, and does not contain large logarithms 
because it is evaluated at the collinear scale.
For the LL accuracy, it is then enough to choose the initial condition 
$\OpII^{B1}_{c}(w; \mu_c) = 0$. In the double-logarithmic approximation,
we could further replace 
\begin{align}
\label{eq:mixingDL}
  \frac{1}{\beta_{0,\rm em}} \ln\eta_{\rm em} &
  \;\; \xrightarrow{\text{DL}} \;\; \frac{\alE}{2\pi}\ln \frac{\mu}{\mu_c} \,. 
\end{align}

%
%--------+---------+---------+---------+---------+---------+---------+---------+
\subsubsection{Complete \SCETII{} operator and evolution}

For convenience, we summarize the renormalization and RGEs for the full
\SCETII{} operators $\OpII_{m\chi}^{A1}$ and $\OpII_{\calA\chi}^{B1}$. The
operator mixing in the collinear sector leads to a $2 \times 2$ renormalization
matrix
\begin{align}
  \begin{pmatrix}
    \left[ \OpII_{m\chi}^{A1}     \right]_\text{ren} \\[0.2cm]
    \left[ \OpII_{\calA\chi}^{B1} \right]_\text{ren}
  \end{pmatrix} &
  = \begin{pmatrix}
    Z_{m\chi,\, m\chi}     &  0 \\[0.2cm]
    Z_{\calA\chi,\, m\chi} & Z_{\calA\chi,\, \calA\chi}
  \end{pmatrix} \otimes_{\omega',w'} \begin{pmatrix}
    \left[ \OpII_{m\chi}^{A1}     \right]_\text{bare} \\[0.2cm]
    \left[ \OpII_{\calA\chi}^{B1} \right]_\text{bare}
  \end{pmatrix} ,
\end{align}
where appropriate convolutions are indicated by $\otimes$. The 
renormalization constants are the products of the soft, collinear and 
anti-collinear factors discussed before,
\begin{align}
  \label{eq:SCET2-Z-mChi}
  Z_{m\chi,\, m\chi}(\omega, \omega') & 
  = Z_s(\omega,\omega') \, Z_\oL{c} \, Z^c_{m\chi} \,, 
\\
  \label{eq:SCET2-Z-aChi}
  Z_{\calA\chi,\, \calA\chi}(\omega, \omega'; w, w') &
  = Z_s(\omega, \omega') \, Z_\oL{c} \, Z^c_{\calA\chi}(w, w') \,, 
\\
  \label{eq:SCET2-Z-ZAm}
  Z_{\calA\chi,\, m\chi} (\omega,\omega'; w) &
  = Z_s(\omega,\omega') \, Z_\oL{c}\,  Z_{\calA\chi,\, m\chi}^{c}(w) \,,
\end{align} % CB: checked
and the anomalous dimension matrix becomes the sum of the soft, collinear,
and anti-collinear anomalous dimensions. We can write this in the 
form 
\begin{align}
  \label{eq:SCETII-adm}
  \Gamma^\text{II} &
  = \Gamma^s(\omega, \omega') \begin{pmatrix} 1 & 0 \\
    0 & \;\; \delta(w - w') \end{pmatrix}
 % \nonumber\\&
  + \delta(\omega-\omega') \begin{pmatrix} 
      \Gamma^{\oL{c}} + \Gamma_{m\chi}^c    & 0 \\ 
      \Gamma_{A\chi,\, m\chi}^c(w) & 
      \quad  \delta(w-w')\,\Gamma^{\oL{c}} + \Gamma_{A\chi}^c (w, w')
\end{pmatrix} ,
\end{align} % CB checked
with entries defined in \refeq{eq:Gamma_s-LCDA}, \refeq{eq:SCET2-ADM_ac} and 
\refeq{eq:SCET2-ADM-mChi}--\refeq{eq:SCET2-ADM-aChi}.
Since the soft part is independent of the collinear and anti-collinear building
blocks, it enters only the diagonal entries. Both operators
$\OpII_{m\chi}^{A1}$ and $\OpII_{\calA\chi}^{B1}$ contain the cusp anomalous
dimension parts $\Gamma_s$ and $\Gamma_c$ from \refeq{eq:cusp-ADMs}, which
appeared already in the anomalous dimension of the \SCETI{} operators. 
Contrary to \SCETI{}, in \SCETII{} $\Gamma_s$ enters with the opposite 
sign and is multiplied by $\ln (\omega/\mu)$, but $\Gamma_c$ is 
multiplied by $\ln (m_{B_q}/\mu)$ as in \SCETI{}. Finally we note that the
soft and (anti-)collinear anomalous dimensions are separately gauge invariant.

We collect at this point all evolution factors, including the 
evolution in \SCETI{}. From $H_i(\mu_b) \OpI_i(\mu_b) = H_i(\mu) \OpI_i(\mu)$ 
and \refeq{eq:sol-RGE-Hi}, we obtain
\begin{eqnarray}
\OpI_i(\mu_b) 
&=& \exp\big[S_{\ell}(\mu_b,\, \mu)\, + 
\, S_{q}(\mu_b,\, \mu) 
\big]
\OpI(\mu)
\nonumber\\
&=& \exp\big[S_{\ell}(\mu_b,\, \mu)\, + \, S_{q}(\mu_b,\, \mu) 
\big] \,\Big[\OpII_{s} 
  \otimes \OpII_{c}^{B1} 
  \otimes \OpII_{\oL{c}}\Big](\mu)\,.
\end{eqnarray}
In passing to the second line, we have matched the \SCETI{} operator 
at the scale $\mu\ll \mu_b$ at tree level to the \SCETII{} operator. 
We also omitted the \SCETII{} matching coefficient, which does not 
change the structure of the result. (The precise matching relation will be given in the following subsection.)  
Next we use the \SCETII{} evolution factors to write 
\begin{eqnarray}
\OpI_i(\mu_b)
&=& \exp\big[S_{\ell}(\mu_b,\, \mu)\, + 
\, S_{q}(\mu_b,\, \mu) 
\big] \, U_s(\mu,\mu_s)\OpII_{s}(\mu_s)
\,\otimes U_{\oL{c}}(\mu,\mu_c)\OpII_{\oL{c}}(\mu_c)
\nonumber\\
&&
\otimes \,U_{c}(\mu,\mu_c)\,\left[\OpII^{B1}_{c}(w; \mu_c) 
     -  \,   \frac{Q_\ell \, \oL{w}}{\beta_{0,\rm em}} 
\ln\eta_{\rm em}\,
     \OpII^{A1}_{c}(\mu_c) \right]
\nonumber\\
&=& \exp\big[S_{\ell}(\mu_b,\, \mu_c)\big]\, 
\exp\big[S_{q}(\mu_b,\, \mu) 
\big] \, U_s(\mu,\mu_s)\OpII_{s}(\mu_s)
\OpII_{\oL{c}}(\mu_c)
\nonumber\\
&&
\otimes \,\left[\OpII^{B1}_{c}(w; \mu_c) 
     -  \,   \frac{Q_\ell \, \oL{w}}{\beta_{0,\rm em}} 
\ln\eta_{\rm em}\,
     \OpII^{A1}_{c}(\mu_c) \right].
\label{eq:opevolved}
\end{eqnarray}
In the final expression we combined the part of the hard-collinear 
evolution contained in $S_{\ell}(\mu_b,\, \mu)$ with 
the collinear and anti-collinear factors making use of\footnote{To
obtain the following identity exactly, we make use of the freedom 
to replace $m_{B_q}$ by $\mu_b$ in \refeq{eq:SCET2_Uc} in the 
LL approximation. Alternatively, we may choose $\mu_b = m_{B_q}$.}
\begin{equation}
  \exp\big[S_{\ell}(\mu_b,\, \mu)\big] \, 
  U_{\oL{c}}(\mu,\mu_c) \, U_{c}(\mu,\mu_c) 
  = \exp \big[ S_{\ell}(\mu_b,\, \mu_c) \big] \,.
  \label{eq:SUUrelation}
\end{equation}
This shows that the logarithms proportional to $Q_\ell^2$ involving 
only the final-state leptons arise from uniform evolution from 
the hard scale $m_{B_q}$ to the collinear scale $m_\ell$. In the 
final expression \refeq{eq:opevolved}, we may drop the term 
$\OpII^{B1}_{c}(w; \mu_c)$, since in the LL approximation the 
initial condition of the B1-operator at the collinear scale can 
be set to zero, as discussed above. When one takes the matrix element 
of \refeq{eq:opevolved}, no large logarithms appear in the 
matrix elements of the $\OpII$, since they are evaluated at their 
natural scale, and all large logarithms are already summed.

%
%--------+---------+---------+---------+---------+---------+---------+---------+
\subsection{Matching}
\label{sec:SCET-2-matching}

We match the \SCETI{} operators $\OpI_i$ at the hard-collinear
scale $\mu_{hc}$ on the \SCETII{} operators in momentum space. The
matching equations read 
\begin{align}
  \OpI_i (u) & 
  = \int\! d\omega \left[ J_m (u; \omega)\, \OpII_{m\chi}^{A1} (\omega)
  + \int\! dw \, J_A (u; \omega, w)\, \OpII_{\calA\chi}^{B1}(\omega, w) \right] ,
\label{eq:SCETItoIImatch}\\
  \OpI_\oL{i} (u) & 
  = \int\! d\omega \left[ J_\oL{m} (u; \omega)\, \OpII_{m\oL{\chi}}^{A1} (\omega)
  + \int\! dw \, J_\oL{A} (u; \omega, w)\, \OpII_{\calA\oL{\chi}}^{B1}(\omega, w) \right] ,
\end{align} % CB: checked
with perturbative matching coefficients $J_i$, also called ``jet functions'', 
which account for the (anti-) hard-collinear modes. There are no 
leading-power interactions between soft and hard-collinear fermions in 
\SCETI{}, hence to obtain the power-enhanced contribution we must include 
a single insertion of the power-suppressed Lagrangian 
$\mathcal{L}^{(1)}_{\xi q}(x)$, given in \refeq{eq:SCET-Lag-xiq1}, to 
convert the hard-collinear quark into a soft quark. 
The jet functions $J_{m,\, \oL{m}} (u;\omega)$ start at the one-loop order, 
while $J_{A,\, \oL{A}} (u; \omega, w)$ coincide at tree level  
with expressions 
\begin{align}
  \label{eq:JA0}
  J_A^{(0)}(u; \omega, w, \mu) 
  = J_\oL{A}^{(0)}(u; \omega, w, \mu) & 
  = - \frac{Q_q}{\omega} \,\delta(u - w)  
\end{align} % CB: checked
{}from  the lower diagram depicted in the column labelled ``\SCETI{}'' 
in \reffig{fig:heuristic-scheme}.
The explicit calculation shows that both operators $\OpI_i$ for $i = 9, 10$
contribute equally to $\OpII_{\calA\chi}^{B1}$, whereas $i = \oL{9}, \oL{10}$
contribute to $\OpII_{\calA\oL{\chi}}^{B1}$ with an opposite sign.
Summarizing, we find
\begin{align}
  \label{eq:SCET-1-1}
  H_9 \otimes_u \OpI_9 + H_{10} \otimes_u \OpI_{10} & & 
  \to & & &
  (H_9 + H_{10}) \otimes_u  J_A \otimes_{\omega,w} \OpII_{\calA\chi}^{B1} \,, 
\\  
  \label{eq:anti-SCET-1-2}
  H_\oL{9} \otimes_u \OpI_\oL{9} + H_\oL{10} \otimes_u \OpI_\oL{10} & &
  \to & & & 
  (H_\oL{9} - H_\oL{10}) \otimes_u J_\oL{A} \otimes_{\omega,w} \OpII_{\calA\oL{\chi}}^{B1} \, .
\end{align} % CB: checked
Here we anticipate that the relative minus sign in front of $H_{10}$ 
and $H_\oL{10}$ to the right-hand side of the arrows, 
together with~\refeq{eq:SCETI-hard-matching},
is the origin of the cancellation of the Wilson coefficient $C_{10}$ at 
the amplitude level after adding the collinear and anti-collinear 
contributions. Thus we reproduce in the SCET approach our previous finding 
\cite{Beneke:2017vpq} that the power-enhanced contribution 
\refeq{eq:mainresult} does not depend on~$C_{10}$. 

As an aside, we note that for the charged $B$-meson leptonic decay 
$B_u\to \mu\bar\nu_\mu$, the anti-collinear 
parts are not present because the anti-lepton is replaced by the chargeless
neutrino, thus \refeq{eq:anti-SCET-1-2} will not contribute. Further, there
is only a single operator with Wilson coefficient $C$ and the Dirac structure 
$\gamma_\mu (1 - \gamma_5)$ in the lepton current. This implies the 
replacements $C_9 = C$ and $C_{10} = -C$, such that $H_9 + H_{10} = 0$ 
and no power-enhanced contribution arises for this process.

Returning to $B_q \to \mu^+\mu^-$, let us briefly discuss also the 
one-loop matching of the coefficients $J_{m,\, \oL{m}} (u;\omega)$.
The one-loop matrix elements of the left-hand sides of 
\refeq{eq:SCET-1-1} and \refeq{eq:anti-SCET-1-2} in a 
$\langle\ell\bar{\ell}|...|\bar{q} b\rangle$ 
state used to extract $J_{m,\, \oL{m}} (u;\omega)$ contain an 
IR divergence. This divergence is reproduced on the right-hand side 
in \SCETII{} by the scaleless one-loop matrix element and the renormalization
constant \refeq{eq:ZAm} convoluted with the tree-level (anti-) 
hard-collinear jet function~\refeq{eq:JA0}. After we include these 
contributions in the matching, the IR divergence cancels and we find
\begin{align}
  \label{eq:Jm1}
  J_m^{(1)} (u; \omega; \mu) &
  = \frac{\alE}{4\pi} Q_\ell Q_q \frac{1-u}{\omega} 
    \left[ \ln\left(\frac{\omega\, \np p_\ell}{\mu^2} \right)
         + \ln\left[u (1-u)\right] \right]
    \theta(u) \, \theta(1-u) .
\end{align} % CB: checked
The result for $J_\oL{m}^{(1)}$ is obtained by the replacement 
$\np p_\ell \to \nm p_\oL{\ell}$. The cancellation of the IR divergence 
in the matching of \SCETI{} on 
\SCETII{} confirms the short-distance nature of the jet function, and 
serves as a check of the EFT setup. We note that when $\mu \ll \mu_{hc}$, 
the one-loop expression above contains a large logarithm. This is 
precisely the logarithm that is generated by RG evolution and 
correctly taken into account by the LL result \refeq{eq:SCET2-RG-sol-B1},   
\refeq{eq:mixingDL}. The non-logarithmic term $\ln\left[u (1-u)\right]$ 
enters \refeq{eq:mainresult} together with the non-logarithmic 
contributions from the collinear matrix element \refeq{eq:M_A-1} below.

%--------+---------+---------+---------+---------+---------+---------+---------+
%
%
%
%--------+---------+---------+---------+---------+---------+---------+---------+
\section{\boldmath
  QED effects and the $B$-meson decay constant and LCDA}
\label{sec:fB-LCDA-defs}

Before turning to the factorized matrix element for the 
power-enhanced part of the $B_q\to \mu^+\mu^-$ amplitude, we 
discuss the hadronic matrix element of the soft 
operator $\widetilde{\OpII}_s (v)$~\refeq{eq:def-soft-op}, which 
is related to the $B$-meson decay constant and the leading-twist $B$-meson 
LCDA \cite{Grozin:1996pq,Beneke:2000wa}. 
However, the additional soft Wilson lines in the \SCETII{} operators 
\refeq{eq:def-SCET-II-ops-1}--\refeq{eq:def-SCET-II-ops-4} and 
 $\widetilde{\OpII}_s (v)$ imply that the hadronic matrix element 
does not coincide with the universal $B$-meson LCDA that would 
appear in the absence of electromagnetic interactions, and 
indicate a dependence on the final-state particles of the specific 
process. We discuss these issues in this section. 

We thus define the generalized and process-dependent $B$-meson 
LCDA~$\Phi_+(\omega)$ by the soft matrix element of the operator 
$\widetilde{\OpII}_s (v)$ 
\begin{eqnarray}
  \label{eq:def-B-LCDA}
(-4) \big<0 \big|\widetilde{\OpII}_s (v) \big| \oL{B}_q (p) \big>  
&=& \frac{\big< 0 \big| \oL{q}_s (v \nm) Y(v\nm,0)  \nms \gamma_5 \, h_\vb(0) \, 
    Y^\dagger_+(0) Y_-(0) \big| \oL{B}_q (p) \big> }
{ \big< 0\big|\big[Y_+^\dagger\, Y_- \big](0)\big| 0 \big>} 
\nonumber\\
&\equiv& i m_{B_q} \int_0^\infty \!\!\! d\omega \, 
       e^{-i\omega v} \,\mathscr{F}_{B_q} \Phi_+ (\omega) \,.
\end{eqnarray} 
The analogous definition holds for the anti-collinear case after 
interchanging $n_+ \leftrightarrow n_-$ in the $\oL{q}_s [\ldots ]h_v$
part of the operator, but with the same 
function $\Phi_+ (\omega)$. As an overall factor we include the 
generalized process-dependent $B$-meson decay constant 
$\mathscr{F}_{B_q}$ in the presence of electromagnetic corrections, defined through the local matrix 
element
\begin{align}
\label{eq:def-B-Fb}
  \frac{\big< 0 \big| \oL{q}_s (0) \gamma^\mu \gamma_5\, h_\vb(0) \, 
    Y^\dagger_+(0) Y_-(0) \big| \oL{B}_q (p) \big>}
{ \big< 0\big|\big[Y_+^\dagger\, Y_- \big](0)\big| 0 \big>} &
  = i \mathscr{F}_{B_q}m_{B_q} \vb^\mu ,
\end{align}
where $p=m_{B_q} \vb$, and $\vb^\mu$ is the four-velocity label of 
the heavy-quark field. Since we are working with the heavy
quark field in HQET, $\mathscr{F}_{B_q}$ is the so-called static 
$B$-meson decay constant. It is related to the decay constant in 
full QCD and QED by matching corrections at the hard scale. 
The generalized $B$-meson LCDA satisfies the RGE \refeq{eq:softRGE}, 
\begin{align}
  \frac{d}{d\ln \mu } \big[\mathscr{F}_{B_q}(\mu)\, 
\Phi_+ (\omega; \mu)\big] &
 = - \int_0^\infty \! d\omega' \, \Gamma^s (\omega,\omega')\, 
\mathscr{F}_{B_q}(\mu)\,\Phi_+(\omega'; \mu), 
\end{align}
with the anomalous dimension kernel $\Gamma^s$ given in 
\refeq{eq:Gamma_s-LCDA} at the one-loop order. Note that 
it depends on the charges $Q_\ell$ of the leptons in the 
final state. Keeping only the 
cusp part as before, the solution is 
\begin{align}
\mathscr{F}_{B_q}(\mu)\, \Phi_+ (\omega; \mu) & 
 = U_s(\mu,\mu_s; \omega)\, \mathscr{F}_{B_q}(\mu_s)\, 
 \Phi_+ (\omega; \mu_s) 
\label{eq:UsoftBLCDA}
\end{align}
with $U_s(\mu,\mu_s; \omega)$ from \refeq{eq:softRGEUsolution}.

In practice, owing to the smallness of $\alE$, we can treat QED
effects on hadronic matrix elements perturbatively. Since we wish 
to sum logarithmic QED effects to all orders, the expansion of 
the matrix element in $\alE$ must be done at the soft scale 
$\mu_s \sim \Lambda$, where the matrix element contains no large 
logarithms. We can then use the RGE including the QED 
anomalous dimension to sum the large logarithms between the 
soft and the hard-collinear and hard scale. We therefore define 
the expansions
\begin{align}
    \mathscr{F}_{B_q}(\mu_s) &=
   \sum_{n=0}^{\infty} \left( \frac{\alE(\mu_s)}{4\pi} \right)^{\!n}  
    F^{(n)}_{B_q}(\mu_s)\,,
\label{eq:QEDexpansionFB}\\
  \mathscr{F}_{B_q}(\mu_s) \Phi_+(\omega; \mu_s) &
  = \sum_{n=0}^{\infty} \left( \frac{\alE(\mu_s)}{4\pi}\right)^{\!n}  
    F^{(n)}_{B_q}(\mu_s)\, \phi^{(n)}_+ (\omega; \mu_s)\,,
\label{eq:QEDexpansionPhiB}
\end{align}
of the $B$-meson decay constant and LCDA. The leading terms in the 
expansion coincide with the standard $B$-meson decay 
constant $F_{B_q}(\mu)$ and LCDA $\phi_+(\omega; \mu)$ defined 
in the absence of QED {\em at the soft scale}, that is, 
$F^{(0)}_{B_q}(\mu_s) \equiv F_{B_q}(\mu_s)$ and 
$\phi^{(0)}_+(\omega; \mu_s) \equiv \phi_+(\omega; \mu_s)$, 
respectively. However, they evolve differently to $\mu\gg \mu_s$, 
since the RGE for $\phi_+(\omega; \mu)$ does not include 
QED effects. To be specific, write 
\begin{equation}
U_s(\mu,\mu_s; \omega,\omega') = 
U_s^{\rm QCD}(\mu,\mu_s; \omega,\omega')\,
U_s^{\rm QED}(\mu,\mu_s; \omega,\omega')\,,
\end{equation}
where $U_s^{\rm QCD}(\mu,\mu_s; \omega,\omega')$ is defined as 
$U_s(\mu,\mu_s; \omega,\omega')$ with the electromagnetic 
coupling $\alE$ set to zero, 
and $U_s^{\rm QED}(\mu,\mu_s; \omega,\omega')$ as the 
rest.\footnote{Note that this definition implies that 
in general $U_s^{\rm QED}(\mu,\mu_s; \omega,\omega')$ depends 
on the strong coupling, although not at the LL accuracy. It is 
defined as the additional evolution caused by  
QED, and therefore includes mixed QED-QCD effects. We also 
added the second argument $\omega'$, such that these general  
definitions are valid beyond the LL approximation.} In other words, 
$U_s^\text{QED}(\mu,\mu_s)$ fulfils the RGE 
\begin{equation}
\label{eq:def-U^QED}
\frac{d}{d\ln\mu} U_s^\text{QED}(\mu,\mu_s; \omega,\omega') 
 = - \left[\Gamma^s - \Gamma^s \big|_{\alE \to 0}\right] 
U_s^\text{QED}(\mu,\mu_s; \omega,\omega')
\end{equation}
with initial condition  $U_s^\text{QED}(\mu_s,\mu_s; \omega,\omega') 
= \delta(\omega-\omega')$. Since  
$U_s^{\rm QCD}(\mu,\mu_s; \omega,\omega')$ is the evolution 
factor for the standard $B$-meson LCDA in the absence of QED, 
we have the relation 
\begin{eqnarray}
F_{B_q}^{(0)}(\mu)\, \phi^{(0)}_+(\omega; \mu) & = &
U_s^{\rm QCD}(\mu,\mu_s; \omega,\omega')\,
U_s^{\rm QED}(\mu,\mu_s; \omega,\omega')\, \otimes_{\omega'}
\big[F_{B_q}^{(0)}(\mu_s)\, \phi^{(0)}_+(\omega; \mu_s)\big] 
\nonumber\\
 & = &
U_s^{\rm QED}(\mu,\mu_s; \omega,\omega')\,
U_s^{\rm QCD}(\mu,\mu_s; \omega,\omega')\, \otimes_{\omega'}
\big[F_{B_q}(\mu_s)\, \phi_+(\omega; \mu_s)\big] 
\nonumber\\
&=& U_s^\text{QED}(\mu,\mu_s; \omega, \omega') \otimes_{\omega'}
   \big[F_{B_q}(\mu) \, \phi_+(\omega'; \mu)\big] , 
  \label{eq:phiQCD}
\end{eqnarray} 
at an arbitrary scale. 

Higher-order terms in the expansion \refeq{eq:QEDexpansionFB}, 
\refeq{eq:QEDexpansionPhiB} define non-universal, non-local QCD 
(more precisely, HQET) matrix elements that have to be evaluated 
nonperturbatively. Since $\alE$ is small, only a few terms will 
be needed in practice. For example, the computation of the 
time-ordered product of the electromagnetic current 
$j^{\text{em}}_\mu(x)$ with the soft quark fields contained in the
\SCETII{} operators contributes to $\phi^{(1)}_+(\omega)$ and 
$F^{(1)}_{B_q}$. The decay constants $F^{(n)}_{B_q}(\mu_s)$ and  
LCDAs $\phi^{(n)}_+(\omega)$ at the scale $\mu_s$ provide a basis 
of initial conditions for the systematic inclusion and 
resummation of QED effects. At the leading and 
next-to-leading logarithmic (NLL) accuracy, 
only the universal objects $F_{B_q}(\mu_s)$
and $\phi_+(\omega)$ need to be known. For N${}^{k+1}$LL or 
fixed-order N${}^k$LO accuracy, the expansions 
\refeq{eq:QEDexpansionFB}, \refeq{eq:QEDexpansionPhiB}
can be truncated at $n=k$.

The above discussion, applicable to $B_q\to\ell^+\ell^-$, 
illustrates some complications related to the factorization of QED
corrections for exclusive $B$-meson decays. Only the leading 
and next-to-leading QED logarithms can be computed without  
introducing new QED-specific nonperturbative hadronic matrix 
elements. To be more explicit on the process dependence of the $B$-meson 
LCDA and the decay constant in the presence of QED, we consider 
defining the QED gauge-invariant generalization of the 
standard LCDA by 
\begin{align}
  \label{eq:def-B0-LCDA}
  \big< 0 \big| \oL{q}_s (v \nm) Y(v\nm, 0) \nms \gamma_5 \, h_\vb(0)  
  \big| \oL{B}_q (p) \big>  &
  \equiv i  m_{B_q} \int_0^\infty \!\!\! d\omega \, 
       e^{-i\omega v} \, \mathscr{F}^0_{B_q}\Phi^0_+ (\omega) \,,
\end{align} 
where the matrix element should be evaluated with the QCD and 
QED Lagrangians. At least the local matrix elements, defining 
$\mathscr{F}^0_{B_q}$, could be computed with lattice QCD. 
This is indeed a valid definition, however, it would be 
relevant in factorization theorems for processes like 
$B_q \to \gamma \gamma$ or $B_q \to \nu \oL{\nu}$ with no 
charged particles in the final state.
It cannot be used for $B_q \to \ell^+\ell^-$. In fact, the 
functions $\Phi^0_+$ and $\Phi_+$, when evolved to scales 
$\mu\gg \mu_s$  differ already in the LL approximation, since they 
have different cusp anomalous dimensions. The one for $\Phi^0_+$
does not contain the terms proportional to $Q_q Q_\ell$; in 
particular, at the one-loop order, the diagrams
3--5 in \reffig{fig:SCET2-adm-soft} are absent. In general, 
the presence of non-local Wilson lines even in the definition of 
naively local objects such as the $B$-meson decay 
constant, see \refeq{eq:def-B-Fb}, provides a serious
obstacle to any attempt to include QED effects in lattice 
computations of hadronic matrix elements for processes with 
energetic, charged particles in the final state.

Another interesting example is the leptonic charged $B$-meson 
decay $B_u \to \ell\bar\nu_\ell$. In this case, we need to 
introduce an auxiliary Wilson line to achieve 
soft-collinear factorization. The LCDA is then defined via the 
soft matrix element
\begin{align}
  \label{eq:def-Bpm-LCDA}
  \frac{\big< 0 \big| \oL{q}'_s (v \nm) \widetilde{Y}(v\nm,0) \nms \gamma_5 \, 
        h_\vb(0) Y^\dagger_+(0) \big| \oL{B}_u (p) \big> }
       {\big< 0 \big| Y_v(0) Y^\dagger_+(0)  \big|0 \big>} &
  \equiv i  m_{B_u} \int_0^\infty \!\!\! d\omega \, 
       e^{-i\omega v} \mathscr{F}^\pm_{B_u}\Phi^\pm_+ (\omega)\,.
\end{align} 
The QCD+QED Wilson line that ensures gauge invariance 
is now given by 
\begin{align}
  \widetilde{Y}(v\nm, 0) & 
  = \oL{Y}_{q'+}(v\nm) \, \oL{Y}_{\text{QCD}+}(v\nm)\,
    \oL{Y}_{\text{QCD}+}^\dagger(0) \, \oL{Y}_{q+}^\dagger(0),
\end{align}
where $Q_{q'}$ is the charge of the soft $u$-quark denoted 
by $q_s'$ in the $B_u$ meson. The explicit definitions of soft 
Wilson lines can be found in \refapp{app:SCET-Lag}. The new 
auxiliary Wilson line $Y_v(0)$ is defined with the time-like vector
$v$ and carries the charge of the $B_u$ meson. The dependence of 
the LCDA on the arbitrary vector $v$ cancels after convolution 
with the collinear matrix element. It is clear that the arbitrary 
vector $v$ breaks the boost invariance of the collinear matrix 
element, which includes the factor $R_{v+} \equiv 
\big< 0 \big| Y_v(0) Y^\dagger_+(0)  \big|0 \big>$ that was 
removed above from the soft matrix element. 
The same breaking occurs also for 
$B_q\to \ell^+\ell^-$ since the boost-invariant vacuum expectation 
value of the Wilson lines is factorized into the boost non-invariant 
quantities $R_\pm$.  This is a consequence of the \SCETII{} 
factorization anomaly~\cite{Beneke:2005, Becher:2010tm, Becher:2011pf, Echevarria:2012js}, which frequently appears when there are 
collinear and soft modes with equal invariant mass, which cannot 
be uniquely separated in dimensional regularization.

Finally, let us comment on the dipole operator contribution 
proportional to $C_7$. From \refapp{sec:operator}, we expect 
that for this case yet another generalized LCDA should 
be defined containing
the soft leptons of the operator in \refeq{eq:softleptonop}. 
Thus, the set of required LCDAs is not only process-dependent 
but also depends on the operator at the hard scale.

%--------+---------+---------+---------+---------+---------+---------+---------+
%
%
%
%--------+---------+---------+---------+---------+---------+---------+---------+
\section{\boldmath Resummed power-enhanced 
$B_q \to \ell^+\ell^-$ amplitude}
\label{sec:factorization}

%
%
%--------+---------+---------+---------+---------+---------+---------+---------+
\subsection{Factorization of the amplitude}
\label{sec:factorization-formula}
 
Having defined the soft matrix element in terms of the generalized 
$B$-meson LCDA, we now focus on the collinear and anti-collinear 
matrix elements. As they involve only the leptons and their 
interactions with collinear/anti-collinear photons, they are 
free of QCD effects at the considered order. As is the case for 
other low-energy electromagnetic quantities, hadronic vacuum 
polarization and other strong interaction effects would become 
relevant in higher orders in the electromagnetic coupling. 
In \SCETII{}, the A1-type operators contain either a single collinear 
(anti-collinear) lepton field, and B1-type operators 
a product of both collinear (anti-collinear) lepton and photon 
fields. In each case, we are interested only in the matrix element 
of $\oL{B}_q(p) \to \ell^+(p_\oL{\ell}) \ell^-(p_\ell)$  with 
only leptons in the final state.\footnote{We recall that the power-enhanced QED corrections are purely virtual. By
assumption, we define the IR finite observable through a 
narrow signal window in the di-muon invariant mass around the 
mass of the $B_q$ meson. This allows photons with ultrasoft 
energy $E_\gamma <\Delta E\ll \Lambda$ in the final state, but 
excludes final-state radiation of real collinear 
photons with virtuality of order $\Lambda^2$. We consider 
ultrasoft photons in~\refsec{sec:decay-width}.} 
Thus, we define the 
\emph{renormalized} collinear and anti-collinear on-shell matrix 
elements related to $\OpII_{m\chi}^{A1}$ as
\begin{align}
  \label{eq:zl}
  \big\langle \ell^-(p_\ell) \big| R_+ \oL{\ell}_c(0)\big| 0\big\rangle 
  & = Z_\ell \, \oL{u}_c(p_\ell), 
& 
  \big\langle \ell^+(p_\oL{\ell}) \big| R_- \ell_\oL{c}(0) \big| 0 \big\rangle
  & = Z_\oL{\ell} \, v_\oL{c}(p_\oL{\ell}),
\end{align} % CB checked
and those related to $\OpII_{\calA\chi}^{B1}$ as
\begin{equation}
\begin{aligned}
  R_+\int\frac{dt}{2\pi} \, e^{-i t \oL{w}\, (\np p_\ell)} \,
  \big\langle \ell^-(p_\ell) \big| \oL{\ell}_c (0) 
     \calA_{c\perp}^\mu (t\np) \big|0 \big\rangle &
  = Z_\ell\, M_A(w) \; m_\ell \left[\oL{u}_c (p_\ell) \gamma_\perp^\mu \right] ,
\\
  R_-\int\frac{dt}{2\pi} \, e^{-i t \oL{w}\, (\nm p_\oL{\ell})} \,
  \big\langle \ell^+(p_\oL{\ell}) \big| 
     \calA_{\oL{c}\perp}^\mu (t\nm)  \ell_\oL{c} (0) \big|0 \big\rangle &
  = Z_\oL{\ell} \, M_\oL{A}(w)\; m_\ell 
    \left[\gamma_\perp^\mu  v_\oL{c}(p_\oL{\ell}) \right] .
\label{eq:MA}
\end{aligned}
\end{equation} % CB checked
We note that the second equation in \refeq{eq:zl} simply defines 
the matrix element of $\OpII_{\oL{c}}$ from \refeq{eq:Jcbar}, while 
the first and \refeq{eq:MA} gives the matrix elements of 
\refeq{eq:opcoll1}, \refeq{eq:opcoll2} after straightforward 
multiplications and contractions. Explicit computation to the 
required order gives 
\begin{align}
  Z_{\ell} = Z_\oL{\ell} &
  = 1 + \mathcal{O}(\alE) ,
\\
  \label{eq:M_A-1}
  M_A^{(1)} (w; \mu) = M_\oL{A}^{(1)} (w; \mu) &
  = - \frac{\alE}{4\pi} Q_\ell \, \oL{w} 
    \left( \ln\frac{\mu^2}{m_\ell^2} - \ln\oL{w}^2 \right) ,
\end{align} % CB checked
with $\oL{w}\equiv 1-w$. In the case of $\OpII_{\calA\chi}^{B1}$, the 
matrix element starts at the one-loop order, as indicated by the 
superscript. The bare matrix element is UV divergent and rendered 
finite by the operator mixing counterterm \refeq{eq:ZAm}. When 
evaluated at the collinear scale $\mu\sim \Lambda$, the 
matrix elements do not contain large logarithmic corrections.

With the above collinear matrix elements and the parametrization 
\refeq{eq:def-B-LCDA} of the soft matrix element at hand, we can 
now derive the factorized expression for the matrix elements 
of the Fourier transforms \refeq{eq:JA1FT}, \refeq{eq:JB1FT} 
of the \SCETII{} operators \refeq{eq:def-SCET-II-ops-1}, 
\refeq{eq:def-SCET-II-ops-2} in the form 
\begin{align}
  \label{eq:SCET2-me-JA1}
  \big\langle \ell^+(p_\oL{\ell}) \, \ell^-(p_\ell) \big|
    \OpII_{m\chi}^{A1} (\omega) \big| \oL{B}_q(p) \big\rangle &
  = T_+ \, m_{B_q} \mathscr{F}_{B_q} \Phi_+(\omega) \,,
\\[0.2cm]
  \label{eq:SCET2-me-JB1}
  \big\langle \ell^+(p_\oL{\ell}) \, \ell^-(p_\ell) \big|
    \OpII_{\calA\chi}^{B1}(\omega, w) \big|\oL{B}_q(p) \big\rangle &
  = T_+ \,  M_A (w) \, m_{B_q} \mathscr{F}_{B_q} \Phi_+(\omega) \,.
\end{align} % CB checked
All scale-dependent quantities are understood to be evaluated 
at the scale $\mu$, and we defined the common factor
\begin{align}
  T_+ (\mu) & 
  \equiv   (-i) \,m_\ell(\mu) \, Z_\ell(\mu) Z_\oL{\ell}(\mu) \,
  [\oL{u}_c (p_\ell) P_R v_\oL{c}(p_\oL{\ell})] \,.
\end{align}
Note that $\langle \OpII_{m\chi}^{A1} \rangle$ contributes at tree 
level, whereas $\langle \OpII_{\calA\chi}^{B1} \rangle$ starts to 
contribute only from the one-loop order. The same result holds 
for the anti-collinear operators $i = m\oL{\chi}, \, 
\calA \oL{\chi}$ owing to \refeq{eq:M_A-1} and the definition 
of the soft matrix element \refeq{eq:def-B-LCDA}. 

The complete expression for the power-enhanced $B_q \to \ell^+\ell^-$ amplitude due to the operators $Q_{9,10}$ of the effective weak 
interaction Lagrangian is now obtained by adding the hard 
($H_{9,10}$) and hard-collinear ($J_{m,A}$) 
matching coefficients according to 
\refeq{eq:SCETImatchrel} and \refeq{eq:SCETItoIImatch}, and by 
summing over all contributions
$i = 9,\oL{9}, 10,\oL{10}$ in the general factorized form 
\begin{equation}
  \label{eq:amp-factorized-gen}
\begin{aligned}
  i \calA_9 
  = T_+ & \,\;\Big[ \;
       (H_9 + H_{10}) \otimes_u \left(J_m + J_A \otimes_w M_A \right)
\\ & + (H_\oL{9} - H_\oL{10}) \otimes_u \left(J_\oL{m} + J_\oL{A} \otimes_w M_\oL{A} \right)
   \Big] \otimes_\omega m_{B_q} \mathscr{F}_{B_q} \Phi_+ \,,
\end{aligned}
\end{equation} % CB: checked
where we have suppressed all arguments, which will be shown 
explicitly below. The formula simplifies considerably when 
accounting for several relations between the matching coefficients 
of the collinear and anti-collinear sectors, which show up at 
tree level: for the hard functions $H_9^{(0)} = H_\oL{9}^{(0)}$ 
and $H_{10}^{(0)} = H_\oL{10}^{(0)}$ from 
\refeq{eq:SCETI-hard-matching} and for the
jet functions $J_A^{(0)} = J_\oL{A}^{(0)}$ from \refeq{eq:JA0}, 
and $J_m^{(0)} = J_\oL{m}^{(0)} = 0$.
In fact, higher-order QED corrections are symmetric under the 
exchange of the collinear and anti-collinear sectors once hard 
fluctuations are decoupled, such that the relations
$H_9 = H_\oL{9}$ and $H_{10} = H_\oL{10}$ are valid even beyond 
tree level. Thus the hard functions in both sectors will exhibit
the same $u$-dependence. Concerning the jet functions and matrix elements of the \SCETII{} operators, the
explicit one-loop results show that $J_m^{(1)} = 
J_\oL{m}^{(1)}$ upon the identification of $\np p_\ell = 
\nm p_\oL{\ell} = m_{B_q}$ in \refeq{eq:Jm1}, while 
$M_A^{(1)} = M_\oL{A}^{(1)}$ according to 
\refeq{eq:M_A-1}. Again we expect these relations to extend 
to higher orders in QED, because of the symmetry between the 
collinear and anti-collinear sectors. Making use of these relations 
we find that \refeq{eq:amp-factorized-gen} simplifies to 
\begin{align}
  \label{eq:amp-factorized}
  i \calA_9 & 
  = T_+ \! \int_0^1 \!\! du \, 2 H_9 (u) \! \int_0^\infty \!\! d\omega 
    \left[ J_m (u;\omega) 
         + \int_0^1 \!\! dw \, J_A (u; \omega, w) \, M_A (w) \right] 
    m_{B_q} \mathscr{F}_{B_q} \Phi_+(\omega)
\end{align} % CB: checked
even beyond leading logarithmic approximation. The contribution 
from the operator $Q_{10}$ has cancelled and a factor of two 
arises for the $Q_9$ term, as anticipated earlier. All momentum 
fraction arguments and convolutions have now been made explicit. 
Every factor is understood to be evaluated at the same scale 
$\mu$. In this form there is no value of $\mu$ in which not 
at least one of the factors contains large logarithms. For 
example, if $\mu$ is chosen of order of the soft and collinear 
scale $\Lambda$, large logarithms occur in the hard and 
hard-collinear coefficients functions. On the other hand, 
if $\mu$ is chosen at the 
hard-collinear scale $\sqrt{m_b\Lambda}$, $H_9(u)$ and the 
matrix element factors $T_+$, $M_A(w)$ and 
$\mathscr{F}_{B_q} \Phi_+(\omega)$ contain large logarithms.

%
%
%
%--------+---------+---------+---------+---------+---------+---------+---------+
\subsection{Resummed amplitude}
\label{sec:SCET2-resummation}

We will now use the solutions to the renormalization group 
equations derived earlier to convert \refeq{eq:amp-factorized} 
into a formula in which large logarithms are summed. The explicit 
result is given in the LL approximation, but the essence of 
the manipulations is general. We shall take the common scale 
to be the hard-collinear scale $\mu_{hc}\sim \sqrt{m_b\Lambda}$, 
hence we have to 
evolve the hard function from $\mu_b\sim m_b$ down to $\mu_{hc}$ 
and the soft and collinear functions up from $\mu_s\sim \mu_c \sim 
\Lambda$ to $\mu_{hc}$.

To implement this procedure into \refeq{eq:amp-factorized}, 
we use \refeq{eq:sol-RGE-Hi}, and include the hard-function 
evolution to $\mu_{hc}$ via the substitution
\begin{equation}
H_9(u) \to \exp\big[S_{\ell}(\mu_b,\, \mu_{hc})\, 
+ \, S_{q}(\mu_b,\, \mu_{hc}) 
\big]\, H_9(u, \mu_b)\,.
\end{equation}
For the soft matrix element, we use 
\refeq{eq:UsoftBLCDA}, \refeq{eq:phiQCD} to obtain 
\begin{eqnarray}
  \mathscr{F}_{B_q}\, \Phi_+ (\omega) 
  & \to & U_s(\mu_{hc},\mu_s; \omega)\, \mathscr{F}_{B_q}(\mu_s) \, 
  \Phi_+ (\omega; \mu_s) 
  \nonumber
\\
  & \to & U_s^\text{QED}(\mu_{hc},\mu_s; \omega) \,
  F_{B_q}(\mu_{hc}) \, \phi_+(\omega; \mu_{hc})\,.
\end{eqnarray}
After the second arrow, we expressed the initial condition for 
the generalized $B$-meson LCDA at the soft scale in terms of 
the standard LCDA in the absence of QED corrections, which 
can be done at LL accuracy, as discussed 
\refsec{sec:fB-LCDA-defs}, and evolved the latter back 
to the hard collinear scale. The advantage of this procedure is 
that while pure QED quantities can be evaluated perturbatively 
at low scales of order $\Lambda \sim m_\ell$, the soft scale 
is generally nonperturbative in QCD. The above form requires only 
that the standard $B$-meson LCDA $\phi_+(\omega; \mu_{hc})$ is 
provided at the hard-collinear 
scale by some nonperturbative method, or by extracting it from 
data directly at this scale \cite{Beneke:2011nf}. Finally, 
for the anti-collinear part we use \refeq{eq:SCET2-U-ac} 
to substitute 
$Z_{\oL{\ell}} \to U_{\oL{c}}(\mu_{hc},\mu_c) Z_{\oL{\ell}}(\mu_c)$, 
which together with  
\refeq{eq:SCET2-RG-sol-A1}, \refeq{eq:SCET2-RG-sol-B1} for the 
collinear part amounts to
\begin{align}
  \label{eq:SCET2-RG-m}
  T_+ & 
  \to U_c(\mu_{hc},\, \mu_c)U_\oL{c}(\mu_{hc},\, \mu_c) \, T_+(\mu_c) ,
\\
  \label{eq:SCET2-RG-M_A}
  T_+ M_A(w) &
  \to U_c(\mu_{hc},\, \mu_c)U_\oL{c}(\mu_{hc},\, \mu_c) \, T_+(\mu_c) \left[
      M_A (w;\mu_c)
    - \frac{Q_\ell \oL{w}}{\beta_{0,\rm em}}\, \ln\eta_{\rm em} \right] ,
\end{align}  % CB: checked
where $\eta_{\rm em} = \alE(\mu_c)/\alE(\mu_{hc})$. After these 
replacements, the result contains the scales $\mu_b$, $\mu_c$ and 
$\mu_s$ where the initial conditions of the various evolutions 
are set. This dependence cancels between the evolution factors, 
matching coefficients and matrix elements up to residual dependence 
of higher order than LL accuracy.

Putting this together in \refeq{eq:amp-factorized} and 
making use of \refeq{eq:SUUrelation} results in the all-order 
LL-resummed amplitude
\begin{align}
  \label{eq:amp-factorized-rearr}
  i  \calA_9  
  & = e^{S_\ell(\mu_b,\, \mu_c)}\, T_+(\mu_c) 
   \nonumber 
\\ & \times
  \int_0^1 \! du \, e^{S_q(\mu_b,\, \mu_{hc})} \, 2 H_9 (u; \mu_b) \, 
  \int_0^\infty \! d\omega \, U^\text{QED}_s (\mu_{hc}, \mu_s; \omega) \,
     m_{B_q} F_{B_q}(\mu_{hc}) \,\phi_+(\omega; \mu_{hc}) 
   \nonumber
\\ & \times 
  \left[ J_m (u;\omega; \mu_{hc}) 
       + \int_0^1 \! dw \, J_A (u; \omega, w; \mu_{hc}) \,
         \left( M_A (w; \mu_c)  - \frac{Q_\ell \oL{w}}{\beta_{0,\rm em}}\, \ln\eta_{\rm em} \right)
    \right] .
\end{align}  
We note that the prefactor $\exp\,[S_\ell(\mu_b,\, \mu_c)]$ 
sums the purely leptonic leading-logarithms proportional to 
$Q_\ell^2$ between the hard scale $\mu_b$ and the collinear scale 
$\mu_c$. They originate from virtual QED corrections in \SCETI{} 
and \SCETII{}. 
In \refsec{sec:decay-width} it will be combined with the
remaining final-state contributions due to ultrasoft photons 
to provide the radiative $B_q \to \ell^+\ell^-$  branching fraction 
including the fully resummed double-logarithmic QED corrections 
to all orders in perturbation theory.
 
The resummation of the leading-logarithmic QED (and QCD) 
corrections to all orders in perturbation theory is achieved 
by keeping the one-loop expressions of the cusp part of the 
anomalous dimensions together with tree-level results for the hard
and jet functions. In addition, due to the presence of operator mixing, 
the leading off-diagonal elements in the anomalous dimension matrix 
must also be kept. Otherwise, $i \calA_9 = 0$,
because $M_{A,\oL{A}}^{(0)}(\mu_c) = 0$ and 
$J_{m,\oL{m}}^{(0)}(\mu) = 0$ for all $\mu$ when the one-loop 
mixing of $\Gamma^c_{\calA\chi,\, m\chi}$ in 
\refeq{eq:SCET2-rge-coll} is neglected. 

In the following we 
obtain from \refeq{eq:amp-factorized-rearr} an expression 
that is both LL-accurate {\em and} NLO-accurate, thus 
generalizing the previous NLO result \refeq{eq:mainresult} to 
include the leading logarithms to all orders. This can be 
achieved by keeping the non-logarithmic one-loop 
corrections to $J_m$ and $M_A$ given in \refeq{eq:Jm1} and 
\refeq{eq:M_A-1}, respectively. First using \refeq{eq:JA0} 
removes the $w$-integral in the second line of 
\refeq{eq:amp-factorized-rearr}, such that the square bracket 
turns into 
\begin{equation}
  J_m (u;\omega; \mu_{hc}) 
  - \frac{Q_q}{\omega} \, J_A (u; \omega, u; \mu_{hc}) \,
  \left( M_A (u; \mu_c) - \frac{Q_\ell \oL{u}}{\beta_{0,\rm em}}\, \ln\eta_{\rm em} \right) .
\end{equation}
With LL accuracy it is also justified to apply \refeq{eq:mixingDL} 
with $\mu=\mu_{hc}$ in the last term. Inserting \refeq{eq:Jm1} and 
\refeq{eq:M_A-1} produces
\begin{eqnarray}
  \frac{\alE}{4 \pi} Q_\ell Q_q \frac{1-u}{\omega} \left[ 
     \ln\frac{\omega \,\np p_\ell}{\mu^2_{hc}} 
   + \ln u\bar{u} 
   + \Big[\ln\frac{\mu_c^2}{m_\ell^2}-2 \ln \bar{u}\Big]
   + \ln\frac{\mu_{hc}^2}{\mu_c^2}  \right] .
\end{eqnarray}
After combining the logarithms and setting $\np p_\ell = m_{B_q}$, 
we recognize the factor that appears in 
\refeq{eq:mainresult}.\footnote{The present analysis clarifies 
that $m_{B_q}$ should appear in the logarithm, because it 
arises from the kinematic lepton momentum rather than the 
bottom quark mass.} This allows us to put 
\refeq{eq:amp-factorized-rearr} into the final form 
\begin{equation}
  \label{eq:final-ampQ9}
\begin{aligned}
  i \calA_9 &
  = \frac{\alE(\mu_c)}{4\pi} Q_\ell Q_q \,  m_\ell\,(-i)  \, m_{B_q} f_{B_q} \, 
    e^{S_\ell(\mu_b,\, \mu_c)} \calN  \left[ \oL{u}_c (1 + \gamma_5) v_\oL{c} \right]
\\ & \qquad \times
    e^{S_{q}(\mu_b,\, \mu_{hc})} \int_0^1 du \,(1-u) \,  
    C_9^\text{eff}(u, \mu_b) 
\\ & \qquad \times
   \int_0^\infty \frac{d\omega}{\omega}  \,
   U_s^\text{QED}(\mu_{hc},\, \mu_s;\omega) \, \phi_+(\omega; \mu_{hc})
   \left[ \ln\frac{\omega m_{B_q}}{m_\ell^2} + \ln \frac{u}{1-u} \right] .
\end{aligned}
\end{equation} % CB: checked
At LL accuracy the scale of the overall factor of $\alE$ is 
arbitrary and we have chosen the collinear scale. Within the same 
LL approximation, we can also replace the HQET decay constant 
by the QCD decay constant $f_{B_q}$.
The one-loop QED result of \cite{Beneke:2017vpq} 
in \refeq{eq:mainresult} is obtained from the above expression 
when setting the Sudakov exponentials and the soft 
evolution factor $U_s^{\rm QED}$ to unity, 
apart from the term proportional to $C_7^\text{eff}$ that 
was not considered up to now. The explicit result for 
$S_\ell(\mu_b,\, \mu_c)$ and $S_{q}(\mu_b,\, \mu_{hc})$ 
can be inferred from \refeq{eq:def-S_ell} and \refeq{eq:def-S_qI}, 
respectively. The residual QED evolution from the $B$-meson LCDA 
is obtained from \refeq{eq:softRGEUsolution} or the simpler 
version \refeq{eq:SCET2-softRG-ana-sol} by setting $\alS = 0$, 
see \refeq{eq:def-U^QED}. Explicitly
\begin{align}
  U_{s}^{\rm QED}(\mu_{hc},\mu_s;\omega) & = 
  \exp\!\bigg[ \frac{4\pi}{\alE(\mu_s)} 
    \frac{Q_q\left(2 Q_\ell+Q_q\right)}{\beta_{0,\rm em}^2}
    \Big( g_0(\eta_{\rm em}) 
        + \frac{\alE(\mu_s)}{2\pi}  \beta_{0, \rm em} \ln\eta_{\rm em} \, \ln\frac{\omega}{\mu_s}
    \Big) \bigg]\quad
  \nonumber
\\ 
  \label{eq:softRGEUQEDsolution}
  & \xrightarrow{\text{DL}} 
  \exp\left[\frac{\Gamma_s^{\rm QED}}{2} 
  \left(\ln^2\frac{\omega}{\mu_s} - \ln^2\frac{\omega}{\mu_{hc}} \right) \right],
\end{align}
where for soft evolution $\eta_{\rm em}$ stands for  
$\eta_{\rm em}(\mu_s,\mu_{hc}) \equiv \alE(\mu_s)/\alE(\mu_{hc})$, 
or, more simply, by dropping the $\mathcal{O}(1)$ logarithm 
of $\omega/\mu_s$,  
\begin{align}
  U_s^{\rm QED}(\mu_{hc},\mu_s) & =
  \exp\!\bigg[ \frac{4\pi}{\alE(\mu_s)} \frac{Q_q(2 Q_\ell + Q_q)}
     {\beta_{0,\rm em}^2} \, g_0(\eta_{\rm em}) \bigg]
  \nonumber
\\
  \label{eq:SCET2QED-softRG-ana-sol}
  & \xrightarrow{\text{DL}}
  \exp\!\bigg[-\frac{\alE}{2\pi} \big[Q_q(2 Q_\ell + Q_q)\big] 
     \ln^2 \frac{\mu_s}{\mu_{hc}} \bigg]\,.
\end{align} 
Here, similarly to \refeq{eq:def-U^QED}, $\Gamma_s^\text{QED} \equiv 
\Gamma_s(\alS, \alE) - \Gamma_s(\alS, \alE=0)$ is obtained at the 
one-loop order from 
$\Gamma_s$ in \refeq{eq:cusp-ADMs} by setting $\alS = 0$.

%--------+---------+---------+---------+---------+---------+---------+---------+
%
%
%
%--------+---------+---------+---------+---------+---------+---------+---------+

\section{\boldmath $B_q \to \mu^+\mu^-$ decay width}
\label{sec:decay-width}

In the decay $B_q \to \mu^+\mu^-$ we encounter the peculiar 
situation that the numerically leading amplitude at tree-level in 
$\alE$, discussed in more detail below as $\calA_{10} \propto 
C_{10}$, is power-suppressed compared to the amplitude $\calA_9$ 
in \refeq{eq:amp-factorized}, which on the other hand is suppressed
by $\alE$, hence
\begin{align}
  \calA_{10} & \sim 1    \cdot \lambda^{12} , &
  \calA_9    & \sim \frac{\alE}{\pi} \cdot \lambda^{10}
                    \cdot \ln\frac{\mu_{hc}}{\mu_c} .
\end{align}
Indeed the hierarchy $\alE/\pi \sim 1/420$ compared to
 $\lambda^2 \sim \LambQCD/m_b \sim 1/20$ confirms that $\calA_{10}$ 
is numerically the most relevant amplitude, but in an 
imaginary world with a much larger value of $m_b$ or a much larger
electromagnetic coupling, the amplitude $\calA_9$ would be 
largest. As the decay width is proportional to 
$|\calA_{10} + \calA_9|^2$, the dominant effect of
$\calA_9$ is the interference with $\calA_{10}$. The investigation 
of the full QED effects at the subleading power in $1/m_b$, 
as would be required for $\calA_{10}$ in the 
SCET approach, is a rather daunting task and we leave it for the 
future. However, based on our derivations in the previous sections, 
we discuss the leading effect, which requires only 
tree-level matching and leading-logarithmic resummation. 

%
%
%
%--------+---------+---------+---------+---------+---------+---------+---------+
\subsection[Tree-level amplitude to $B_q \to \mu^+\mu^-$]
{\boldmath  Tree-level amplitude to 
$B_q \to \mu^+\mu^-$}
\label{sec:calA_10}

Here we derive the LL resummation of  the formally 
power-sup\-pressed but numerically dominant amplitude $\calA_{10}$ for
$B_q \to \ell^+\ell^-$. At this accuracy it is sufficient 
to match at tree level and to employ the one-loop cusp anomalous 
dimensions of the relevant operators. The fact that we restrict 
ourselves to the operators obtained from tree-level matching in 
$\alE$ simplifies the operator structure; in particular it 
implies that in the \SCETI{} operator  the light quark field 
must be soft, since otherwise the operator could not overlap 
with the $B$-meson state at tree level. Due to the chiral structure 
of the weak EFT operators, the lepton-mass term
appears already after the hard matching leading to the 
well-known helicity suppression of the amplitude. 
Hence, the relevant \SCETI{} operator is 
\begin{align}
   \widetilde{\OpI}_{m} &
  = m_\ell \, 
    \big[\oL{q}_s (0)\, P_R \, h_\vb(0)   \big]
    \big[\oL{\ell}_C (0)\, \gamma_5 \, \ell_\oL{C}(0) \big] ,
\label{eq:Omop}
\end{align} % CB: checked
with the matching coefficient
\begin{align}
  \label{eq:Hm-matching}
  H_m (\mu_b) & 
  = \calN \, \frac{2\, C_{10}(\mu_b)}{m_{B_q}}
\end{align} % CB: checked
at the hard scale $\mu_b$. As one works to subleading order in 
$\lambda$, one must use the
$\mathcal{O}(\lambda)$ relation between  the full theory fields and 
SCET fields, as derived for example in \cite{Beneke:2003pa}, to 
obtain the above result. The \SCETI{} RGE 
of the coefficient $H_m$ at LL accuracy, i.e. neglecting non-cusp 
parts of the anomalous dimension and possible
operator mixing, reads
\begin{align}
  \label{eq:SCET1-RG-Hm}
  \frac{d}{d\ln \mu} H_m (\mu) &
  = \Gamma_c \,\ln \frac{m_{B_q}}{\mu}\, H_m(\mu) \,.
\end{align} % CB: checked
It is governed by the collinear cusp anomalous 
dimension \refeq{eq:cusp-ADMs} encountered previously for the 
power-enhanced amplitude in \SCETI{} and \SCETII{}. 

The decoupling of the hard-collinear modes of \SCETI{} is trivial 
and yields the \SCETII{} operator (in position space)
\begin{align}
  \widetilde{\OpII}_m^{A1} &  
  = m_\ell \, \oL{q}_s (0)  P_R h_\vb (0)
    \big[Y_+^\dagger Y_- \big]  (0)
    \big[\oL{\ell}_c (0) \gamma_5 \, \ell_\oL{c} (0) \big] ,
\end{align} % CB: checked
with unit matching coefficient. Note that the jet function 
cannot depend on the soft quark position along the light-cone 
at tree level, hence the operator remains local, unlike for 
\SCETI{} to \SCETII{} matching of the power-enhanced amplitude 
$\calA_9$. The RG evolution of the matrix element of $\OpII_m^{A1}$ 
is governed by the same cusp anomalous dimension as in \SCETI{} 
in \refeq{eq:SCET1-RG-Hm}. The
amplitude in \SCETII{} at the collinear scale is then given by 
\begin{align}
  \nonumber %\label{eq:calA_10}
  i\, \calA_{10} &
  = \frac{1}{2} (-i)\,m_{B_q} \mathscr{F}_{B_q}(\mu_{hc}) \, 
  m_\ell  [Z_\ell Z_\oL{\ell}](\mu_{hc}) \, 
    H_m(\mu_{hc}) [\oL{u}_c (p_\ell) \gamma_5 v_\oL{c}(p_\oL{\ell})] 
\\ &
  \label{eq:calA_10-rearr}
  = \frac{1}{2} (-i)\,m_{B_q} m_\ell \,f_{B_q}  \,
    e^{S_\ell(\mu_b,\, \mu_c)} \, 
    H_m(\mu_b) [\oL{u}_c (p_\ell) \gamma_5 v_\oL{c}(p_\oL{\ell})] .
\end{align} % CB: checked
In the first line the hard function is meant to be evaluated at 
the hard-collinear scale by means of its \SCETI{} RGE and further 
a \SCETII{} RG evolution is implied for the matrix element of 
$\langle \OpII_{m}^{A1} \rangle \propto 
Z_\ell Z_\oL{\ell} \mathscr{F}_{B_q}$ from the soft and collinear 
scale to the hard-collinear scale.
% in the spirit of \refsec{sec:SCET2-resummation}. 
In the second line, 
we make use of the LL solution of the RGEs discussed in
\refsec{sec:SCET2-resummation} and set 
$[Z_\ell Z_\oL{\ell}](\mu_c) = 1$.
In particular, the same Sudakov factor $e^{S_\ell(\mu_b,\, \mu_c)}$ 
between the hard and collinear scales as in the power-enhanced 
amplitude \refeq{eq:amp-factorized-rearr} appears as an overall 
factor. Moreover the hard function $H_m$ enters now at the hard 
scale with value given in \refeq{eq:Hm-matching}. 
The static $B$-meson decay constant $\mathscr{F}_{B_q}(\mu_{hc})
= F_{B_q}^{(0)}(\mu_{hc}) + \alE/(4\pi) F_{B_q}^{(1)} (\mu_{hc})
+ \calO(\alE^2)$ contains the term $F_{B_q}^{(1)}$ 
that contributes at the same order in $\alE$ as $\calA_9$, but 
is power-suppressed by $\lambda^2$ compared to $\calA_9$, 
and for this reason will be omitted. Further we replace 
$F_{B_q}^{(0)}(\mu_{hc})$  by $F_{B_q}(\mu_{hc})$, the 
static decay constant in the absence of QED, because the 
difference in RG evolution does not contribute double logarithms. 
For the same reason, we equate $F_{B_q}(\mu_{hc})$ 
to the full QCD decay constant $f_{B_q}$,
which is usually calculated on the lattice within 
QCD.\footnote{The lattice
calculation \cite{Bazavov:2017lyh} includes electromagnetic
contributions to meson masses to fix quark masses, which leads to some isospin
breaking for the $B_q$ meson decay constants.} This is exact to the considered
accuracy of tree-level matching at the hard scale $\mu_b$.

Since both, $\calA_{9}$ and $\calA_{10}$, share the same 
overall leptonic Sudakov factor, it proves advantageous for 
later purposes to factor it from the sum of both amplitudes. 
We write 
\begin{align}
  \label{eq:def-red-ampl}
  i (\calA_{10} + \calA_9) & \equiv 
  e^{S_\ell(\mu_b,\, \mu_c)} \Big( A_{10} \left[ \oL{u}_c \gamma_5 v_\oL{c} \right] 
+ A_9 \left[ \oL{u}_c (1 + \gamma_5) v_\oL{c} \right] \Big)\,,
\end{align}
where we introduced the scalar reduced amplitudes $A_{9,10}$, 
which can be extracted from \refeq{eq:amp-factorized-rearr} and
\refeq{eq:calA_10-rearr}. An analogous reduced amplitude $A_7$ is  
defined for the part of the amplitude proportional to 
$C_7^\text{eff}$. Its non-resummed one-loop 
expression can be read off from \refeq{eq:mainresult}. Moreover,
the resummation of the leading logarithmic QCD (but not QED) corrections 
in \SCETI{} is also possible for the $A_7$ 
contribution, because the operators which give rise to the $A_7$ 
part in \SCETI{} have the same QCD anomalous dimension as the 
corresponding operators of the $A_9$ part. 
Therefore, the amplitude $A_7$ also receives 
the factor $e^{S_{q}(\mu_b,\, \mu_{hc})}$ defined 
in \refeq{eq:def-S_qI}, but its QED part should be dropped.

%
%
%
%--------+---------+---------+---------+---------+---------+---------+---------+
\subsection{Decay width and ultrasoft photons}
\label{sec:decay-width-2}

So far we ignored ultrasoft photons below the soft scale $\mu_s$. 
We now turn to the radiative $B_q\to \ell^+\ell^-$ decay amplitude, 
and consider the matrix element with an arbitrary ultrasoft 
state $X_s$ consisting of photons and possibly electrons and 
positrons. It factorizes into the non-radiative amplitude $\calA_i$ 
discussed before and an ultrasoft matrix element
\begin{align}
  \label{eq:Ausoft}
 \EWnorm C_i \,\big 
< \ell \oL{\ell} X_s\big| Q_i \big | \oL{B}_s\big> &
= \calA_i \, \big\langle X_s \big| 
S^\dagger_{v_\ell}(0) S_{v_\oL{\ell}}(0) 
    \big| 0 \big\rangle \,, & i & = 9, 10.
\end{align}
To prove this factorization formally we should match \SCETII{} at 
a scale of order $\LambQCD \sim m_\ell$ to an  effective theory 
that contains the $B$-meson field and heavy lepton fields with 
fixed velocity label, in analogy with heavy-quark 
effective theory. In this theory ultrasoft photons with virtuality 
much below $m_\ell^2 \sim \LambQCD^2$ have leading-power couplings 
to the charged leptons but not to the electrically neutral 
$B$-meson. The decoupling of the ultrasoft photons from the heavy 
leptons, $\ell_C \to S_{v_\ell} \ell_C^{(0)}$, gives rise to the 
ultrasoft Wilson lines $S_{v_\ell}$ in \refeq{eq:Ausoft}. The 
lepton-velocity $v_\ell$ is defined via $p_\ell = 
E_\ell v_\ell$ and similarly for $v_\oL{\ell}$. At leading power 
in the $1/m_b$ expansion, the radiation originates only from the 
final-state leptons as the ultrasoft photons do not couple to 
the neutral initial state. Formally, the matching of \SCETII{} 
with quark fields to the EFT with point-like meson fields is 
nonperturbative. We can 
nevertheless sum the leading logarithms, because the $B$-meson 
is neutral and decoupled in the far infrared, so we know that 
the IR logarithms arise from perturbative QED only. 

The partial decay width is obtained after squaring the full amplitude
\refeq{eq:Ausoft} and summing over all ultrasoft final states with 
total energy less than $\Delta E$
\begin{eqnarray}
\Gamma[B_q \to \mu^+\mu^-](\Delta E) 
&=& \frac{m_{B_q}}{8\pi} \beta_\mu \,
 \Big(\big|A_{10} + A_9 + A_7\big|^2 + \beta_\mu^2 \, 
\big|A_9 + A_7\big|^2 \Big)
\nonumber \\ 
&& \times\,
    \Big|e^{S_\ell(\mu_b,\, \mu_c)} \Big|^2
    \mathcal{S}(v_\ell, v_\oL{\ell}, \Delta E) ,
% CB: checked normalization
\label{eq:GBs}
\end{eqnarray}
where $\beta_\mu = \sqrt{1 - 4 m_\mu^2/m_{B_q}^2}$. We include here 
the amplitude $A_7$ even though we do not attempt to sum QED 
corrections for this amplitude. However we compute the leading 
logarithmic QCD corrections to $A_7$ and comment on this in 
\refsec{sec:numeric}. The terms proportional to $|A_9 + A_7|^2$ 
are formally of $\mathcal{O}(\alE^2)$. 
The first term in the parenthesis is due to the pseudo-scalar lepton 
current $[\oL{u}_c \gamma_5 v_\oL{c}]$ in \refeq{eq:def-red-ampl}, 
the second term $\beta_\mu^2 \, |A_9 + A_7|^2$ due to the scalar term
$[ \oL{u}_c v_\oL{c}]$. The ultrasoft function 
\begin{align}
  \mathcal{S}(v_\ell, v_\oL{\ell}, \Delta E) &
  = \sum_{X_s} \big| \big\langle X_s \big|
       S^\dagger_{v_\ell}(0) S_{v_\oL{\ell}}(0) \big| 0 \big\rangle \big|^2 \; 
    \theta (\Delta E - E_{X_s})
\end{align} % CB: checked
accounts for the emission of an arbitrary number of ultrasoft photons with
total energy $E_{X_s} < \Delta E$. 

The ultrasoft function should be further factorized to sum
large logarithmic corrections with the RG technique. This could be achieved 
by introducing another EFT below the muon-mass scale similar to 
the SCET treatment of soft radiation in top-quark jets \cite{Fleming:2007qr}.
Instead, to avoid further technical complications, we use the QED 
exponentiation theorem to write the full soft function as the exponent
of the one-loop result
\begin{align}
  \mathcal{S}(v_\ell, v_\oL{\ell}, \Delta E) &
  = \exp \left[\frac{\alE}{4\pi} Q_\ell^2 \,S^{(1)}(v_\ell, v_\oL{\ell}, 
\Delta E) \right].
\end{align}
The one-loop result in the appropriate limit $E_\ell \gg m_\mu \gg \Delta E$
is given by (for a result in dimensional regularization, 
see e.g. \cite{vonManteuffel:2014mva})
\begin{align}
  S^{(1)}(v_\ell, v_\oL{\ell}, \Delta E) &
  = 8\left(1 + \ln \frac{m_\mu^2}{s_{\ell \oL{\ell}}} \right) 
    \ln\left(\frac{\mu}{2\Delta E}\right) 
  - 2 \left(2 + \ln \frac{m_\mu^2}{s_{\ell\oL{\ell}}} \right) \ln \frac{m_\mu^2}{s_{\ell \oL{\ell}}}
  - \frac{4}{3} \pi^2 \,,
\label{eq:oneloopultrasoft}
\end{align}
where $s_{\ell\oL{\ell}}$ denotes the invariant mass squared of the 
lepton pair. The $\mu$ dependence of the ultrasoft function is 
cancelled by the explicit $\mu_c$ dependence of the non-radiative 
amplitude, as seen in \refeq{eq:ultraexp} below to the accuracy 
considered here, hence we set $\mu=\mu_c$.  
The second line of \refeq{eq:GBs}, which multiplies the reduced 
amplitude squared, can be rewritten as the single exponential.
In the leading approximation we neglect the constant factor 
$-\frac{4}{3}\pi^2$ in $S^{(1)}$. Then we find 
\begin{eqnarray}
&& \Big|e^{S_\ell(\mu_b,\, \mu_c)} \Big|^2
    \mathcal{S}(v_\ell, v_\oL{\ell}, \Delta E,\mu_c)
\nonumber\\[0.1cm]
&&=\,\exp\left\{\frac{\alE}{4\pi}Q^2_\ell
\left[ 8\, \left(1 + \ln \frac{m_\mu^2}{s_{\ell \oL{\ell}}} \right) 
\ln\left(\frac{\mu_c}{2\Delta E}\right) 
  - 2  \left(2 + \ln \frac{m_\mu^2}{s_{\ell\oL{\ell}}} \right) 
\ln \frac{m_\mu^2}{s_{\ell \oL{\ell}}} -\frac{4}{3}\pi^2
  - 8\ln^2 \frac{\mu_b}{\mu_c}\right]\right\}
\nonumber\\[0.2cm]
 && =\, \exp\Bigg\{\frac{2\alE}{\pi}Q^2_\ell\,
\bigg[  \bigg(1 + \ln \frac{m_\mu^2}{m_{B_q}^2} \bigg) 
    \ln\left(\frac{m_{B_q}}{2\Delta E}\right)
\nonumber\\
&& \hspace*{1.6cm}+\,2 \,\ln\frac{m_{B_q}}{\mu_b}\ln\frac{\mu_b}{\mu_c}\,
+\ln^2 \frac{m_{B_q}}{\mu_b}-\left(1+\ln \frac{m_\mu}{\mu_c}\right)
\ln \frac{m_\mu}{\mu_c} - \frac{\pi^2}{6}\bigg] \Bigg\} .
\label{eq:ultraexp}
\end{eqnarray}
Notice that since \refeq{eq:oneloopultrasoft} is given only in the 
one-loop and not the formal LL approximation, we use the 
double-logarithmic approximation \refeq{eq:def-S_ell} for 
$S_\ell(\mu_b,\, \mu_c)$, which gives the last term in 
square brackets after the first equality. Within the same 
DL approximation, all terms in the last line should be dropped, 
since they are at most single-logarithmic, given 
$\mu_c\sim m_\mu$, $\mu_b\sim m_{B_q}$. We also set 
$s_{\ell \oL{\ell}}=m_{B_q}^2$,\footnote{The dependence of 
$s_{\ell \oL{\ell}}$ on $\Delta E$ 
is a negligible power-suppressed effect.}  since the dependence
of the RG evolution factors and ultrasoft function on the hard scale arises  
from the kinematic variables entering the cusp anomalous dimension, 
rather than from $m_b$ quark mass factors. This allows us to 
rewrite~\refeq{eq:GBs} in the conventional form ($Q_\ell^2=1$)
\begin{align}
  \label{eq:GBs2}
  \Gamma[B_q \to \mu^+\mu^-](\Delta E) & 
  = \Gamma^{(0)}[B_q \to \mu^+\mu^-]
    \left( \frac{2\Delta E}{m_{B_q}} \right)^{
       -\frac{2\alE}{\pi}\left(1+\ln\frac{m_\mu^2}{m_{B_q}^2}\right)} ,
\intertext{with the non-radiative decay width}
  \label{eq:decay-width}
  \Gamma^{(0)}[B_q \to \mu^+\mu^-] &
  \equiv \frac{m_{B_q}}{8\pi} \beta_\mu \,
  \Big(\big|A_{10} + A_9 + A_7 \big|^2 + \beta_\mu^2 \, \big|A_9 + A_7 \big|^2 \Big)\, .
\end{align}
The universal ultrasoft photon corrections, which depend on $\Delta E$,
are now explicitly factorized in the usual manner \cite{Buras:2012ru}. 
In \cite{Buras:2012ru} and other works based on
Yennie-Frautschi-Suura exponentiation \cite{Yennie:1961ad}, the scale 
$m_{B_q}$ in the base $2\Delta E/{m_{B_q}}$ of the exponential is obtained 
by extrapolating the cutoff of the point-like meson theory, which 
should be below $\Lambda_{\rm QCD}$, to $m_{B_q}$. The 
EFT framework developed in this paper justifies this extrapolation 
in part, but it also shows that there are additional, structure-dependent 
double logarithms. In \refeq{eq:GBs2}, these process-specific resummed 
leading-logarithmic QED corrections that depend
on the soft/collinear and hard-collinear scales are included in the amplitudes 
$A_{9,10}$.  The decay rate is then written as the  
product of the resummed non-radiative decay width 
$\Gamma^{(0)}[B_q \to \mu^+\mu^-]$
and the exponentiated ultrasoft photon corrections. The SCET framework 
can in principle be systematically extended to cover next-to-leading 
and higher-order logarithms, as well as power corrections $m_\mu/m_{B_q}$.

%--------+---------+---------+---------+---------+---------+---------+---------+
%
%
%
%--------+---------+---------+---------+---------+---------+---------+---------+
\section{Numerical results}
\label{sec:numeric}

The framework of SCET allows for a systematic factorization and 
resummation of leading logarithmic QED and QCD corrections to the amplitude 
\refeq{eq:def-red-ampl}. In particular it allowed the identification 
of three resummed contributions: 
\begin{itemize}
\item[$i)$] common virtual \SCETI{} and \SCETII{} QED corrections 
among the final-state leptons to the amplitudes 
$\calA_9$ and $\calA_{10}$ between 
hard and soft/collinear scales in the Sudakov factor 
$e^{S_\ell(\mu_b,\,\mu_c)}$, which are combined with the 
contributions of ultrasoft photons at the level
of the decay width in \refsec{sec:decay-width-2}.
\item[$ii)$] virtual \SCETI{} QED and QCD corrections to the 
power-enhanced amplitude $\calA_9$ between the hard and 
hard-collinear scales given by the overall Sudakov factor 
$e^{S_q(\mu_b,\,\mu_{hc})}$ in \refeq{eq:final-ampQ9}.
\item[$iii)$] virtual QED and QCD corrections within \SCETII{} 
between the hard-collinear and soft/collinear scales, for which the 
RG equation was used such as to arrange that the input of the 
nonperturbative quantities is required at the hard-collinear scale, 
avoiding in this way the necessity of QCD RG evolution below the 
hard-collinear scale, as explained \refsec{sec:SCET2-resummation}. 
This part is given by the $\omega$-dependent soft
Sudakov factor $U^\text{QED}_s(\mu_{hc},\,\mu_s;\,\omega)$ in
\refeq{eq:softRGEUQEDsolution} or the  $\omega$-independent version 
\refeq{eq:SCET2QED-softRG-ana-sol}, which are both equivalent at the 
LL accuracy.
\end{itemize}
We will start with the impact of points $ii)$ and $iii)$ on the power-enhanced
amplitude $\calA_9$ in \refsec{sec:num-ampl} and turn to point $i)$ afterwards
in \refsec{sec:num-Br} when considering the branching fraction of $B_q \to
\mu^+\mu^-$.

Throughout $\alS$ and $\alE$ denote the running couplings in the $\oL{\text{MS}}$
scheme with RGEs given in \refapp{eq:coupl-RG-func}. We use as initial
value $\alS(m_Z) = 0.1181$, with $m_Z = 91.1876$~GeV and number of quark
flavours $n_f = 5$, and perform the running with the four-loop expressions, 
including threshold corrections from quark masses ($\oL{\text{MS}}$ scheme) at
the threshold crossings at $\mu_4 = \mu_b$ ($n_f = 4$) and 
$\mu_3 = 1.2$~GeV ($n_f = 3$). The RGEs for the hard function
in \SCETI{} \refeq{eq:SCET1-RG-ana-sol} and the matrix elements in \SCETII{}
\refeq{eq:SCET2-RG-m}--\refeq{eq:SCET2-RG-M_A} had been solved to LL accuracy.
In the numerical evaluation we will use values of $\alE$ at the typical scales
of \SCETI{} and \SCETII{}. For this purpose, we use as
initial value $1/\alE(m_Z) = 127.955$, and perform the RG evolution to lower
scales with the one-loop expression. In addition to the quark thresholds given
above, the $\tau$-lepton is decoupled at its mass $\mu_\tau \approx 1.777$~GeV.

%
%
%
%--------+---------+---------+---------+---------+---------+---------+---------+
\subsection{Resummation effects for power-enhanced amplitude}
\label{sec:num-ampl}

The resummation of virtual QED and QCD corrections in \SCETI{} to 
$\calA_9$ in \refeq{eq:final-ampQ9} is given by the overall Sudakov 
factor $e^{S_q(\mu_b,\,\mu_{hc})}$ from \refeq{eq:def-S_qI}. 
Evaluating this factor thus provides a direct measure of the size 
of these corrections compared to the fixed-order
result \refeq{eq:mainresult}. 

We calculate the Sudakov factor via numerical integration of the 
part of \refeq{eq:RG-SCET1-H} which contains $Q_q$, i.e. that 
corresponds to $e^{S_q(\mu_b, \mu_{hc})}$. The numerical integration 
includes the scale dependence and threshold crossings of both gauge 
couplings. Since the $\mu_{hc}$ dependence of the Sudakov factor is 
cancelled in large parts by the one of the $B$-meson 
LCDA $\phi_+ (\omega; \mu_{hc})$, which in turn is mainly driven by 
the scale dependence of its first inverse moment 
$\lambda_B(\mu_{hc})$, 
\begin{align}
\label{eq:lambdaB}
  \frac{1}{\lambda_B(\mu)} &
  \equiv \int_0^\infty \frac{d\omega}{\omega} \,\phi_+(\omega;\,\mu),
\end{align}
the relevance of resummation is better assessed by multiplying 
$e^{S_q(\mu_b, \mu_{hc})}$ with the ratio $\lambda_B(\mu_0) / 
\lambda_B(\mu_{hc})$, where $\mu_0 = 1\,$GeV is a fixed reference 
scale. The $\mu_{hc}$~dependence of $\lambda_B(\mu_{hc})$ due to 
QCD is approximated following \cite{Beneke:2011nf},
using as numerical inputs $\lambda_B(\mu_0)$ and $\sigma_1(\mu_0)$ 
given in \reftab{tab:num-input} below. 

%%%%%%%%%%%%%%%%%%%%%%%%%%%%%%%%%%%%%%%%%%%%%%%%%%%%%%%%%%%%%%%%%%%
\begin{table}
\centering
\renewcommand{\arraystretch}{1.3}
\begin{tabular}{|c|cc|c|cc|}
\hline
  $\mu_{hc}$  
& \multicolumn{2}{c|}{$\frac{\lambda_B(\mu_0)}{\lambda_B(\mu_{hc})} e^{S_q(\mu_b,\,\mu_{hc})}$}
& \multicolumn{1}{c|}{$e^{S_q(\mu_b,\,\mu_{hc})}$}
& $\alS$
& $1/\alE$
\\
  $[$GeV$]$
& QCD+QED & only QCD
& QCD+QED  
& 
&
\\
\hline
  1.0  &  0.815  &  0.817  &  0.815  &  0.474  &  134.05  \\
  1.5  &  0.815  &  0.817  &  0.904  &  0.350  &  133.65  \\
  2.0  &  0.769  &  0.769  &  0.946  &  0.302  &  133.29  \\ 
\hline
\end{tabular}
\renewcommand{\arraystretch}{1.0}
\caption{\label{tab:SCET-1-RG}
  \small
  The size of the \SCETI{} Sudakov factor for fixed $\mu_b = 5.0$~GeV and 
  three different choices of $\mu_{hc}$. The column ``only QCD'' shows the
  effect when setting $\alE = 0$ in the cusp anomalous dimensions in 
  \refeq{eq:RG-SCET1-H}.
  For convenience  we provide $\alS$ and $1/\alE$ ($\oL{\text{MS}}$ scheme)
  at the scale $\mu_{hc}$.
}
\end{table}
%%%%%%%%%%%%%%%%%%%%%%%%%%%%%%%%%%%%%%%%%%%%%%%%%%%%%%%%%%%%%%%%%%%

\begin{figure}
\centering
  \includegraphics[width=0.45\textwidth]{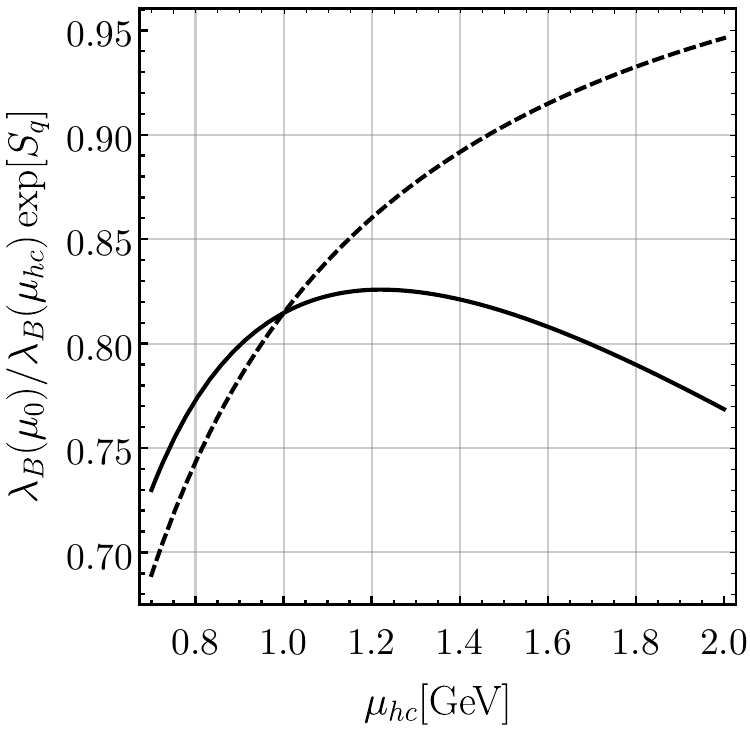}
\caption{\label{fig:U-scet1}
  \small
  The \SCETI{} Sudakov factor $\frac{\lambda_B(\mu_0)}{\lambda_B(\mu_{hc})} 
  e^{S_q(\mu_b,\,\mu_{hc})}$~(solid) and
  $e^{S_q(\mu_b,\,\mu_{hc})}$~(dashed).
}
\end{figure}

The numerical impact of the resummation in \SCETI{} is a suppression 
of the fixed-order result of $\calA_9$ by about $20$\% as tabulated 
in the second column of \reftab{tab:SCET-1-RG}. The dependence on 
$\mu_{hc}$ is also shown in \reffig{fig:U-scet1} (solid line), 
where it reaches 
a maximal value of about $0.82$ around $\mu_{hc} \approx 1.2\,$GeV. 
In the relevant hard-collinear scale range $\mu_{hc} \in [1,2]$~GeV, the 
scale variation is relatively small.   
Note that for $\mu_{hc} < 1\,$GeV the strong coupling $\alS$ 
increases rapidly, reaching for example $\alS \approx 0.75$ at
$\mu_{hc} \approx 0.7\,$GeV, such that the perturbative evaluation 
becomes unreliable. As expected, the Sudakov factor itself has a 
larger $\mu_{hc}$ dependence, varying in the larger range 
($0.80-0.95$), as listed in the fourth column of 
\reftab{tab:SCET-1-RG} and shown in \reffig{fig:U-scet1} (dashed 
line). The resummation effect is by far dominated by the QCD 
evolution as is evident from comparing the columns ``QCD+QED'' 
and ``only QCD''. The difference of both implies that the 
resummation of only QED effects yields tiny suppression of 
($0.1 - 0.3$)\% in agreement with the
natural expectation of the size of a logarithmically enhanced QED correction 
$Q_q Q_\ell \times \alE/\pi\, \times\, \ln^2(\mu_b /\mu_{hc}) \sim 0.2$\%.

The residual dependence on $\mu_{hc}$ displayed by the 
results in \reftab{tab:SCET-1-RG} appears due to the missing 
next-to-leading logarithmic corrections, as well as the 
approximation made to compute the scale-dependence of $\lambda_B$. 
In addition, the QED correction in \refeq{eq:final-ampQ9}
depends also on the first logarithmic moment of the LCDA, 
\begin{align}
\label{eq:sigman}
  \frac{\sigma_1(\mu)}{\lambda_B(\mu)} &
  \equiv \int_0^\infty \frac{d\omega}{\omega} 
  \ln\left(\frac{\mu_0}{\omega}\right)\phi_+(\omega;\,\mu),
\end{align}
where the latter involves the reference scale 
$\mu_0= 1 \; {\rm GeV}$. The scale dependence due to $\sigma_1$ is 
not captured by the results in \reftab{tab:SCET-1-RG}.
Further, a small residual QED  $\mu_{hc}$-dependence is 
compensated by the  \SCETII{} QED evolution 
$U_s^\text{QED} (\mu_{hc}, \mu_s; \omega)$ in 
\refeq{eq:final-ampQ9}.  

%%%%%%%%%%%%%%%%%%%%%%%%%%%%%%%%%%%%%%%%%%%%%%%%%%%%%%%%%%%%%%%%%%%%%%%%%%%%%%%%
%%%%%%%%%%%%%%%%%%%%%%%%%%%%%%%%%%%%%%%%%%%%%%%%%%%%%%%%%%%%%%%%%%%%%%%%%%%%%%%%

Turning to  point $iii)$, we recall that the resummation of virtual 
QED and QCD corrections in \SCETII{} to $\calA_9$ has been organized 
in \refeq{eq:final-ampQ9} such that the nonperturbative input to 
the $B$-meson LCDA is required at the scale $\mu_{hc}$ to avoid 
QCD evolution below $\mu_{hc}$. The resummation of QED effects from
the soft/collinear to the hard-collinear scales are contained in 
the Sudakov factor $B$-meson LCDA momentum-fraction dependent 
factor $U_s^\text{QED}(\mu_{hc}, \mu_s; \omega)$ given in 
\refeq{eq:softRGEUQEDsolution} or the LL-equivalent 
momentum-fraction independent version 
\refeq{eq:SCET2QED-softRG-ana-sol}. 

To estimate the effect of QED resummation in \SCETII{}, we use the 
simple exponential model for the LCDA (see e.g. \cite{Grozin:1996pq})
\begin{align}
  \phi_+ (\omega) & 
  = \frac{\omega}{\omega_0^2} e^{-\omega/\omega_0},
\end{align}
where $\omega_0 = \lambda_B$ and evaluate the ratio of the amplitude 
\refeq{eq:final-ampQ9} with the evolution factor to the amplitude 
without it, 
\begin{align}
  r_\omega & 
  \equiv \frac{\displaystyle\int_0^\infty \frac{d\omega}{\omega}\, 
  U_s^\text{QED} (\mu_{hc},\mu_s; \omega) \,
          \phi_+(\omega) \left[\ln \frac{\omega m_{B_q}}{m_\ell^2} - 1\right]}
         {\displaystyle\int_0^\infty \frac{d\omega}{\omega}\,
          \phi_+(\omega) \left[\ln \frac{\omega m_{B_q}}{m_\ell^2} - 1 \right]} \,.
\label{eq:romegaQED}
\end{align}
For simplicity we assumed that $C_9^\text{eff}$ is constant. 
On the other hand, if we use the $\omega$-independent and 
equally valid expression \refeq{eq:SCET2QED-softRG-ana-sol} for 
the evolution factor, we simply obtain 
$r = U_s^\text{QED} (\mu_{hc},\mu_s)$, independent of the 
form of the $B$-meson LCDA. Evaluation of \refeq{eq:romegaQED} 
results in very small deviations of $r_\omega$ from unity.
For $\mu_s = \lambda_B$ and $\mu_{hc} = 1\,$GeV, the $r_\omega$ 
is a unity minus $(1-3)$\textperthousand{} depending
on the value of $\lambda_B$. For $\mu_{hc} = 2\,$GeV, 
the $r_\omega$ differs from unity
by about $(3-5)$\textperthousand{}. The simpler 
expression $r$ agrees with these deviations from unity within 
about 25\%, independent of whether one uses the LL or the DL 
versions of \refeq{eq:softRGEUQEDsolution} and
\refeq{eq:SCET2QED-softRG-ana-sol}. We conclude that the numerical 
impact of the resummation of QED corrections in
\SCETII{} compared to the fixed-order result is very small. 
As reference value we quote $U_s^\text{QED} (\mu_{hc}, \mu_s) = 
1-0.0015$ at  $\mu_{hc} = 1\,$GeV for $\lambda_B =275\,$MeV.

In summary, the discussion of points $ii)$ and $iii)$ has shown 
that the main numerical effect of resummation on the power-enhanced 
amplitude $\calA_9$ comes from the resummation of QCD corrections 
in \SCETI{} together with QCD running of the LCDA, constituting
an overall suppression factor
\begin{align}
  S_9 &
  \equiv \frac{\lambda_B(\mu_0)}{\lambda_B(\mu_{hc})} e^{S_q(\mu_b,\, \mu_{hc})} \in [0.77,0.82]\,,
\end{align}
that leads to a reduction of $\calA_9$ of about~$20\,\%$, whereas 
\SCETII{} QED resummation from \refeq{eq:softRGEUQEDsolution} can be
safely neglected in \refeq{eq:final-ampQ9}. Thus, from the 
phenomenological perspective it is justified to consider only QCD
resummation on top of the one-loop QED correction, 
once the leptonic Sudakov factor $e^{S_\ell(\mu_b,\,\mu_c)}$ 
is extracted, see point $i)$. 
 
This observation allows us to give a result for the 
$\calA_7$ amplitude including QCD resummation, while the combined 
QCD and QED resummation is not yet feasible. Indeed, the
endpoint divergences, which spoil the factorization of the 
$\calA_7$ amplitude are only related to the QED effects.\footnote{
The operator basis for the amplitude $\calA_7$ is 
more complicated than for $\calA_9$, see \refapp{sec:operator}, 
but one can prove that additional soft, collinear and hard-collinear 
fields that enter the \SCETI{} and \SCETII{} operators relevant 
to $\calA_7$ do not carry color charge.} From the QCD perspective, 
the problem is equivalent to resummation for the heavy-to-light 
tensor current instead of the (axial-) vector current 
relevant to  $\calA_9$. This implies that the QCD cusp anomalous 
dimension is the same as in the $\calA_9$
case\footnote{The non-cusp QCD anomalous dimensions are different for $\calA_9$
and $\calA_7$, due to different Dirac structure and normalization, but in principle
next-to-leading logarithmic resummation of QCD effects could be performed as well
for both amplitudes, see for example \cite{Bell:2010mg}.} and 
accordingly the leading-logarithmic Sudakov factors are equal, 
$S_7 = S_9$. The result is then a uniform reduction of the 
power-enhanced QED correction ``$\calA_9 + \calA_7$''
relative to the fixed-order result \cite{Beneke:2017vpq}.

%
%
%
%--------+---------+---------+---------+---------+---------+---------+---------+
\subsection[Branching fractions $B_q \to \mu^+\mu^-$]
{\boldmath  Branching fractions $B_q \to \mu^+\mu^-$}
\label{sec:num-Br}

We first provide the so-called non-radiative branching
fraction for $B_q \to \mu^+\mu^-$ that is found from the non-radiative decay
width \refeq{eq:decay-width} as
\begin{align}
  \label{eq:non-rad-BR}
  \oL{\text{Br}}_{q\mu}^{(0)} &
  \equiv \frac{\Gamma^{(0)}[B_q \to \mu^+\mu^-]}
{\Gamma_q^\text{tot}}\,.
\end{align}
For the $B_d$ meson the total decay width $\Gamma_d^\text{tot}$ 
is given in by the average width of the light and heavy $B_d$ mass 
eigenstates. In case of the $B_s$ meson the large decay-width 
difference requires a time-integration that can be accounted for 
by using instead of $\Gamma_s^\text{tot}$ the decay width 
$\Gamma_s^H$ of the heavy $B_s$-mass
eigenstate \cite{DeBruyn:2012wk}. According to the results of 
\refsec{sec:num-ampl}, the non-radiative part of the 
amplitude~\refeq{eq:def-red-ampl} is
\begin{align}
   A_{10} \left[ \oL{u}_c \gamma_5 v_\oL{c} \right]
  + \left(S_9 A_9^\text{fix} + S_7 A_7^\text{fix} \right)
  \left[ \oL{u}_c (1 + \gamma_5) v_\oL{c} \right] ,
\end{align}
where the numerically relevant resummation is now factored out 
explicitly as $S_{7,9}$ from the one-loop QED  
amplitudes  $A_{7,9}^\text{fix}$, which were found in 
\cite{Beneke:2017vpq}, see \refeq{eq:mainresult}. 
We have added the fixed-order result of $A_7^\text{fix}$ in the 
numerical analysis and keep $S_7$ separately, although at leading 
logarithmic order in QCD $S_7 = S_9$ as mentioned above.

%%%%%%%%%%%%%%%%%%%%%%%%%%%%%%%%%%%%%%%%%%%%%%%%%%%%%%%%%%%%%%%%%%%%%%%%%%%%%%%%
\begin{table}
\centering
\renewcommand{\arraystretch}{1.3}
\resizebox{\columnwidth}{!}{
\begin{tabular}{|lll|lll|}
\hline
  Parameter  
& Value
& Ref.
&  Parameter  
& Value
& Ref.
\\
\hline
  $G_F$                            & $1.166379 \cdot 10^{-5}$ GeV$^{-2}$  & \cite{Tanabashi:2018oca} % ok
& $m_Z$                            & $91.1876(21)$ GeV                    & \cite{Tanabashi:2018oca} % ok
\\
  $\alS^{(5)}(m_Z)$                & $0.1181(11)$                         & \cite{Tanabashi:2018oca} % ok
& $m_\mu$                          & $105.658\ldots$ MeV                  & \cite{Tanabashi:2018oca} % ok
\\
  $\alE^{(5)}(m_Z)$                & $1/127.955(10)$                      & \cite{Tanabashi:2018oca} % ok 
& $m_t$                            & $173.1(6)$ GeV                       & \cite{Tanabashi:2018oca}
\\
\hline
  $m_{B_s}$                        & $5366.89(19)$ MeV                    & \cite{Tanabashi:2018oca} % ok
& $m_{B_d}$                        & $5279.63(15)$ MeV                    & \cite{Tanabashi:2018oca} % ok
\\
  $f_{B_s}|_{N_f = 2+1}$           & $228.4(3.7)$ MeV                     & \cite{Aoki:2019cca} % ok
& $f_{B_d}|_{N_f = 2+1}$           & $192.0(4.3)$ MeV                     & \cite{Aoki:2019cca} % ok
\\
  $f_{B_s}|_{N_f = 2+1+1}$         & $230.3(1.3)$ MeV                     & \cite{Aoki:2019cca} % ok
& $f_{B_d}|_{N_f = 2+1+1}$         & $190.0(1.3)$ MeV                     & \cite{Aoki:2019cca} % ok
\\
  $1/\Gamma_H^s$                   & $1.615(9)$ ps                        & \cite{Amhis:2016xyh} % ok
& $2/(\Gamma_H^d + \Gamma_L^d)$    & $1.520(4)$ ps                        & \cite{Amhis:2016xyh} % ok
\\
\hline
  $|V_{cb}|_\text{incl}$           & $0.04200(64)$                        & \cite{Gambino:2016jkc} % ok               
& $\lambda_B(\mu_0)$               & $275(75)$ MeV                        & \cite{Beneke:2011nf}
\\ 
  $|V_{tb}^{} V_{ts}^*/V_{cb}^{}|$ & $0.982(1)$                           & \cite{Charles:2015gya, Bona:2016dys}
& $\sigma_1(\mu_0)$                & $1.5(1.0)$                           & \cite{Beneke:2011nf}
\\ 
  $|V_{tb}^{} V_{td}^*|$           & $0.0087(2)$                          & \cite{Charles:2015gya, Bona:2016dys}
& $\sigma_2(\mu_0)$                & $3(2)$                               & \cite{Beneke:2011nf}
\\ 
  $V_{ub}^{} V_{ud}^*/V_{tb}^{} V_{td}^*$ & $0.018 - i\, 0.414$           & \cite{Charles:2015gya, Bona:2016dys}
& & & 
\\
\hline
\end{tabular}
}
\renewcommand{\arraystretch}{1.0}
\caption{\label{tab:num-input}
  \small
  Numerical input values for parameters: Note that $\alE^{(5)}(m_Z)$ has been
  determined with $\alS^{(5)}(m_Z) = 0.1187(16)$ in \cite{Tanabashi:2018oca} and
  the corresponding $\Delta{\alE^{(5)}}_\text{,hadr}(m_Z) = 0.02764(7)$.
  The $B$-meson decay constants are averages from the FLAG group for $N_f = 2 + 1$
  from \cite{McNeile:2011ng, Bazavov:2011aa, Na:2012kp, Christ:2014uea, Aoki:2014nga} 
  and for $N_f = 2+1+1$ from \cite{Bazavov:2017lyh, Bussone:2016iua, Dowdall:2013tga,
  Hughes:2017spc}.  
  The $B_{s,d}$ lifetimes are prepared by HFLAV for the PDG 2018 review 
  \cite{Tanabashi:2018oca}. The same numerical values of $\lambda_B$ and $\sigma_{1,2}$
  at the reference scale $\mu_0 = 1$~GeV are used for $B_s$ and $B_d$ mesons. 
  The values of the Wilson coefficients at $\mu_b = 5.0$~GeV are $C_{1-6} = 
  \{-0.25,\, 1.01,\, -0.005,\, -0.077,\, 0.0003,\, 0.0009 \}$, $C_7^\text{eff} = -0.302$,
  $C_9 = 4.344$ and $C_{10} = -4.198$ from \cite{Beneke:2017vpq}.
}
\end{table}
%%%%%%%%%%%%%%%%%%%%%%%%%%%%%%%%%%%%%%%%%%%%%%%%%%%%%%%%%%%%%%%%%%%%%%%%%%%%%%%%

The non-radiative time-integrated branching fraction of $B_s \to \mu^+\mu^-$
for central values of the parameters collected in \reftab{tab:num-input} and
using the $N_f = 2+1+1$ lattice result of $f_{B_s}$ is
\begin{align}
  \nonumber
  \oL{\text{Br}}_{s\mu}^{(0)} &
  = 3.677 \cdot 10^{-9} \times \Big(1
     + \frac{\text{GeV}}{10^3 \cdot \lambda_B} \big[ 
               S_9\, (-6.46 + 1.27 \sigma_1) 
             + S_7\, (4.74 - 1.54 \sigma_1 + 0.15 \sigma_2) \big] \Big)
\\ &
  = 3.677 \cdot 10^{-9} \times \big( 1 - 0.0166\, S_9 + 0.0105\, S_7 \big)
  = 3.660 \cdot 10^{-9}.
\end{align}
In the first line we keep the $B$-meson LCDA parameters and the 
Sudakov factors unevaluated. The second line shows that $\calA_9$ 
interferes destructively with $\calA_{10}$ whereas $\calA_7$ 
interferes constructively. The separate contributions to the 
branching fraction due to $\calA_9$ and $\calA_7$ are rather large, 
about $-1.7\, S_9\,$\% and $+1.0\, S_7\,$\% as found 
previously~\cite{Beneke:2017vpq}. Both contributions cancel 
in part and lead to an overall reduction of the branching fraction 
of $0.5$\%, when accounting for the Sudakov factor 
$S_9 \approx S_7 \approx 0.8$. The non-radiative branching 
fraction of $B_d \to \mu^+\mu^-$ for central values
of the parameters is
\begin{align}
  \nonumber
  \oL{\text{Br}}_{d\mu}^{(0)} &
  = 1.031 \cdot 10^{-10} \times \Big(1 
           + \frac{\text{GeV}}{10^{3} \cdot \lambda_B} \big[ 
               S_9\, (-6.04 + 1.18 \sigma_1) 
             + S_7\, (4.67 - 1.51 \sigma_1 + 0.15 \sigma_2) \big] \Big)
\\ &
  = 1.031 \cdot 10^{-10} \times \big( 1 - 0.0155\, S_9 + 0.0103\, S_7 \big)
  = 1.027 \cdot 10^{-10}.
\end{align}
The numerical difference between $B_s$ and $B_d$ decays for the
contribution proportional to $S_9$ is due to the terms proportional 
to the CKM factor $V_{ub}^{} V_{uq}^\ast$ in \refeq{eq:C_9^eff}.

Let us for completeness provide also a detailed error budget of 
the non-radiative branching fractions $B_q \to \mu^+\mu^-$. We find
\begin{align}
  \nonumber
  \oL{\text{Br}}^{(0)}_{s\mu}  
  = \begin{pmatrix} 3.599 \\ 3.660 \end{pmatrix} \Big[
    1 &
    + \begin{pmatrix} 0.032 \\ 0.011 \end{pmatrix}_{f_{B_s}} 
    + 0.031|_\text{CKM}
    + 0.011|_{m_t}
\\ & 
    + 0.006|_\text{pmr} 
    + 0.012|_\text{non-pmr}
    \; {}^{+0.003}_{-0.005}|_\text{LCDA}
    \Big] \cdot 10^{-9},
  \label{eq:Br-Bs-errors}
\\[0.2cm]
  \nonumber
  \oL{\text{Br}}^{(0)}_{d\mu} 
  = \begin{pmatrix} 1.049 \\ 1.027 \end{pmatrix} \Big[
    1 &
    + \begin{pmatrix} 0.045 \\ 0.014 \end{pmatrix}_{f_{B_d}} 
    + 0.046|_\text{CKM}
    + 0.011|_{m_t}
\\ &
    + 0.003|_\text{pmr} 
    + 0.012|_\text{non-pmr}
    \; {}^{+0.003}_{-0.005}|_\text{LCDA}
    \Big] \cdot 10^{-10},
  \label{eq:Br-Bd-errors}
\end{align}
where we group uncertainties as follows: 
\begin{itemize}
\item[$i)$] main parametric long-distance ($f_{B_q}$) and 
short-distance (CKM and $m_t$); 
\item[$ii)$] remaining non-QED parametric ($\Gamma_q,\, \alS$) 
and non-QED non-parametric 
   ($\mu_W,\, \mu_b$ and higher order, see \cite{Bobeth:2013uxa});
\item[$iii)$] from the $B$-meson LCDA parameters $\lambda_B$ and $\sigma_{1,2}$ entering
     the power-enhanced QED correction.
\end{itemize}
We provide two values of the branching fraction depending on the 
choice of the lattice calculation of $f_{B_q}$ for $N_f = 2 + 1$ 
(upper) and $N_f = 2 + 1 + 1$ (lower), employing averages from 
FLAG 2019 \cite{Aoki:2019cca}. Note that the small uncertainties 
of the $N_f = 2 + 1 + 1$ results are currently dominated by a single group \cite{Bazavov:2017lyh}
and confirmation by other lattice groups in the future is desirable. It can
be observed that in this case the largest uncertainties are due to CKM parameters,
such that in principle they can be determined, provided the experimental accuracy of the measurements is 
at the one-percent level. Still fairly large errors are due to the top-quark mass
$m_t = (173.1 \pm 0.6)$~GeV, here assumed to be in the pole scheme \cite{Bobeth:2013uxa},
where an additional non-parametric uncertainty of 0.2\% is included (in ``non-pmr'') for
the conversion to the $\oL{\text{MS}}$ scheme as in \cite{Bobeth:2013uxa}. 
Further ``non-pmr'' contains a 0.4\% uncertainty from $\mu_W$ variation and 0.5\% 
further higher order uncertainty, all linearly added, see also \cite{Bobeth:2013uxa}.

%%%%%%%%%%%%%%%%%%%%%%%%%%%%%%%%%%%%%%%%%%%%%%%%%%%%%%%%%%%%%%%%%%%%%%%%%%%%%%%%
\begin{figure}
\centering
  \includegraphics[width=0.45\textwidth]{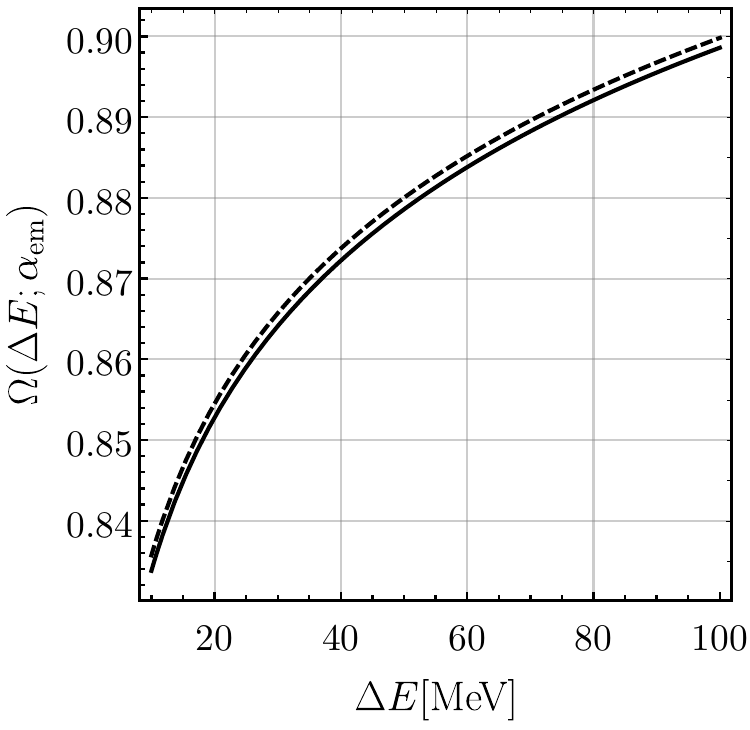}
\caption{\label{fig:DelE}
  \small
  The radiative factor in \refeq{eq:rad-fac} for $B_s\to \mu^+\mu^-$ 
  in the range $\Delta E \in [10,\; 100]$~MeV for the two values 
  $\alE^{-1} = 134.28$~(solid) and $\alE^{-1} = 136.0$~(dashed).
}
\end{figure}
%%%%%%%%%%%%%%%%%%%%%%%%%%%%%%%%%%%%%%%%%%%%%%%%%%%%%%%%%%%%%%%%%%%%%%%%%%%%%%%%

The radiative $B_q \to \mu^+\mu^-$ branching fraction including 
ultrasoft radiation with total energy  $E_{X_s} < \Delta E$ is 
obtained from \refeq{eq:decay-width} and \refeq{eq:non-rad-BR} as 
\begin{align}
  \label{eq:rad-BR}
  \oL{\text{Br}}_{q\mu}(\Delta E) &
  \equiv \oL{\text{Br}}_{q\mu}^{(0)} \times \Omega(\Delta E; \alE) 
  \,, 
\intertext{with radiative factor}
  \label{eq:rad-fac}
  \Omega(\Delta E; \alE) &
  \equiv \left( \frac{2\Delta E}{m_{B_q}} \right)^{
       -\frac{2\alE}{\pi} \left(1+\ln\frac{m_\mu^2}{m_{B_q}^2} \right)} 
\end{align}
from \refeq{eq:GBs2}. $\Delta E$ corresponds to a window in the 
dilepton invariant mass $s_{\ell\oL{\ell}} = 
(p_\ell + p_\oL{\ell})^2$ around $s_{\ell\oL{\ell}} = m_{B_q}^2$ 
defining the signal region in the experimental analysis for which 
our effective theory framework is set up. The dependence of the 
radiative factor $\Omega$ on 
$\Delta E = (m_{B_q}^2 - s_{\ell\oL{\ell}})^{1/2}$ is shown in 
\reffig{fig:DelE} for $B_s$ mesons. One might consider 
$\Delta E \simeq 60$~MeV as an example of a larger value 
for the theory framework that requires $\Delta E \ll \LambQCD$. 
In this case the signal window contains about~88\% of the non-radiative rate,
whereas for example the smaller signal window with $\Delta E \simeq 10$~MeV 
still contains~84\%. The sizes of the signal windows in the first LHCb 
\cite{Aaij:2012nna, Aaij:2013aka} and CMS \cite{Chatrchyan:2013bka} analyses of
$B_s \to \mu^+\mu^-$ were about $\Delta E \simeq \pm 60$~MeV and $\Delta E \simeq 
{}_{-67}^{+83}$~MeV around $m_{B_s}$, respectively.\footnote{
In the experimental analysis the signal window extends also above the 
upper boundary $(s_{\ell\oL{\ell}})_\text{max} = m_{B_q}^2$ of the phase space 
for $B_q \to \ell^+\ell^- + n\gamma$ due to resolution effects in the
reconstruction of $s_{\ell\oL{\ell}}$.} In comparison, 
the experimental resolution in the dilepton-invariant mass is reported to be about
$25$~MeV in LHCb and depending on the muon direction about $(32-75)$~MeV in CMS.
More recent experimental analyses do not use signal windows, but rather
fit the modelled signal components of $B_s\to\mu^+\mu^-$ and $B_d\to\mu^+\mu^-$ 
simultaneously with background components over a wide range of $s_{\ell\oL{\ell}}$.
The modelling of the components involves also the simulation of photon radiation
with the help of {PHOTOS} \cite{Golonka:2005pn}. We note that 
the systematic framework developed in this work makes it 
advantageous to 
compare experimental data in a signal window to corresponding 
theoretical predictions without the need for modeling or simulation.

Finally, we point out that  the dependence on the value of $\alE$ 
in the exponent of $\Omega(\Delta E; \alE)$ slightly changes the 
value of the radiative factor. The associated parametric uncertainty
for $\Delta E = 60$~MeV and the two values $\alE^{-1} = 
\{134.28,\; 136.0\}$
\begin{align}
  \Omega(60\,\text{MeV}; \alE) & =
  \left\{ \begin{array}{ccc}
    \{0.8838,\; 0.8852\} & \qquad \text{for} \qquad & B_s \\[0.2cm]
    \{0.8848,\; 0.8862\} & \text{for} & B_d
  \end{array} \right.
\end{align}  
amounts to less than 0.2\% for the two choices of $\alE$. The first 
choice corresponds to $\alE(1.0\,\text{GeV})$, whereas $1/\alE(\mu_c) = 136.0$ represents the running
at one-loop to the muon mass scale, spanning a range of values that covers the
renormalization scheme dependence of $\alE$. This uncertainty is
comparable to the parametric QED uncertainties due to the $B$-meson LCDA parameters
in \refeq{eq:Br-Bs-errors} and \refeq{eq:Br-Bd-errors}. It must be added when
comparing the predictions of $\oL{\text{Br}}_{q\mu}(\Delta E)$ with measurements,
i.e. it must be accounted for when extracting the non-radiative rate by experiments.

%
%
%
%--------+---------+---------+---------+---------+---------+---------+---------+
\subsection[Rate asymmetries in $B_q \to \mu^+\mu^-$]
{\boldmath  Rate asymmetries in $B_q \to \mu^+\mu^-$}
\label{sec:def-CP-asy}

The decay of neutral $B_q$ mesons into a muon pair $\mu^+_\lambda\mu^-_\lambda$
in a helicity configuration $\lambda = L,R$ allows to investigate further
observables in measurements of its time-dependent rate asymmetry
\begin{align}
  \frac{  \Gamma[B_q(t)     \to \mu^+_\lambda\mu^-_\lambda] 
        - \Gamma[\oL{B}_q(t)\to \mu^+_\lambda\mu^-_\lambda]}
       {  \Gamma[B_q(t)     \to \mu^+_\lambda\mu^-_\lambda] 
        + \Gamma[\oL{B}_q(t)\to \mu^+_\lambda\mu^-_\lambda]} &
  = \frac{C_q^\lambda \, \cos(\Delta m_{B_q} t) + S_q^\lambda \, \sin(\Delta m_{B_q} t)}
         {\cosh(y_q\, t/\tau_{B_q}) 
         + A_q^\lambda \, \sinh(y_q\, t/\tau_{B_q})} \,,
\end{align}
provided the initial flavour of the $B_q$ is tagged. The decay-width difference 
$\Delta \Gamma_q$  enters via $y_q = \tau_{B_q} \Delta\Gamma_q/2$, which is
rather sizeable for the $B_s$ system, $y_s = 0.066\pm 0.004$ \cite{Tanabashi:2018oca}, 
but can be neglected for the $B_d$ system $y_d \lesssim 0.005$ \cite{Tanabashi:2018oca}.
We refer to \cite{DeBruyn:2012wk} for definitions and further details.

The three observables are related by $|C_q^\lambda|^2 + |S_q^\lambda|^2 +
|A_q^\lambda|^2 = 1$. Note that a muon pair $\mu^+_\lambda\mu^-_\lambda$ with
definite helicity $\lambda = L,R$ is not a CP eigenstate, but both are related
under CP transformation as $CP(|\mu^+_L\mu^-_L \rangle) = e^{i \delta_{CP}(\mu^+\mu^-)} 
|\mu^+_R\mu^-_R \rangle$ with a convention-dependent phase $\delta_{CP}(\mu^+\mu^-)$.
Thus the observables $C_q^\lambda$, $S_q^\lambda$ and $A_q^\lambda$ are not CP 
asymmetries. The SM predictions based on the LO QED amplitude $\calA_{10}$ alone
lead to vanishing $C_q^\lambda = S_q^\lambda = 0$. The mass-eigenstate
rate asymmetry $A_q^\lambda = 1$ because only the heavier $B_q$ 
mass-eigenstate can decay into leptons. The NLO QED amplitude 
$\calA_9$ contains in addition to the dominant amplitude involving 
$V_{tb}^{} V_{tq}^*$ the term proportional to the CKM element product 
$V_{ub}^{} V_{uq}^*$, and hence a second weak (CP-violating) 
phase, as well 
as scattering (CP-conserving) phases through $C_9^\text{eff}$ in 
\refeq{eq:C_9^eff}, potentially changing these predictions. For 
the future it is important to know at which level QED
corrections in the SM induce a deviation from $C_q^\lambda = S_q^\lambda = 0$
and $A_q^\lambda = 1$ to disentangle them from new physics effects. 

The suppression factors $\alE$ and 
$(V_{ub}^{} V_{uq}^*)/(V_{tb}^{} V_{tq}^*)$ in $\calA_9$ suggest 
that the deviations from the LO SM predictions will be very 
small. Due to the presence of a scattering phase we must generalize 
the results of \cite{DeBruyn:2012wk} that were based on the 
assumption of only additional weak phases. The muon-helicity 
dependent observables are given as
\begin{align}
  \label{eq:def-rate-asy}
  C_q^\lambda & = \frac{1 - |\xi_q^\lambda|^2}{1 + |\xi_q^\lambda|^2} \,,
&
  S_q^\lambda & = \frac{2 \im \xi_q^\lambda}{1 + |\xi_q^\lambda|^2} \,,
&
  A_q^\lambda &
  = \frac{2 \re\xi_q^\lambda}{1 + |\xi_q^\lambda|^2} \,,
\end{align}
where
\begin{equation}
  \xi_q^\lambda 
  = - \frac{S + \eta_\lambda P}{\oL{S} - \eta_\lambda \oL{P} } \,,
  \qquad\mbox{with}\qquad 
  \eta_{L/R}  = \pm 1\,.
\end{equation}
The barred quantities $\oL{P} = P[\varphi_W \to -\varphi_W]$ and
$\oL{S} = S[\varphi_W \to -\varphi_W]$ are obtained from the 
unbarred ones after reverting the signs of all CP-violating 
phases $\varphi_W$.
We introduced here
\begin{align}
  P & = \frac{A_{10} + A_9 + A_7}{m_\ell m_{B_q} f_{B_q} \calN} \,, 
&
  S & = \beta_\mu \frac{A_9 + A_7}{m_\ell m_{B_q} f_{B_q} \calN} \,.
\end{align}

The generalized expressions \refeq{eq:def-rate-asy} in terms of 
$P$, $\oL{P}$ and $S$, $\oL{S}$ can be derived straightforwardly. 
We find
\begin{align} 
  C_q^\lambda & = \frac{\oL{B}_\lambda - B_\lambda}{\oL{B}_\lambda + B_\lambda} \,,
&
  S_q^\lambda & = \frac{2 \im \widetilde{B}_\lambda}{\oL{B}_\lambda + B_\lambda} \,,
&
  A_q^\lambda &
  = \frac{2 \re \widetilde{B}_\lambda}{\oL{B}_\lambda + B_\lambda} \,,
\end{align}
where
\begin{align}
  B_\lambda & 
  = |P|^2 + |S|^2 + \eta_\lambda \big( PS^* + SP^* \big) ,
\\
  \oL{B}_\lambda & 
  = |\oL{P}|^2 + |\oL{S}|^2 - \eta_\lambda \big(\oL{P}\oL{S}^* + \oL{S}\oL{P}^* \big) ,
\\
  \widetilde{B}_\lambda &
  = P \oL{P}^* - S \oL{S}^* + \eta_\lambda \big(S \oL{P}^* - P \oL{S}^* \big) .
\end{align}
The linear combinations $C_q \equiv \frac{1}{2}(C_q^L + C_q^R)$ and 
$S_q \equiv \frac{1}{2}(S_q^L + S_q^R)$ are CP-odd, while 
 $\Delta C_q \equiv \frac{1}{2}(C_q^L - C_q^R)$ and 
$\Delta S_q \equiv \frac{1}{2}(S_q^L - S_q^R)$ are CP-even 
quantities, see for example the related discussion 
\cite{Beneke:2003zv} on time-dependent rates in the 
$B^0 \to \pi^\mp \rho^\pm$ systems. 

The assumption of only weak phases implies $\oL{P} = P^*$ and 
$\oL{S} = S^*$, and leads to significant 
simplifications as shown in \cite{DeBruyn:2012wk}. In the following 
we do not make this general assumption, but we use that 
the amplitudes $A_{10}$ and $A_7$ are real within 
the approximations adopted in this paper, i.e. do not contain 
neither weak nor scattering phases, and further that the 
amplitudes $A_{7,9} \ll A_{10}$ are
suppressed by a factor $\alE$. The expansion in $\alE$ yields
\begin{align}
  \nonumber
  C_q^\lambda & 
  = -\frac{\re\big[A_9 - \oL{A}_9 + \eta_\lambda \big(2 A_7 + A_9 + \oL{A}_9 \big) \big]}{A_{10}} \,,
\\
  \label{eq:sol-rate-asy}
  S_q^\lambda & 
  = \frac{\im\big[ A_9 - \oL{A}_9 + \eta_\lambda \big( A_9 + \oL{A}_9 \big) \big]}{A_{10}} \,,
\\
  \nonumber
  A_q^\lambda & 
  = 1
  - \frac{2 (A_7)^2 + (1 + \eta_\lambda) |A_9|^2 + (1 - \eta_\lambda) |\oL{A}_9|^2
         +2 A_7 \re\big[ (1 + \eta_\lambda) A_9 + (1 - \eta_\lambda) \oL{A}_9 \big]}{(A_{10})^2} \,,
\end{align}  
where we further use $\beta_\mu \approx 1$ in $S$, which is 
numerically well justified for muons. Note that for $A_q^\lambda$ 
the first non-vanishing deviation from unity appears 
only at $\calO{(\alE^2)}$.

The expressions show that in the absence of a weak phase 
difference (i.e. when $\oL{A}_9 = A_9$), the QED 
corrections imply a modification of the observables
\begin{align}
  \label{eq:no-strong-ph}
  C_q^\lambda & 
  = - \eta_\lambda \frac{2 \re[A_7 + A_9]}{A_{10}} \,,
&
  S_q^\lambda & 
  = \eta_\lambda \frac{2 \im[A_7 + A_9]}{A_{10}} \,, 
&
  A_q^\lambda & 
  = 1
  - 2 \frac{|A_7 + A_9|^2}{(A_{10})^2} \,.
\end{align}
from their naive values 0, 0, 1. We 
added $\im A_7 = 0$ in the numerator of $S_q^\lambda$ 
to make the
result appear in line with $C_q^\lambda$ and $A_q^\lambda$. 
As expected, the CP-odd observables $C_q$, $S_q$ vanish, 
since there is no CP-violating phase, while 
$\Delta C_q=C_q^L$, $\Delta S_q=S_s^L$ are non-zero though small.
$C_q^\lambda$ and $S_q^\lambda$ still depend on the 
muons' helicity through $\eta_\lambda$, whereas $A_q^\lambda$ 
becomes independent on the helicity. The expressions 
\refeq{eq:no-strong-ph} are very good approximations for the  
case of $B_s$ mesons, where the $V_{ub}^{} V_{us}^*$ term in the 
amplitude is negligible due to the strong suppression 
$|(V_{ub}^{} V_{us}^*)/(V_{tb}^{} V_{ts}^*)| \lesssim 0.005$ in 
\refeq{eq:C_9^eff}. In consequence \cite{Beneke:2017vpq}\footnote{
Note the missing factor $\eta_\lambda$ in Eq.~(20) of 
\cite{Beneke:2017vpq}.}
\begin{align}
  \label{eq:Bs-rate-asy}
  C_s^\lambda & = + \eta_\lambda \,0.6\% \,,
&
  S_s^\lambda & = - \eta_\lambda \,0.1\% \,, 
&
  A_s & = 1 - 2.0 \cdot 10^{-5} \,.
\end{align}
While the first measurement of $A_s = 8.2 \pm 10.7$ from LHCb 
\cite{Aaij:2017vad} suffers from huge errors, a future deviation 
from the SM prediction 
$A_s = 1$ at LO in QED can be safely attributed to non-standard 
effects in view of the tiny QED contributions.

On the other hand, for the $B_d$ system the expressions 
\refeq{eq:sol-rate-asy} must be used. Here the weak phase
in $(V_{ub}^{} V_{ud}^*)/(V_{tb}^{} V_{td}^*)$ leads to differences 
among the two helicities, and we therefore provide 
\begin{equation}
\begin{aligned}
\label{eq:Bd-rate-asy}
  C_d & = - 0.08\% \,,
&
  S_d & = + 0.03\% \,, 
&
  A_d^L & = 1 - 1.4 \cdot 10^{-5} \,,
\\
\Delta C_d & = + 0.60\% \,,
&
  \Delta S_d & = - 0.13\% \,, 
&
  A_d^R & = 1 - 2.4 \cdot 10^{-5} \,.
\end{aligned}
\end{equation}
We find similar magnitudes as for the $B_s$ system, but 
non-vanishing CP asymmetries $C_d$ and $S_d$, which are 
suppressed by a factor of a few compared to the CP-conserving 
quantities $\Delta C_d$, $\Delta S_d$.

In summary we find tiny contributions from the power-enhanced NLO 
QED amplitudes $\calA_{7,9}$ to the rate asymmetries in 
$B_q \to \mu^+\mu^-$ decays, leaving these observables as   
``null tests'' at the percent level of the SM. The generated CP 
asymmetries in $B_d \to \mu^+\mu^-$ are about 
$|C_d|, |S_d| \approx 0.1\%$ and are strongly suppressed in 
$B_s \to \mu^+\mu^-$. Note that there
are other, even smaller higher-order corrections to the rate 
asymmetries in the SM, such as for example neutral Higgs boson 
penguin diagrams, which give rise to higher-dimensional operators 
than dimension six at the electroweak scale, suppressed 
by $(m_b/m_W)^2\sim 10^{-3}$.

%--------+---------+---------+---------+---------+---------+---------+---------+
%
%
%
%--------+---------+---------+---------+---------+---------+---------+---------+
\section{Summary and conclusions}
\label{sec:summary}

We have developed a systematic treatment of virtual and real QED 
effects for the power-enhanced QED contribution to 
$B_q\to \mu^+\mu^-$, previously reported in \cite{Beneke:2017vpq}, 
for the case when the energy of undetected photons 
$\Delta E$ is small compared to the typical scale of the 
QCD binding energy and the muon mass. The treatment includes 
the resummation of large QED logarithms from various scales. 
The effects from the process-specific energy scales set by the 
external kinematics and internal dynamics 
of $B_q\to \mu^+\mu^-$ are factorized with the help of 
soft-collinear effective theory (SCET) starting at the hard
scale of order of the $B$-meson mass $m_{B_q}$ in a 
two-step matching. First hard fluctuations on the heavy meson 
mass scale are decoupled in the matching on \SCETI{},
and subsequently hard-collinear modes with virtuality 
$m_{B_q} \Lambda$ are decoupled in the matching onto \SCETII{}, 
which contains collinear and soft degrees of freedom of order 
$\Lambda$. Finally we treat the remaining
ultrasoft QED interactions
in the limit of static heavy leptons.

Due to the double helicity and annihilation suppression of the 
$B_q\to \mu^+\mu^-$ amplitude in the absence of QED corrections, 
the power-enhanced amplitude analyzed in this paper is actually 
an example of subleading-power resummation in SCET. 
The SCET framework allows us to resum the large QCD and QED 
logarithmic corrections systematically and reveals a lepton-mass 
induced operator mixing between so-called A-type and
B-type SCET collinear operators as the origin of the 
leading logarithmic correction. We derive a \SCETII{} 
factorization theorem for the non-radiative amplitude, for 
the contributions due to the weak semileptonic operators $\Op_9$ 
and $\Op_{10}$. The formula includes ``structure-dependent'' 
logarithms beyond the standard Yennie-Frautschi-Suura (YFS)
exponentiation. After squaring the amplitude, the result is 
organized such that the standard exponentiated logarithms 
are factorized and the process-specific terms are made 
explicit. We provide the relevant operator definitions, which 
facilitates the reuse of existing results within our formalism 
avoiding double-counting. We emphasize that the standard YFS 
approach contains the implicit assumption that the cutoff on virtual 
photon effects, which should be below the scale of the inverse 
size of the $B$-meson for the latter to be treated as point-like,  
can be raised to the $B$-meson mass scale. Our result justifies 
this procedure to a certain extent and gives a precise expression 
for the corrections to this extrapolation.

On the quantitative side, we performed the resummation at the 
leading-logarithmic level in both, the QCD and the QED coupling. 
We find that once the standard YFS exponent is extracted, the 
relevant effect comes from QCD logarithms. For practical purposes 
it is therefore sufficient to improve the previous 
one-loop QED calculation \cite{Beneke:2017vpq} by QCD logarithms, 
which results in an approximately 20\% reduction of the 
power-enhanced amplitude. We updated the $B_{s,d}\to \mu^+\mu^-$ 
branching fractions and provided an estimate of the present 
theoretical uncertainty. We also discussed various rate 
asymmetries, including CP violation at the permille level 
in the $B_d$ decay mode that arises entirely from the 
power-enhanced QED correction. These results quantify the 
SM background to New Physics searches in what would be ``null 
observables''  in the absence of QED. 

The derivation of the \SCETII{} factorization theorem shows the 
need of generalizing the concepts of the $B$-meson decay constant 
and light-cone distribution amplitudes (LCDA) in the presence of 
QED. We establish that even in very simple processes QED corrections 
are non-universal beyond the leading logarithmic approximation. 
The coupling of soft photons to final-state charged particles 
renders the $B$-meson matrix elements dependent on the soft 
Wilson lines, which carry information on the charge and direction 
of the final-state particles. Upon expansion in the QED coupling, 
this necessitates the inclusion of hadronic matrix elements, 
which are non-local time-ordered products. This has to be 
considered when these quantities are evaluated with nonperturbative 
methods, such as lattice gauge theory. 
The situation becomes even more complicated for exclusive semileptonic
decays, where different process-dependent nonperturbative objects have
to be defined for different phase-space regions. The number of nonperturbative
objects will proliferate, as in this case it will also 
be necessary to include
QED effects for the final-state hadrons.

%--------+---------+---------+---------+---------+---------+---------+---------+
\subsubsection*{Acknowledgements}

This work is supported by the DFG 
Sonderforschungsbereich/Transregio 110  ``Symmetries and the 
Emergence of Structure in QCD''. The work of CB is 
supported by DFG under grant BO-4535/1-1.

%--------+---------+---------+---------+---------+---------+---------+---------+
%
%
%
%--------+---------+---------+---------+---------+---------+---------+---------+
\appendix

%
%
%
%--------+---------+---------+---------+---------+---------+---------+---------+
\section{SCET conventions and results}
\label{app:SCET}

Throughout we use the definitions
\begin{align}
   g_{\mu\nu}^\perp &
  \equiv g_{\mu\nu} - \frac{n_+^\mu n_-^\nu}{2} - \frac{n_-^\mu n_+^\nu}{2} ,
&
  \varepsilon_{\mu\nu}^\perp &
  \equiv \varepsilon_{\mu\nu\alpha\beta} \, \frac{n_+^\alpha n_-^\beta}{2} ,
\end{align}
and the convention $\varepsilon_{0123} = -1$ such that 
\begin{align}
  \mbox{Tr}[\gamma_\mu \gamma_\nu \gamma_\alpha \gamma_\beta \gamma_5] &
  = - 4 i \varepsilon_{\mu\nu\alpha\beta} ,
&
  \sigma_{\mu\nu} \gamma_5 &
  = \frac{i}{2} \varepsilon_{\mu\nu\alpha\beta} \, \sigma^{\alpha\beta} .
\end{align}
%\cb{Differs by sign from TRACER - manual eq.(24), but compatible with package X.}

The running QCD and QED couplings in the $\oL{\text{MS}}$ scheme are denoted
as $\alS$ and $\alE$, respectively. They obey the RG equations 
\begin{align}
  \label{eq:coupl-RG-func}
  \frac{d\alpha_i}{d\ln \mu} & 
  = \beta_i(\alS, \alE) 
  = -2 \alpha_i \frac{\alpha_i}{4\pi} \beta_{0,i}  + \mathcal{O}(\alpha_i^3), &
  i & = (s, \text{em}) ,
\end{align}
which decouple at the leading order. The one-loop contributions are
\begin{align}
  \beta_0 & = \frac{11}{3} N_c -\frac{2}{3} n_f , &
  \beta_{0,\rm em}& = - \frac{4}{3} \left[N_c (n_u Q_u^2 + n_d Q_d^2) + n_\ell Q_\ell^2 \right] ,
\end{align}
where $N_c = 3$ and $n_f=n_u+n_d$ is the total number of active quark 
flavours. The separate up-type quark ($Q_u = + 2/3$), down-type quark 
($Q_d = -1/3$) and charged lepton ($Q_\ell = -1$) numbers are denoted 
as $n_u$, $n_d$ and $n_\ell$, respectively.

%
%--------+---------+---------+---------+---------+---------+---------+---------+
\subsection{Lagrangians}
\label{app:SCET-Lag}

The leading-power Lagrangian of a hard-collinear
fermion $f_C$  in \SCETI{} reads 
\cite{Bauer:2000yr, Beneke:2002ni}
\begin{align}
  \label{eq:SCET-Lag-xi0}
  \mathcal{L}_f^{(0)} &
  = \bar{f}_C \left( i\nm D 
      + i \slashed{D}_{C\perp} \frac{1}{i\np D_C} i \slashed{D}_{C\perp} 
\right) \frac{\nps}{2} f_C .
\end{align}
The capital subscript $C$ is used in \SCETI{} to denote collinear 
fields with hard-collinear and collinear scaling.
The field $f_C$ represents either the light quark or the lepton fields in 
\reftab{tab:fields-scaling}. The corresponding anti-hard-collinear field
is denoted by $C \to \oL{C}$ and its Lagrangian is obtained by the replacement
$\np \leftrightarrow \nm$. The covariant derivatives are
\begin{align}
  \label{eq:cov-dev-SCETI-nm}
  i \nm D & 
  = i \nm \partial + e Q_f \big[\nm A_C + \nm A_s(x_-)\big] + g_s 
  \big[\nm G_C + \nm G_s(x_-)\big]\, ,
\\
  \label{eq:cov-dev-SCETIC}
  i D_C & 
  = i \partial + e Q_f A_C + g_s G_C ,
\end{align} 
where $A_\mu$ and $G_\mu = G_\mu^A T^A$ denote the photon and gluon 
fields, respectively, and their subscripts $C$ and $s$ distinguish 
the hard-collinear  and soft fields. The $Q_f$ denotes the electric 
charge of $f$, whereas the generators of QCD for the fundamental 
representation are denoted as $T^A$.
The  QED and QCD coupling constants are $e = \sqrt{4 \pi \alE}$ and 
$g_s = \sqrt{4 \pi \alS}$, respectively. Fields without argument are 
taken at position $x$. For the soft fields, their multipole expansion 
in \SCETI{} interactions with collinear fields \cite{Beneke:2002ni} 
is made explicit by the argument 
$x_\mp \equiv (n_\pm x) \,n_\mp/2$. 

The operators in \SCETI{} have to be gauge-invariant under 
hard-collinear QED and QCD gauge transformations, which is 
achieved by combining
$f_C$ with appropriate Wilson lines of hard-collinear photons and gluons
\begin{align}
  \label{eq:def-C-wl-QED}
  W_{f C} (x) &
  \equiv \exp\left[ie \, Q_f \! \int_{-\infty}^{0} \!\! ds \, \np 
    A_C \left(x+s \np \right)\right] ,
\\
  \label{eq:def-C-wl-QCD}
  W_{C} (x) &
  \equiv \mathcal{P} \exp\left[i g_s \! \int_{-\infty}^{0} \!\! ds \, \np 
    G_C \left(x+s \np \right)\right] ,
\end{align}
respectively. The Wilson lines $W_{fC}$ in QED depend on the charge $Q_f$ 
of $f_C$, whereas the QCD Wilson lines $W_{C}$ involve the path-ordering 
operator $\mathcal{P}$. There are analogous anti-hard-collinear Wilson 
lines $W_{f\oL{C}}$ and $W_\oL{C}$, obtained by $\np \to \nm$. Depending 
on $f_C$, the following invariant building blocks under hard-collinear 
gauge transformations appear
\begin{align}
  \label{eq:def-ell_C}
  \text{lepton :} & & f_C & = l_C & & \to & \ell_C & = W_{\ell C}^\dagger l_C , 
\\
  \text{quark :}  & & f_C & = \xi_C & & \to & \chi_C & 
  = \big[W_{\xi C} W_C\big]^\dagger \xi_C ,
\end{align} 
with analogous building blocks for anti-hard-collinear leptons $\ell_\oL{C}$ 
and quarks $\chi_\oL{C}$ that involve $W_{f\oL{C}}$ and $W_\oL{C}$. The 
collinear fields in the main text refer to these collinear-gauge invariant 
fields including these collinear Wilson lines.

The leading-power Lagrangian of the soft light quark $q_s$ with mass $m_q$ and
the heavy quark $h_\vb$ 
\begin{align}
  \mathcal{L}_s &
  = \bar q_s (i \slashed{D}_s - m_q) q_s
  + \bar h_\vb ( i v \cdot D_s) h_\vb
\end{align}
contains the covariant derivative with soft gauge fields only. It is the
same for \SCETI{} and \SCETII{}.

In addition we need the subleading \SCETI{} interaction involving both, the
soft and hard-collinear light quarks \cite{Beneke:2002ph}
\begin{align}
  \label{eq:SCET-Lag-xiq1}
  \mathcal{L}^{(1)}_{\xi q} &
  = \oL{q}_s(x_-) \, \big[W_{\xi C} W_C\big]^\dagger(x) \, 
    i\slashed {D}_{C\perp } \, \xi_C(x) + \text{h.c.}  
\end{align}
and analogously for anti-hard-collinear fields with the replacements 
$C \to \oL{C}$, $\np \leftrightarrow \nm$ and $x_- \to x_+$.
Further the mass-suppressed Lagrangian 
\cite{Leibovich:2003jd} for the hard-collinear leptons is needed 
\begin{align}
  \mathcal{L}^{(1)}_{m} &
  = m_\ell \, \oL{l}_C \left[i \slashed D_{C\perp}, \, 
  \frac{1}{i\np D_C} \right] \frac{\nps}{2} l_C .
\end{align}

In \SCETI{} the hard-collinear and soft fields are decoupled at 
leading power by the field redefinitions \cite{Bauer:2001yt}
\begin{align}
\label{eq:decoupling-trafo}
  f_C(x)      & = Y_{f+}(x_-) Y_{\text{QCD}+}(x_-) \, f_C^{(0)}(x) , & 
  f_\oL{C}(x) & = Y_{f-}(x_+) Y_{\text{QCD}-}(x_+) \, f_\oL{C}^{(0)}(x) ,
\end{align}
if $f_C$ ($f_\oL{C}$) creates an outgoing antiparticle. If it 
destroys an incoming particle, $\overline{Y}$ instead of $Y$ 
is used.  Here the QED Wilson lines with soft gauge fields 
are defined as follows:
\begin{align}
  \text{outgoing particles:} & &
  Y_{f\pm}^\dagger(x) & 
  = \exp\left(+ieQ_f \int_0^{\infty}  ds\,n_\mp A_s(x+sn_{\mp})\, e^{-\epsilon s}\right),
\\
  \text{outgoing antiparticles:} & &
  Y_{f\pm}(x) & 
  = \exp\left(-ieQ_f \int_0^{\infty}  ds\,n_\mp A_s(x+sn_{\mp})\, e^{-\epsilon s}\right),
\\
  \text{incoming particles:} & & 
  \oL{Y}_{f\pm}(x) &
  = \exp\left(+ieQ_f \int_{-\infty}^0 ds\,n_\mp A_s(x+sn_{\mp})\, e^{ \epsilon s}\right),
\\
  \text{incoming antiparticles:} & &
  \oL{Y}_{f\pm}^\dagger(x) & 
  = \exp\left(-ieQ_f \int_{-\infty}^0 ds\,n_\mp A_s(x+sn_{\mp})\, e^{ \epsilon s}\right),
\end{align}
where $\epsilon$ is introduced to ensure the convergence of the integral.
Similarly, for QCD,
\begin{align}
  \text{outgoing particles:} & &
  Y_{\text{QCD}\pm}^\dagger(x) & = \mathcal{P} 
  \exp\left(+ig_s \int_0^{\infty} ds\,n_\mp  G_s(x+sn_\mp)\, e^{-\epsilon s}\right),
\\
  \text{outgoing antiparticles:} & &
  Y_{\text{QCD}\pm}(x) & = \oL{\mathcal{P}}
  \exp\left(-ig_s \int_0^{\infty} ds\,n_\mp  G_s(x+sn_\mp)\, e^{-\epsilon s}\right),
\\
  \text{incoming particles:} & & 
  \oL{Y}_{\text{QCD}\pm}(x) & = \mathcal{P} 
  \exp\left(+ig_s \int_{-\infty}^0 ds\,n_\mp G_s(x+sn_\mp)\, e^{\epsilon s}\right),
\\
  \text{incoming antiparticles:} & &
  \oL{Y}_{\text{QCD}\pm}^\dagger(x) & = \oL{\mathcal{P}}
  \exp\left(-ig_s \int_{-\infty}^0 ds\,n_\mp G_s(x+sn_\mp)\, e^{\epsilon s}\right).
\end{align}
The new hard-collinear fields with superscript $(0)$ do not have 
interactions with soft fields at leading power, and after the
decoupling transformation the superscript is dropped. Further, in 
the main text  we omit the label $f$ on $Y_{f\pm}$, whenever $f_C = \ell_C$ is a lepton or anti-lepton, see \refeq{eq:SCETII-def-soft-Y}. 

The \SCETII{} is obtained from \SCETI{} by integrating out the 
modes with hard-collinear virtuality. The leading-power  Lagrangian 
of a collinear fermion $f_c$ in \SCETII{} now includes
its mass $m_f$ \cite{Beneke:2002ni, Leibovich:2003jd}, 
\begin{align}
  \label{eq:SCET2-Lag-xi0}
  \mathcal{L}_f^{(0)} &
  = \oL{f}_c \left[ i\nm D_c + (i \slashed{D}_{c\perp} - m_f)
      \frac{1}{i\np D_c}(i \slashed{D}_{c\perp} + m_f) \right] \frac{\nps}{2} f_c \,.
\end{align}
The \SCETII{} covariant derivative 
$i D_c = i \partial + e Q_f A_c + g_s G_c$ 
does not contain the soft gauge field, in distinction to the case of 
\SCETI{} in \refeq{eq:cov-dev-SCETI-nm}, because the interaction between a 
single soft mode and collinear modes would necessarily create a mode with 
hard-collinear virtuality.  Thus in \SCETII{} there is no need to perform 
a decoupling transformation to achieve factorization of soft and 
collinear sectors. The gauge-invariance of the 
\SCETII{} operators under collinear gauge transformations is achieved 
analogously to the case of \SCETI{} with the help of collinear QED and 
QCD Wilson lines $W_{fc}$ and $W_c$, respectively, which are obtained 
from \refeq{eq:def-C-wl-QED} and \refeq{eq:def-C-wl-QCD} by the replacement 
$C \to c$. The collinear gauge-invariant building block $\ell_c$ for 
leptons is then formed analogously to \refeq{eq:def-ell_C}. 
 
The collinearly gauge-invariant building block 
\begin{align}
  \calA_{c\perp}^\mu &
  = e \left[A_{c\perp}^\mu 
    - \frac{i \partial_\perp^\mu \np A_c}{i \np\partial} \right] ,
\end{align}
of the collinear QED gauge field appears in \SCETII{} operators 
as well as the anti-collinear version 
defined through the replacements $c \to \oL{c}$ and $\np \to \nm$.
As a consequence of the decoupling of hard-collinear modes, the 
\SCETII{} soft fields are dressed by soft Wilson lines. For the 
$\oL{q}_s [\ldots]h_v$ bilinear they combine to the finite-distance
Wilson line $Y(x,y)$ introduced in \refeq{eq:SCETII-def-soft-finite-Y}. 
The relation of this finite-distance Wilson line $Y(v\nm,0)$ 
that appears in \refeq{eq:def-SCET-II-ops-1}, \refeq{eq:def-SCET-II-ops-2}
to the infinite-distance 
Wilson lines defined above can be seen from the identity 
\begin{align}
  \oL{q}_s(v \nm) Y(v\nm,0)  h_\vb(0) & 
  = \Big[\oL{q}_s(v \nm) \oL{Y}_{q+}(v\nm) \oL{Y}_{\text{QCD}+}(v\nm)\Big]
    \Big[\oL{Y}_{\text{QCD}+}^\dagger(0) \oL{Y}_{q+}^\dagger(0) h_\vb(0)\Big] \,.
\end{align}
These soft Wilson lines are necessary to maintain invariance of 
the non-local soft
operators under soft QCD and QED gauge transformations.

%
%
%--------+---------+---------+---------+---------+---------+---------+---------+
\subsection{Renormalization conventions}
\label{app:SCET-renorm}

The convention of the operator renormalization follows 
\cite{Beneke:2017ztn}. The renormalization condition of the matrix 
element of an operator $\OpI_{P}$ with a suitable choice of external 
states denoted by $\big< \ldots \big>$ is given as
\begin{align}
  \label{eq:renorm-cond}
  \left\langle \OpI_P(\{\phi_\text{ren}\},\{g_\text{ren}\}) \right\rangle_\text{ren} &
  = \sum_Q Z_{PQ}\prod_{\phi\in Q} Z_\phi^{1/2}\prod_{g\in Q} Z_g 
  \langle \OpI_{Q, \text{bare}}(\{\phi_\text{ren}\},\{g_\text{ren}\}) \rangle \,.
\end{align}
Here $\phi_\text{ren}$ and $g_\text{ren}$ denote the renormalized 
fields and parameters, such as coupling constant or masses, out of 
which the operator $\OpI_Q$
is composed. The mixing of operators $\OpI_Q$ into $\OpI_P$ is given
by the corresponding entries $Z_{PQ} = ({\bf Z})_{PQ}$ of the renormalization
matrix of the operators. The anomalous dimension matrix ${\bf \Gamma}$ is
defined as
\begin{align}
  \label{eq:gamma}
  {\bf \Gamma} & 
  = -\left(\frac{d}{d\ln\mu}{\bf Z}\right)\,{\bf Z}^{-1} 
  = {\bf Z} \frac{d}{d\ln\mu}{\bf Z}^{-1},
\end{align} 
which implies renormalization group equations for the operators and Wilson
coefficients as 
\begin{align}
  \label{eq:def-RG-op_and_C}
  \frac{d}{d\ln\mu} {\cal O}_{P} & 
  = - \sum_Q \Gamma_{PQ} {\cal O}_{Q} , &
  \frac{d}{d\ln\mu} C_P &
  =   \sum_Q \Gamma_{QP} C_Q .
\end{align}
Operators in SCET depend on continuous variables such as light-cone positions 
or their Fourier-conjugates, the momentum fractions.
These are included in the index $P$. When necessary, the sums in the above 
equations should therefore be understood as convolutions either in position 
or in momentum space. 

The renormalization condition of the one-loop matrix element of an operator 
$\OpI_P$ is determined from
\begin{align}
  \text{finite} &
  = \langle \OpI_{P,\text{bare}} \rangle_\text{1-loop} 
  + \sum_Q \left[ \delta Z_{PQ} 
      + \delta_{PQ} \left( \frac{1}{2} \sum_{\phi\in P} \delta Z_\phi
                         + \sum_{g\in P} \delta Z_g \right)
    \right] \langle \OpI_{Q,\text{bare}} \rangle_\text{tree} \,,
\end{align}
where at one-loop the renormalization constants are expanded as
$Z_{PQ} = \delta_{PQ} + \delta Z_{PQ}$ and $Z_i = 1 + \delta Z_i$ for
$i = \phi, g$.

%
%
%--------+---------+---------+---------+---------+---------+---------+---------+
\subsection[\SCETI{} renormalization]
{\boldmath  \SCETI{} renormalization}
\label{app:SCET-1-adm}

The one-loop anomalous dimension of the \SCETI{} operators $\OpI_{9,10}$ and
$\OpI_{\oL{9},\oL{10}}$ of \refsec{sec:SCET-1-ops} can be obtained
by adapting \cite{Beneke:2017ztn, Beneke:2018rbh, Beneke:2019kgv} to the 
case of QED. These operators contain two collinear sectors and one soft 
sector, the latter consisting of the heavy quark field $h_\vb$. For example, 
in $\OpI_{9,10}$, the collinear sector in $\np$ direction comprises the 
light quark field $\chi_C$ and the lepton field $\ell_C$, thus representing 
an $F=2$ operator of the form considered in \cite{Beneke:2017ztn}, 
whereas the second collinear sector (called anti-collinear)
in the $\nm$ direction contains the single anti-lepton $\ell_\oL{C}$.
The one-loop diagrams that determine the renormalization constant, fall 
into two classes: $i)$ with soft photon exchange between all four external 
lines\footnote{Soft
photon exchange within a single collinear direction vanishes.} and 
$ii)$ from collinear photon exchange that is restricted to each collinear 
sector separately. The dependence of the soft and collinear
contributions on the infrared regulator cancels in their sum. 

The \SCETI{} operators do not mix due to their particular Dirac
structure. Besides the cusp part of the anomalous dimension given 
in \refeq{eq:SCET1-ADM-cusp}, the remainder of the QED part of the 
one-loop anomalous dimension in \refeq{eq:SCET1-ADM-remd} reads 
\begin{equation}
  \label{eq:SCET1-ADM-remainder}
\begin{aligned}
  \gamma_i(x,& \, y) =
  \delta(x - y) \, \Big[
    Q_\ell^2 (4\ln x - 6) + Q_\ell Q_q 4\ln \oL{x} + Q_q^2 (4 \ln \oL{x} - 5)
  \Big]
\\ & + 4 Q_\ell Q_q \left[
     \theta(x - y) \frac{\oL{x}}{\oL{y}} \left(\left[\frac{1}{x - y}\right]_+ - 1 \right) 
   + \theta(y - x) \frac{x}{y} \left(\left[\frac{1}{y - x}\right]_+ - 1 \right)
  \right] .
\end{aligned}
\end{equation} % CB: checked %RS STILL TO CHECK
We use bar-notation for momentum fractions, $\oL{x}\equiv 1-x$ etc.
The plus distribution is defined as
\begin{equation}
  \left[f(x, y)\right]_+ 
  = f(x, y) -\delta(x - y)\, \int_0^1 dz f(z, y) \,.
\end{equation}
This result can be obtained from \cite{Beneke:2017ztn}. However, there 
the $N$-jet operator does not contain soft fields. Introducing the heavy 
quark field as a possible building block modifies the
soft, but not the collinear one-loop contribution. Since soft loops do
not change the momentum fraction of the collinear fields, only the part 
proportional to $\delta(x-y)$ is modified. For diagrams with soft photon 
exchange with the soft heavy quark the coefficient of the cusp logarithm 
is one half of that for soft loops connecting two different collinear 
directions, and one has to replace $s_{ij} \to \mu (\nm \vb) (\np p_j)$. 
In addition, the colour generators are replaced by appropriate electric charge 
factors and one has to contract spinor indices according to the 
definition of the operator. It is permitted to use four-dimensional 
identities to reduce the basis of Dirac matrices. We also checked 
this procedure by explicit computation of the anomalous dimension 
of the \SCETI{} operators without using results of \cite{Beneke:2017ztn}.

The general solution of the RGE \refeq{eq:SCET1-rge} of the hard
functions due to the cusp anomalous dimension, neglecting the running of
$\alE$, can be written as 
\begin{align}
  \label{eq:general-RG-sol-SCETI}
  \frac{H_i(u,\mu)}{H_i(u, \mu_b)} & 
  = \left( \frac{m_{B_q}}{\mu_b} \right)^{\!a_\Gamma}
    \exp\left[
      -\int_{\alS(\mu_b)}^{\alS(\mu)} d\alS 
        \frac{\Gamma_\text{cusp}^\text{I}(\alS, \alE)}{\beta_s(\alS)}
        \int_{\alS(\mu_b)}^{\alS} \frac{d\alS'}{\beta_s(\alS')}
    \right],
\end{align}
with 
\begin{align}
  a_\Gamma &
  = \int_{\alS(\mu_b)}^{\alS(\mu)} d\alS
       \frac{\Gamma_\text{cusp}^\text{I}(\alS, \alE)}{\beta_s(\alS)}
\end{align}
and $\beta_s(\alS)$ the beta function of the strong coupling 
as defined in \refeq{eq:coupl-RG-func}.

%
%
%--------+---------+---------+---------+---------+---------+---------+---------+
\subsection[\SCETII{} renormalization]
{\boldmath  \SCETII{} renormalization}
\label{app:SCET-2-adm}

The regularization of UV and IR divergences in \SCETII{} is not 
as simple as in \SCETI{}. First we note that the presence of the 
lepton mass in the collinear and 
anti-collinear contributions in \reffig{fig:SCET2-adm-coll} 
and \reffig{fig:SCET2-adm-coll-aChi} does not regularize
all IR divergences. We therefore use an IR regulator inspired by introducing
an off-shellness for external lines in corresponding diagrams in \SCETI{} 
before performing the decoupling transformation. In the diagrams with 
soft photon
exchange in \reffig{fig:SCET2-adm-soft}, the IR regulator has to be 
introduced by modifying the $i 0^+$ prescription in a manner consistent 
with \SCETI{}, i.e.~the appropriate Wilson lines are regularized 
by a parameter related to the off-shell
momentum $p_\ell^2$ of the original hard-collinear fields before the decoupling
of the soft modes. This implies a regulator in the propagators that originate
from soft Wilson lines $Y_\pm$ in the following way:
\begin{equation}
\begin{aligned}
  \text{diagram 3}: & &
  [\np\ell - i 0^+]  & \;\; \to \;\;
  [\np\ell - \delta_\oL{\ell} - i 0^+]
\\ 
                    & &
  [\nm\ell - i 0^+] & \;\; \to \;\;
  [\nm\ell - \delta_\ell - i 0^+] 
\\
  \text{diagram 4}: & &
  [\np\ell + i 0^+] & \;\; \to \;\;
  [\np\ell + \delta_\oL{\ell} + i 0^+]  
\\
                    & &
  [\nm\ell + i 0^+] & \;\; \to \;\;
  [\nm\ell + \delta_\ell + i 0^+] 
\\
  \text{diagram 5}: & &
  [\np\ell - i 0^+] & \;\; \to \;\;
  [\np\ell - \delta_\oL{\ell} - i 0^+] 
\\
  \text{diagram 6}: & &
  [\np\ell - i 0^+] [\nm\ell + i 0^+] & \;\; \to \;\;
  [\np\ell - \delta_\oL{\ell} - i 0^+] [\nm\ell + \delta_\ell + i 0^+] \,,
\end{aligned}
\end{equation} % CB: checked
where $\ell$ denotes the loop momentum, and 
$\delta_\ell \equiv p_\ell^2/\np p_\ell$ and $\delta_\oL{\ell} \equiv p_\oL{\ell}^2/\nm p_\oL{\ell}$. The cancellation
of the IR regulators then takes place between the soft-photon exchange 
diagrams 3, 4 and 5 of \reffig{fig:SCET2-adm-soft}. Another cancellation 
occurs between diagram~6 of \reffig{fig:SCET2-adm-soft} and the 
collinear/anti-collinear diagrams in \reffig{fig:SCET2-adm-coll}
and \reffig{fig:SCET2-adm-coll-aChi}.

The results for the renormalization constants 
entering \refeq{eq:SCET-2-soft-RC} are 
\begin{equation}
  \label{eq:Z_s^QED}
\begin{aligned}
  Z_s^\text{QED} (\omega, \omega') &
  = \delta(\omega' - \omega) \left[ 
    Q_q^2 \left( \frac{1}{\epsilon^2} + \frac{1}{\epsilon} \ln\frac{\mu^2}{\omega^2}
               - \frac{1}{\epsilon} \frac{5}{2} \right)
  + 2 Q_\ell Q_q \left( \frac{1}{\epsilon^2} + \frac{1}{\epsilon}\ln\frac{\mu^2}{\omega^2} 
    \right) \right]
\\ & 
  - Q_q (Q_q + Q_\ell) \frac{2}{\epsilon} F(\omega, \omega')
\end{aligned}
\end{equation} % CB: checked
for the QED contribution, while the universal QCD part coincides with 
the well-known expression\footnote{See erratum at the end of the paper.}
\cite{Lange:2003ff}
\begin{align}
  \label{eq:Z_s^QCD}
  Z_s^\text{QCD} (\omega, \omega') &
  = \delta(\omega' - \omega)\, C_F \left[
     \frac{1}{\epsilon^2} + \frac{1}{\epsilon}\ln\frac{\mu^2}{\omega^2}
    - \frac{1}{\epsilon}\frac{5}{2} \right] 
  - \frac{2 C_F}{\epsilon} F(\omega, \omega')\,.
\end{align} % CB: checked
The common function $F$ is
\begin{align}
  \label{eq:F(om,om')}
  F(\omega, \omega') &
  \equiv \left[
      \frac{\theta(\omega-\omega')}{(\omega-\omega')}
    + \frac{\omega\, \theta(\omega'-\omega)}{\omega'(\omega'-\omega)}\right]_+ .
\end{align} % CB: checked
The plus distribution for the soft convolutions is defined as
\begin{equation}
 \int_0^\infty d\omega' \left[f(\omega,\omega')\right]_+ \, g(\omega') = \int_0^\infty d\omega' f(\omega,\omega')\left[g(\omega')-g(\omega)\right]\,.
 \end{equation}
The collinear renormalization constants in \refeq{eq:SCET2-Z-mChi}
and \refeq{eq:SCET2-Z-aChi} are 
\begin{align}
  \label{eq:Z^1_ac}
  Z_\oL{c}^{(1)} & 
  =  -\frac{\alE}{4\pi} Q_\ell^2 \left[
       \frac{1}{\epsilon^2}
     + \frac{1}{\epsilon} \ln\frac{-\mu^2}{(\nm p_\oL{c})  (\np p_{c})} 
     + \frac{3}{2\epsilon} \right] ,
\\
  \label{eq:Z^1_chi}
  Z_{m\chi}^{c,(1)} &
  = -\frac{\alE}{4\pi} Q_\ell^2 \left[
       \frac{1}{\epsilon^2} 
     + \frac{1}{\epsilon} \ln\frac{-\mu^2}{(\nm p_\oL{c})  (\np p_{c})} 
     - \frac{3}{2\epsilon} \right] ,
  % CB checked
\\ 
  \label{eq:Z^1_Achi}
  Z_{\calA\chi}^{c,(1)}(w, w') &
  = -\frac{\alE}{4\pi} Q_\ell^2 \left\{
     \left[ \frac{1}{\epsilon^2} 
         + \frac{1}{\epsilon} \ln\frac{-\mu^2}{w^2 (\nm p_\oL{c})  (\np p_{c})}
         + \frac{3}{2\epsilon} \right] \delta(w-w') 
    + \frac{1}{\epsilon} \gamma_{\calA\chi,\calA\chi}(w, w') \right\}  .
\end{align}
We note that $Z_{m\chi}^{c,(1)}$ receives also a contribution due 
to the renormalization of the lepton mass appearing in the 
definition $\OpII_{m\chi}^{A1}$. The anomalous dimension 
$\gamma_{\calA\chi,\, \calA\chi}$,
\begin{align}
  \label{eq:gam-calAchi}
  \frac{\gamma_{\calA\chi,\calA\chi}(w,w')}{2} &
  = - w
    + \theta[w' - \oL{w}] \frac{w' - \oL{w}}{w'}
    + \theta[\oL{w} - w'] \frac{1}{\oL{w'}} \left(\frac{w'}{\oL{w}} - w - w' \right),
\end{align} % CB checked % RS checked
can be extracted from the results given in \cite{Beneke:2017ztn, 
Beneke:2018rbh}, and is due to the two diagrams in \reffig{fig:SCET2-adm-coll-aChi}
with $A_{c\perp}$ in the loop.

%
%
%
%--------+---------+---------+---------+---------+---------+---------+---------+
\section{\boldmath SCET operators}
\label{sec:operator}

In this Appendix, we discuss the construction of the \SCETII{} 
operator basis for the power-enhanced correction to 
$B_q\to \mu^+\mu^-$. We first note that classifying the 
operators in \SCETI{} $\to$ \SCETII{} matching is substantially 
more complicated than classifying \SCETI{} operators in the 
matching of the effective weak Hamiltonian to \SCETI{}. 
The reason is that although \SCETI{} operators are non-local, 
the non-locality of collinear fields is related to the 
$\mathcal{O}(1)$ inverse derivative $1/(i\np \partial)$. 
For a \SCETI{} operator that scales as $\lambda^n$, the power of 
$n$ can therefore never be smaller than the scaling of the 
products of fields contained in the operator. In \SCETI{} $\to$ 
\SCETII{} matching, however, integrating out the hard-collinear 
modes leads to non-locality of soft fields related to 
$1/(i\nm \partial_s) \sim 1/\lambda^2$ from the hard-collinear 
propagators, hence the above statement does not hold. 
Part of the construction of the operator basis therefore consists 
in constraining the number of times such inverse derivatives 
can occur in the operator. A systematic procedure for 
constructing the \SCETII{} operator basis when there is a single 
collinear and a soft sector, has been described in 
\cite{Beneke:2003pa}. We adapt this procedure developed for 
heavy-to-light form factors to the present 
situation. Here we are interested in \SCETII{} operators with 
the scaling $\lambda^{10}$ of the power-enhanced $B_q\to\mu^+\mu^-$ 
amplitude, which can arise in the matching 
of $Q_{9,10}$ and $Q_7$.

We first consider the possibility of a \SCETI{} operator without 
hard-collinear or hard-anti-collinear fields. Since the collinear, 
anti-collinear and soft fields do not interact, such 
operators must have the flavour quantum numbers of the 
external state to have non-vanishing overlap. The operator 
with the smallest $\lambda$ scaling is 
$\oL{q}_s \Gamma_s h_v \,\oL{\ell}_c\Gamma \ell_{\oL{c}} \sim 
\lambda^{10}$. However, the chiral structure of the effective 
weak Hamiltonian implies that the operator has overlap with a
pseudoscalar $B$ meson only after a chirality flip by the 
lepton mass term. Since in matching to \SCETI{} the 
$\lambda^2$-suppression cannot be compensated by 
$1/(i\nm \partial_s) \sim 1/\lambda^2$, this results in 
a $\lambda^{12}$ operator, which is, in fact, the operator
\refeq{eq:Omop} that contributes to the standard non-enhanced 
$B_q\to\mu^+\mu^-$ amplitude in the absence of QED effects. 
We conclude that any  \SCETI{} operator relevant to the 
power-enhanced amplitude must contain 
at least one hard-collinear field, which leads to non-trivial 
\SCETI{} $\to$ \SCETII{} matching. 

Any relevant \SCETI{} operator must contain the heavy quark field 
$h_v$, at least one hard-collinear field as concluded above, and 
at least one anti-collinear or anti-hard-collinear field. As always, 
the corresponding operators with (hard-) collinear and 
anti (hard-collinear) modes interchanged also exist. 
The following discussion is phrased for 
the first case and holds with obvious modifications for the second. 
With the power counting of fields as 
given in \reftab{tab:fields-scaling}, the field content of the 
operators with the smallest $\lambda$ scaling is 
\begin{equation} 
\label{oplist} 
\begin{array}{ll} 
\lambda^5: \quad & \oL{\chi}_{ hc}h_v 
\mathcal{A}_{{\oL{hc}}\perp}\,,\\[0.2cm] 
\lambda^6: & 
1)\hspace*{0.2cm}  \oL{\chi}_{ hc}h_v \oL{\ell}_{hc}
\ell_{\oL{hc}}, \quad  
2)\hspace*{0.2cm} \oL{\chi}_{c}h_v \mathcal{A}_{{\oL{hc}}\perp},\quad 
3)\hspace*{0.2cm}  \oL{\chi}_{ hc}h_v 
\mathcal{A}_{{\oL{c}}\perp},\\[0.2cm]  
& 4)\hspace*{0.2cm} 
\oL{\chi}_{ hc}h_v \mathcal{A}_{{\oL{hc}}\perp} \times 
\{ \mathcal{A}_{{hc}\perp},  \mathcal{A}_{{\oL{hc}}\perp} \},
\end{array} 
\end{equation} 
and so on. For every operator there is another operator 
with (hard-) collinear and anti(-hard)-collinear fields 
exchanged. We then need to determine the $\lambda$ suppression  
factors incurred when the hard-collinear fields convert into  
soft and collinear fields through \SCETI{} time-ordered products. 
From the discussion below it will become clear that operators 
that scale as $\lambda^7$ cannot match to
\SCETII{} operators with $\lambda^{10}$ scaling.
 
We begin with the derivation of the \SCETII{} operator basis 
for the matching of the semi-leptonic operators  $Q_{9,10}$ 
in the effective weak Hamiltonian, which is the case discussed 
in the main text. We first note that the operator with field 
content  $\bar\chi_{ hc}h_v \mathcal{A}_{{\oL{hc}}\perp}$ 
cannot be generated at $\mathcal{O}(\lambda^5)$ from the 
operators $Q_{9,10}$. To obtain  $\bar\chi_{ hc}h_v 
\mathcal{A}_{{\oL{hc}}\perp}$ from hard matching to \SCETI{} 
the lepton fields in  $Q_{9,10}$ must be contracted and a 
anti-hard-collinear photon field attached. For the vectorial 
operators $Q_{9,10}$ this results in $\partial^\mu A_\mu$ or 
$m_l \,\sigma^{\mu\nu} F_{\mu\nu}$, which both lead to 
\SCETI{} operator that count as $\lambda^7$. We will come 
back to the $\mathcal{O}(\lambda^5)$ operator below, where we 
briefly discuss the operator basis in the matching of $Q_7$, 
but for now we proceed with the $\mathcal{O}(\lambda^6)$ operators 
in the above list. 

The operators 2) and 3) can immediately be discarded, since 
the collinear (case 2)) or anti-collinear (case 3)) field 
content does not have the correct lepton flavour number to 
overlap with the $\ell^+\ell^-$ state. For example, 
the collinear antiquark field $\oL{\chi}_c$ can never turn 
into a collinear state with lepton number one. In case of 
operator 4) adding more fields to the $\mathcal{O}(\lambda^5)$ 
operator does not allow one to overcome the above suppression.
This leaves the operator $\oL{\chi}_{ hc}h_v \oL{\ell}_{hc}
\ell_{\oL{hc}}$ as the only candidate  $\mathcal{O}(\lambda^6)$ 
\SCETI{} operator.

It is easy to see that this operator does contribute at 
$\mathcal{O}(\lambda^{10})$ to the amplitude in question. 
At tree-level, this involves the \SCETI{} $\to$ \SCETII{} matching 
relations 
\begin{equation}
\ell_{hc} \,\stackrel{\lambda}{\to} \,\ell_{c}, 
\qquad 
\ell_{\oL{hc}} \,\stackrel{\lambda}{\to} \,\ell_{\oL{c}},
\qquad 
 \chi_{hc} \,\stackrel{\lambda^2}{\to}\,
\frac{1}{in_-\partial}\,
Q_q e \Slash A_{c \perp }\,\frac{\slash2 n_-}{2}\,q_s
\label{eq:Q9treesplit}
\end{equation}
and their hermitian conjugates.
The last relation is taken from Section~3.2.1 
of~\cite{Beneke:2003pa}, and describes the splitting of the 
hard-collinear quark field into a collinear photon and 
a soft quark.\footnote{For simplicity of notation, all collinear 
and soft Wilson lines are set to unity here. They can be restored 
unambiguously by making the operator invariant under the 
\SCETII{} gauge symmetries.} The power of $\lambda$ indicated 
above the arrow gives the $\lambda$ suppression from the left- 
to the right-hand side as a consequence of \SCETI{} $\to$ 
\SCETII{} matching. 
Thus, the $\mathcal{O}(\lambda^6)$ 
\SCETI{} operator turns into a $\mathcal{O}(\lambda^{10})$ \SCETII{} 
operator with the field content of  
$\widetilde{\OpII}_{\calA\chi}^{B1}$ defined in 
(\ref{eq:def-SCET-II-ops-2}). The present discussion 
in fact corresponds to the matching shown in the 
\SCETI{} column of Figure~\ref{fig:heuristic-scheme} and 
shows that one must obtain an inverse derivative 
factor $1/in_-\partial \sim 1/\lambda^2$, which corresponds 
to the factor $1/\omega$ in the tree-level matching 
coefficient (\ref{eq:Jm1}). Another way to obtain a 
$\mathcal{O}(\lambda^{10})$ \SCETII{} operator consists of 
\begin{equation}
\ell_{hc} \,\stackrel{1}{\to} \,\ell_{hc}, 
\qquad 
\ell_{\oL{hc}} \,\stackrel{\lambda}{\to} \,\ell_{\oL{c}},
\qquad 
 \chi_{hc} \,\stackrel{\lambda}{\to}\,
\frac{1}{in_-\partial}\,
Q_q e  \Slash A_{hc \perp }\,\frac{\slash2 n_-}{2}\,q_s
\label{eq:Q9treesplit2}
\end{equation}
followed by the fusion 
\begin{equation}
\oL{\ell}_{hc} + \Slash A_{hc \perp } \,\stackrel{\lambda^2}{\to}\, 
m_\ell \,\oL{\ell}_c
\end{equation}
through a hard-collinear one-loop diagram.

To find the basis of the \SCETII{} operators, these considerations 
have to be generalized to arbitrary loop order. To this end 
we follow \cite{Beneke:2003pa} and classify all possible building 
blocks according to their scaling in $\lambda$, canonical dimension 
$d$ and transformation property under type-III reparametrization 
transformations 
\begin{align}
  \nm & \to \alpha \nm , &
  \np & \to \alpha^{-1} \np,
\end{align}
of the reference vectors, which must be preserved in the matching 
due to reparametrization symmetry of SCET. Accounting for the 
flavour quantum numbers of the initial and final state in the 
$B_q\to\ell^+\ell^-$ process, we can write the possible \SCETII{} 
operators in the form 
\begin{equation}
  \calO^{(\alpha)}  = 
\left[\mbox{objects}\right]\times
\big(\oL{q}_s \, \Gamma^{(\alpha)}_s \, h_\vb   \big)
    \big(\oL{\ell}_c\, \Gamma_c \, \ell_{\oL{c}}\big), 
\qquad\quad
  \alpha  = \pm 1, 0
\label{eq:SCETIIprimary}
\end{equation}
with Dirac structures given in \reftab{tab:Dirac}. The 
``objects'' can be chosen from the factors, operators and 
field products listed in Table~\ref{tab:Collinear-sector}.

%%%%%%%%%%%%%%%%%%%%%%%%%%%%%%%%%%%%%%%%%%%%%%%%%%%%%%%%%%%%%%%%%%
\begin{table}[t]
\centering{}%
\renewcommand{\arraystretch}{1.4}
\begin{tabular}{|llr|}
\hline
Class & Elements   & $\alpha$  
\\
\hline 
  $\Gamma_c$
& $1,\,\gamma_5,\,\gamma_\perp^\mu$
& 0
\\
  $\Gamma_s^{(0)}$
& $1,\, \gamma_5,\, \gamma_\perp^\mu,\, \gamma_\perp^\mu \gamma_5,\, 
   \sigma_\perp^{\mu\nu}, \, \nps\nms$ 
& 0
\\
  $\Gamma_s^{(+1)}$ 
& $\nms,\, \nms \gamma_5,\, \nms \gamma_\perp^\mu$
& 1
\\
  $\Gamma_s^{(-1)}$
& $\nps,\, \nps \gamma_5,\, \nps \gamma_\perp^\mu$
& $-1$
\\
\hline 
\end{tabular}
\renewcommand{\arraystretch}{1.0}
\caption{\label{tab:Dirac} 
\small 
Dirac matrices and their scaling $\alpha^n$ under ``boost'' or 
type-III RPI transformations. 
}
\end{table}
%%%%%%%%%%%%%%%%%%%%%%%%%%%%%%%%%%%%%%%%%%%%%%%%%%%%%%%%%%%%%%%%

%%%%%%%%%%%%%%%%%%%%%%%%%%%%%%%%%%%%%%%%%%%%%%%%%%%%%%%%%%%%%%%%%
\begin{table}[t]
\centering{}%
\renewcommand{\arraystretch}{1.4}
\begin{tabular}{|c|c|rrr|}
\hline
  $\calO_\text{A}$ & Object &  $\lambda$  &  $\alpha$  &  $d$
\\
\hline 
  I    & $(\nm \partial_s)^{-1}$  & $-2$ & $-1$ & $-1$
\\
  II   & $(\np \partial_c)^{-1}$  & 0    & $+1$ & $-1$
\\
\hline
  III  & $\partial_\perp$, $A_{c\perp}$, $A_{s\perp}$, $m_\ell$, $m_q$  & 2  & 0  & 1
\\
  IV   & $\np \partial_s$, $\np A_s$  & 2  & $-1$  & 1
\\
  V    & $\nm \partial_c,$ $\nm A_c$  & 4  & $+1$  & 1
\\
\hline
  VI   & $\oL{f}_c \frac{\nps}{2} \Gamma_c f_c$  & 4  & $-1$  & 3
\\
  VII  & $\oL{f}_s \Gamma_s^{(-1)} f_s$  & 6  & $-1$  & 3
\\
  VIII & $\oL{f}_s \Gamma_s^{(+1)} f_s$  & 6  & $+1$  & 3
\\
  IX   &  $\oL{f}_s \Gamma_s^{(0)} f_s$  & 6  & 0  & 3
\\
\hline 
\end{tabular}
\renewcommand{\arraystretch}{1.0}
\caption{\label{tab:Collinear-sector} 
  \small 
Building blocks that can be added to (\ref{eq:SCETIIprimary}). 
The column $\lambda$ denotes the $\lambda$ scaling of the object, 
$\alpha$ its scaling under type-III reparametrizations, and 
$d$ its canonical dimension. The Dirac matrices $\Gamma$ are 
defined in \reftab{tab:Dirac}. The counting refers to matching 
of the hard-collinear sector.}
\end{table}
%%%%%%%%%%%%%%%%%%%%%%%%%%%%%%%%%%%%%%%%%%%%%%%%%%%%%%%%%%%%%%%%%%%%%%%%%%%%%%%%

The \SCETI{} operator $\oL{\chi}_{ hc}h_v \oL{\ell}_{hc}
\ell_{\oL{hc}}$, which is generated in the matching of 
$Q_{9,10}$ to \SCETI{}, has $d=6$ and boost scaling $\alpha=0$. 
Requiring these to match those of (\ref{eq:SCETIIprimary}), 
allows us to solve for $n_1$ and $n_2$, the number of times 
the objects I and II appear in (\ref{eq:SCETIIprimary}), in the 
form
\begin{eqnarray} 
\label{eq:n1} 
n_1 &=& \frac{n_3+\alpha}{2} + n_5+n_6+n_7+2 n_8+\frac{3}{2}n_9\,, 
\\[0.1cm] 
\label{eq:n2} 
n_2 &=& \frac{n_3-\alpha}{2} + n_4+2 n_6+2 n_7+
n_8+\frac{3}{2}n_9\,. 
\end{eqnarray} 
Eliminating $n_1, n_2$ in the expression for the 
$\lambda$ scaling $[\lambda]$ of (\ref{eq:SCETIIprimary}), we 
obtain 
\begin{eqnarray} 
\label{eq:nlam} 
[\lambda] &=& 10-\alpha+n_3+2 n_4+2 n_5+2 n_6+4 n_{7}+2 n_8
+ 3 n_9
\end{eqnarray} 
in terms of the number of times $n_i$ the objects III to IX 
appear in (\ref{eq:SCETIIprimary}). Since we are looking 
for solutions with $[\lambda]=10$, the following cases arise:
\begin{itemize}
\item $\alpha=-1$: Since all $n_i$ are positive and non-negative 
integers, solutions to (\ref{eq:nlam}) have at least 
$[\lambda]=12$. This case does not contribute to the 
power-enhanced amplitude. 
\item $\alpha=0$: All $n_i$ must be zero for a  $[\lambda]=10$ 
solution. 
\item $\alpha=+1$: The solution  $[\lambda]=9$, and 
$n_i=0$ for all $n_i$ is excluded, because $n_{1,2}$ must 
be integer. This leaves the $[\lambda]=10$ solution with   
$n_3=1$, $n_i=0$ for $i=4,\ldots,9$. In this case further 
$n_1=1$, $n_2=0$.
\end{itemize}
Thus we identified the following two $\mathcal{O}(\lambda^{10})$ 
\SCETII{} operators:
\begin{align}
  \calO^{(0)} &
  = \big[\oL{q}_s\, \Gamma^{(0)}_s \, h_\vb\big]
    \big[\oL{\ell}_c\, \Gamma_c \, \ell_{\oL{c}}\big] \,,
\\
  \calO^{(+1)} &
  = \frac{\calO_\text{III}}{in_- \partial_s} 
    \big[\oL{q}_s \, \Gamma^{(+1)}_s \, h_\vb \big]
    \big[\oL{\ell}_c\, \Gamma_c \, \ell_{\oL{c}}\big]\, .
\end{align}
Note that this analysis implicitly assumed that in the 
anti-collinear direction the \SCETI{} $\to$ \SCETII{} 
matching was trivial, $\ell_{\oL{hc}} \to \ell_{\oL{c}}$. 
This is justified, since any non-trivial matching would 
lead to further $\lambda$ suppression.

At this point we have to invoke helicity conservation, 
which implies that for $\calO^{(0)}$ only 
$\Gamma_c=\gamma_\perp^\mu$ has a non-vanishing matching 
coefficient to any order in $\alE$. However, this Dirac structure 
cannot contribute to the decay rate of a (pseudo-) scalar 
$B$-meson, which excludes this operator. 

In the case of $\calO^{(+1)}$, the helicity structure 
of the lepton current depends on $\calO_\text{III}$. For 
$\calO_\text{III} =m_\ell$ helicity conservation 
implies $\Gamma_c\in \{1,\gamma_5\} \equiv \Gamma^{S/P}$, 
and the operator has a non-vanishing 
matrix element for the $B_q\to \ell^+\ell^-$ transition. For all 
other members of object class III, only $\Gamma_c = 
\gamma_\perp^\mu$ is allowed by helicity conservation. However, 
only $\calO_\text{III} = A_{c\perp}$ has a non-vanishing
matrix element for the (pseudo-) scalar $B$-meson decay. 
In summary, we find the following \SCETII{} operators
with $\lambda^{10}$ suppression:
\begin{align}
  \calO_-^{A1} &
  = m_\ell\,
    \Big[\,\frac{1}{i\nm\partial_s} \oL{q}_s \, \nms \Gamma^{S/P} h_\vb\Big]
    \Big[\oL{\ell}_c\, \Gamma^{S/P} \, \ell_\oL{c}\Big],
\\
  \calO_-^{B1}  & 
  = T_{\mu\nu}
    \Big[\,\frac{1}{i\nm\partial_s} \oL{q}_s\, \nms \Gamma^{S/P} h_\vb\Big]
    \Big[\calA_{c\perp}^\nu\, \oL{\ell}_c\, \gamma_\perp^\mu \, \ell_\oL{c}\Big],
\end{align}
with $\Gamma^{S/P}\in \{1,\, \gamma_5\}$ and $T_{\mu\nu} \in 
\{ g^\perp_{\mu\nu},\, \varepsilon_{\mu\nu}^\perp\}$. When the 
Wilson lines and non-localities are restored these two 
operators correspond to (\ref{eq:def-SCET-II-ops-1}) and 
 (\ref{eq:def-SCET-II-ops-2}) in the main text. The present 
analysis also implies that to any order in the matching, 
only a single power of $1/i\nm\partial_s$ can appear, modified 
by logarithms, together with the leading-twist $B$-meson 
LCDA. Hence, similar to the case of heavy-to-light form factors 
discussed in \cite{Beneke:2003pa}, the power-enhanced correction 
from the operators $Q_{9,10}$ can be expressed in terms of the 
inverse moment $\lambda_B$ and the logarithmic moments, 
(\ref{eq:lambdaB}) and (\ref{eq:sigman}), respectively.
Upon exchanging the collinear and anti-collinear sectors, 
we obtain the corresponding two operators
\begin{align}
  \calO_+^{\oL{A1}}  & 
  = m_\ell\,
    \Big[\,\frac{1}{i\np\partial_s} \oL{q}_s\, \nps \Gamma^{S/P} \, h_\vb\Big]
    \Big[\oL{\ell}_c\, \Gamma^{S/P} \, \ell_\oL{c}\Big],
\\
  \calO_+^{\oL{B1}} & 
  = T_{\mu\nu}
    \Big[\,\frac{1}{i\np\partial_s} \oL{q}_s\, \nps \Gamma^{S/P} \, h_\vb  \Big]
    \Big[\calA_{\oL{c}\perp}^\nu\, \oL{\ell}_c \, \gamma_\perp^\mu \, \ell_\oL{c}\Big].
\end{align}

The all-order analysis of the  $\mathcal{O}(\lambda^5)$ 
operator  $\bar\chi_{hc}h_v \mathcal{A}_{{\oL{hc}}\perp}$ 
from (\ref{oplist}), which arises in \SCETI{} from the 
tree-level matching of $Q_7$, is substantially more complicated.
This is because the operator involves a hard-collinear quark 
and an anti-hard-collinear photon field, which both 
must undergo not-trivial \SCETI{} $\to$ \SCETII{} matching 
to obtain an operator that overlaps with the external 
state of the $B_q\to \ell^+\ell^-$ decay. Since the collinear 
and anti-collinear sector interact with the same soft 
fields, the analysis above must be extended to keep track 
from which sector a soft field and its corresponding 
non-localities arises. In particular, since the collinear 
final state must have lepton number $+1$, and the 
hard-collinear sector of the \SCETI{} does not carry lepton 
number, likewise for the anti-collinear direction, the 
\SCETII{} operator must necessarily contain a soft lepton 
pair, which is generated in \SCETI{} $\to$ \SCETII{} matching. 
We are therefore led to consider operators of the form 
\begin{align}
\label{eq:softleptonop}
\calO(s, t, u) &
  = \left[\mbox{objects}\right]\times
\big[\oL{q}_s(s n_{\pm}) \, \Gamma_s \, h_\vb(0) \big]
    \big[\oL{\ell}_s(t \np) \, \Gamma^\prime_s \, 
    \ell_s(u \nm) \big]  
    \big[\oL{\ell}_c(0) \, \Gamma_c \, \ell_\oL{c}(0) \big]\,,
\end{align}
where here we kept the position arguments to indicate the 
non-locality of the fields. Not counting the ``objects'', this 
operator has dimension $d=9$ and $[\lambda]=16$, 
while the initial operator has $d=5$. For this operator 
to contribute to the power-enhanced amplitude, it must 
contain a number of inverse derivatives $1/(i n_- \partial_s)$ 
acting on soft fields that arise from hard-collinear matching, 
or  $1/(i n_+ \partial_s)$ from the matching of the 
anti-hard-collinear sector. 

That this possibility is indeed realized can be seen from 
tree-level matching. The relevant splitting of the hard-collinear 
quark field is (\ref{eq:Q9treesplit2}) above, followed by 
\begin{equation} 
A_{hc \perp }^\mu  \,\stackrel{\lambda^2}{\to}\,
 \frac{Q_\ell e }{(in_+\partial_c)(in_-\partial_s)}\,
 \Big\{ \oL{\ell}_c \gamma_\perp^\mu \ell_s + {\rm h.c.} \Big\}
\label{eq:Aperpsplit}
\end{equation}
from Section~3.2.2 of~\cite{Beneke:2003pa}. For the splitting 
of the anti-hard-collinear photon, (\ref{eq:Aperpsplit}) 
adapted to the anti-collinear sector applies. All together this 
provides a $\lambda^5$ suppression of the initial operator, 
resulting in $[\lambda]=10$ and the correct dimension. Putting 
everything together, we obtain 
\begin{eqnarray}
\bar\chi_{hc}h_v \mathcal{A}_{{\oL{hc}}\perp} 
\,&\to&\, 
\oL{q}_s i \overleftarrow{n_-\partial}_s^{-1} h_v\,
\Big[\frac{1}{(in_+\partial_c)(in_-\partial_s)}\,
 \oL{\ell}_c \ell_s\Big] \,
\Big[ \frac{1}{(in_-\partial_{\oL{c}})(in_+\partial_s)}\,
 \oL{\ell}_s \ell_{\oL{c}}\Big]\,.
\label{eq:Q7matchII}
\end{eqnarray}
The QED one-loop calculation of the matrix element of $Q_7$ 
performed in \cite{Beneke:2017vpq} corresponds to evaluating the 
one-loop \SCETII{} matrix element of this operator, obtained 
by contracting the soft lepton fields. The endpoint divergence 
in the one-loop integral found there is a consequence of the 
large number of inverse derivative operators in 
(\ref{eq:Q7matchII}). We performed a general analysis 
of \SCETI{} $\to$ \SCETII{} also for $Q_7$ and find that 
the above operator together with those discussed above 
for $Q_{9,10}$ are the only \SCETII{} operators relevant 
to the power-enhanced $B_q\to\ell^+\ell^-$ amplitude.

\section*{Erratum}

The expression for $F(\omega,\omega')$ entering (A.36) and the QED term in (4.15) is incorrect. 
The correct expression differs
from the QCD one for this function. The corrected anomalous dimension
is given in Section 2.2.4 of \cite{Beneke:2022msp}. The anomalous dimension relevant to
$\overline{B}_q\to\ell^+\ell^-$ is obtained from (2.23)
in this reference by setting $Q_{M_1}=-Q_{M_2}\to Q_{\ell}$ and $Q_{\rm sp}=Q_d\to Q_q$.
In \cite{Beneke:2022msp}, the collinear direction has been chosen for the $M_1^+$ meson, which
corresponds to the anti-collinear sector of the anti-lepton in the present work. Therefore
the statement after (2.30) in the published version and arXiv version 1 of \cite{Beneke:2022msp} should read $Q_{M_2}\to -Q_\ell$ rather than $+Q_\ell$ and applies to $\overline{B}_q$ decay
rather than $B_q$. In particular, the correct anomalous dimension
includes support for $\omega<0$. This correction does not affect the cusp terms and hence does not affect other equations and results reported in the present work.

%--------+---------+---------+---------+---------+---------+---------+---------+
%
% References
%
%--------+---------+---------+---------+---------+---------+---------+---------+

\renewcommand{\refname}{R\lowercase{eferences}}

\addcontentsline{toc}{section}{References}

\bibliographystyle{JHEP}

\small

\bibliography{bibliography}

\end{document}